\documentclass{SciPost}

\hypersetup{
    colorlinks,
    linkcolor={red!50!black},
    citecolor={blue!50!black},
    urlcolor={blue!80!black}
}

\usepackage[backend=biber,style=numeric-comp,sorting=none]{biblatex}
\usepackage[bitstream-charter]{mathdesign}
\allowdisplaybreaks
\DeclareSymbolFont{usualmathcal}{OMS}{cmsy}{m}{n}
\DeclareSymbolFontAlphabet{\mathcal}{usualmathcal}

\fancypagestyle{SPstyle}{
\fancyhf{}
\lhead{\colorbox{scipostblue}{\bf \color{white} ~SciPost Physics }}
\rhead{{\bf \color{scipostdeepblue} ~Submission }}

\fancyfoot[C]{\textbf{\thepage}}
}
\newtheorem{theorem}{Theorem}

\newcommand{\ellh}{\widetilde{\ell}}
\newcommand{\chih}{\widetilde{\chi}}

\newcommand{\efill}{\;\;\;\;\;\;\;\;\;\;}
\newcommand{\be}{\begin{equation}}

\newcommand{\ee}{\end{equation}}
\newcommand{\bthe}{\begin{theorem}}
\newcommand{\ethe}{\end{theorem}}
\newcommand{\bea}{\begin{eqnarray}}
\newcommand{\eea}{\end{eqnarray}}
\newcommand{\al}{\alpha}
\renewcommand{\d}{\delta}
\newcommand{\e}{\epsilon}

\newcommand{\g}{\gamma}
\renewcommand{\k}{\kappa}

\newcommand{\la}{\lambda}
\newcommand{\m}{\mu}
\newcommand{\n}{\nu}
\newcommand{\Om}{\Omega}
\newcommand{\om}{\omega}

\newcommand{\hlf}{\frac{1}{2}}

\newcommand{\non}{\nonumber}

\newcommand{\p}{\partial}

\newcommand{\R}{\mathbb{R}}
\newcommand{\rr}{\rightarrow}

\newcommand{\Z}{\mathbb{Z}}

\newcommand{\lp}{\left(}
\newcommand{\rp}{\right)}
\newcommand{\ls}{\left[}
\newcommand{\rs}{\right]}

\newcommand{\mcL}{\mathcal{L}}

\newcommand{\wtK}{\widetilde{K}}
\newcommand{\wtell}{\widetilde{\ell}}
\newcommand{\wtL}{\widetilde{L}}
\newcommand{\wtmcL}{\widetilde{\mathcal{L}}}
\newcommand{\wtn}{\widetilde{n}}
\newcommand{\wtN}{\widetilde{N}}

\newcommand{\wtS}{\widetilde{S}}
\newcommand{\wtsquare}{\widetilde{\square}}

\newcommand{\wtV}{\widetilde{V}}
\newcommand{\wtZ}{\widetilde{Z}}

\begin{document}

\pagestyle{SPstyle}

\begin{center}{\Large \textbf{\color{scipostdeepblue}{
Non-supersymmetric heterotic strings on $AdS_{4}\times S^{3}\times S^{3}$\\
}}}\end{center}

\begin{center}\textbf{
Ivano Basile\textsuperscript{1$\star$},
Daniel Robbins\textsuperscript{2$\dagger$} and
Hassaan Saleem\textsuperscript{2$\ddag$}
}\end{center}

\begin{center}
{\bf 1} Max-Planck-Institut für Physik (Werner-Heisenberg-Institut)
Boltzmannstraße 8, 85748 Garching, Germany
\\
{\bf 2} Department of Physics, University at Albany,
1400 Washington Avenue, Albany, NY, U.S.A.
\\[\baselineskip]
$\star$ \href{mailto:ibasile@mpp.mpg.de}{\small ibasile@mpp.mpg.de}\,,\quad
$\dagger$ \href{mailto:dgrobbins@albany.edu}{\small dgrobbins@albany.edu}\,,\quad
$\ddag$ \href{mailto:hsaleem@albany.edu}{\small hsaleem@albany.edu}
\end{center}

\section*{\color{scipostdeepblue}{Abstract}}
\textbf{\boldmath{%
We analyze the stability properties of a family of anti-de Sitter flux compactifications of the tachyon-free non-supersymmetric heterotic string in ten dimensions. In contrast with simpler such solutions, the solutions include two independent unbounded fluxes, leading to richer instability phenomena. In particular, when the two fluxes are sufficiently close in magnitude, the perturbative spectrum develops tachyonic modes, which can be projected out by an orbifold action. When the fluxes are far apart, tachyonic modes are absent and the geometry displays inverse scale separation, where a factor of the internal manifold becomes parametrically larger than the anti-de Sitter factor. Still, non-perturbative instabilities in the form of brane nucleation are always available decay channels, and tend to drive the two fluxes closer together, eventually triggering the tachyonic instability when present.
}}

\vspace{\baselineskip}





\vspace{10pt}
\noindent\rule{\textwidth}{1pt}
\tableofcontents
\noindent\rule{\textwidth}{1pt}
\vspace{10pt}
\section{Introduction}

Despite excellent phenomenological motivations and numerous experimental searches, no evidence has been found for supersymmetry at accessible energies.  Hence, any effort to derive low energy effective physics from string theory must include some mechanism for breaking supersymmetry. The traditional approach has been to start with one of the supersymmetric strings in ten dimensions and then break supersymmetry in the process of compactification, for example by including fluxes, branes, or geometrical elements that are incompatible with spacetime supersymmetry.  This typically leads to supersymmetry breaking at the compactification scale (or even lower). This mechanism is certainly not ruled out by experiment, but the ongoing lack of evidence for low energy supersymmetry suggests that we should perhaps consider less well-explored options in which supersymmetry is already broken in ten dimensions, up at the string scale. Then we can study backgrounds of that theory, with the eventual goal of compactifying to get a low-energy effective theory in four dimensions.

Without spacetime supersymmetry at the string scale, one has to be much more careful about stability of a given background solution.  Tachyons, signaling linear instabilities, are a common pathology of non-supersymmetric string theory.  However, there are three ten-dimensional strings in which spacetime supersymmetry is broken or absent, and yet the spectrum (about flat space) is tachyon-free (note that all three models still have some supersymmetry on the worldsheet, which does seem to be a necessary ingredient for quantum consistency).  Two of the models, \cite{Sugimoto:1999tx} and~\cite{Sagnotti:1995ga,Sagnotti:1996qj}, include both open and closed strings.  In the current work we will focus on the third option, the $O(16)\times O(16)$ heterotic string~\cite{Dixon:1986iz,Alvarez-Gaume:1986ghj,Ginsparg:1986wr}, which contains only closed strings.  This theory can be obtained as a $\Z_2$ orbifold of either of the supersymmetric heterotic strings by a quotient that mixes an internal symmetry of the lattice with spacetime fermion number.  The massless bosonic spectrum includes the usual ten-dimensional metric, $B$-field, and dilaton, along with gauge bosons transforming in the adjoint of\footnote{We will not need to worry about the precise global form of the gauge group in this paper. In fact, we will turn off Yang-Mills fields entirely.} $O(16)\times O(16)$.  In addition there are massless spin-$1/2$ fermions transforming in the spinor representation of each gauge factor and in the bifundamental representation of $O(16)\times O(16)$.

Since there is no longer cancelation between bosons and fermions, there are string loop corrections to the effective potential, even in flat space.  The leading correction is a one-loop contribution to the ten-dimensional cosmological constant given by integrating the genus-one partition function $Z(\tau,\bar{\tau})$ over its fundamental domain $\mathcal{F}$,
\begin{equation}
    \Lambda_{10}=-\left(4\pi^2\alpha'\right)^{-5}\int_{\mathcal{F}}\frac{d^2\tau}{2\tau_2^2}Z(\tau,\bar{\tau}),
\end{equation}
leading to an action
\begin{equation}
    S=\frac{1}{2\kappa_{10}^2}\int d^{10}x\sqrt{-g}e^{-2\phi}\left(R+4\left(\partial\phi\right)^2-\frac{1}{12}\left|H_3\right|^2-\frac{1}{4}\left|F_2\right|^2\right)-\int d^{10}x\sqrt{-g}\Lambda_{10}.
\end{equation}
Switching to Einstein frame,
\begin{equation}
    \frac{1}{2\kappa_{10}^2}\int d^{10}x\sqrt{-g}\left[R-\frac{1}{2}\left(\partial\phi\right)^2-\frac{1}{12}e^{-\phi}\left|H_3\right|^2-\frac{1}{4}e^{-\frac{1}{2}\phi}\left|F_2\right|^2-2\kappa_{10}^2e^{\frac{5}{2}\phi}\Lambda_{10}\right].
\end{equation}
we see that the one-loop contribution leads to a runaway exponential potential for the dilaton. This peculiar feature notwithstanding, the absence of spacetime supersymmetry and tachyons in this heterotic setting led to several attempts in phenomenological model building \cite{Abel:2015oxa, Ashfaque:2015vta, Blaszczyk:2015zta, Faraggi:2020wej, Basaad:2024lno, Detraux:2024esd}.

In order to stabilize the dilaton, one can look for a background solution which induces another contribution to the dilaton potential. Several previous works have followed this approach, including~\cite{Lust:1986kj,Dudas:2000ff,Gubser:2001zr,Mourad:2016xbk,Antonelli:2019nar,Basile:2020xwi,Basile:2022ypo,Raucci:2022bjw,Baykara:2022cwj,Fraiman:2023cpa,Raucci:2024fnp,Abel:2024vov,Raucci:2025bev,Leone:2025mwo, Leone:2025eht,Robbins:2025wlm}.  As an example, if we restrict to background solutions in which the gauge fields are set to zero, then the tree-level action is identical to that of the supersymmetric heterotic string, and there are known solutions~\cite{Duff:1996cf,Lima:1999dn,deBoer:1999gea,Gukov:2004ym,Macpherson:2018mif,Eberhardt:2019niq} such as $AdS_3\times S^3\times S^3\times S^1$ and $AdS_3\times S^3\times T^4$, with $H_3$ flux threading $AdS_3$ and the three-spheres, where the radii of $AdS_3$ and the $S^3$s, as well as the expectation value of the dilaton, are fixed in terms of the fluxes, with (uniformly) large fluxes leading to large radii and weak string coupling.  That solves the dilaton runaway problem (at tree level).  However, in the supersymmetric context the flat part of the compactification (either $S^1$ or $T^4$) leads to a collection of additional massless scalar fields (seventeen or eighty, respectively).  In the non-supersymmetric context, the one-loop contribution depends on the values of these moduli, and so introduces a potential for these fields.  Extrema of this potential do exist, but are generically in the regime where the torus is string scale.  As such, the proper approach is to consider the non-supersymmetric string reduced on $S^1$ or $T^4$, and then include the one-loop potential contribution to the effective theory in nine or six dimensions.  The induced potential gives order $g_s$ masses to the scalar fields, which we will call classical moduli, or moduli for short.

For example, the $O(16)\times O(16)$ string on $S^1$ has an extremum (with respect to the moduli; the dilaton again sees a runaway potential) when the circle is at the self-dual radius $R=\sqrt{\alpha'}$ and all sixteen Wilson lines are set to zero~\cite{Ginsparg:1986wr}.  This extremum is a saddle-point, with the radial modulus having a positive mass squared, while the sixteen Wilson line moduli are tachyons with negative mass squared.  A handful of other extrema in this seventeen-dimensional moduli space were found in~\cite{Fraiman:2023cpa}, all of them either maxima or saddle points\footnote{Some of the extrema found in~\cite{Fraiman:2023cpa} were ``knife-edges'', where a small perturbation causes the effective potential to diverge to minus infinity.  These pathological extrema should be excluded from this discussion.}.  With the one-loop contribution included, the resulting nine-dimensional effective action admits an $AdS_3\times S^3\times S^3$ solution~\cite{Baykara:2022cwj} with $H_3$ flux through each of the three factors.  When all the fluxes are large, the one-loop solution can be considered a small perturbation of the tree-level solution.  Somewhat surprisingly, once the one-loop contribution is included, one can find extrema even when the flux through $AdS_3$ (which we'll call electric flux) is small or zero, and these are well-controlled (small curvatures, weak string coupling) as long as the fluxes through the spheres (the magnetic flux) is large.  These background solutions do not exist at tree-level, nor as solutions of the supersymmetric heterotic string, and were called ``intrinsically quantum'' in~\cite{Baykara:2022cwj}. The same is true for the heterotic $AdS_7 \times S^3$ solutions of~\cite{Mourad:2016xbk}.

For all of these extrema, one should next investigate the question of stability.  In the supersymmetric case, fluctuations around the $AdS_3\times S^3\times S^3\times S^1$ vacuum include modes with negative mass squared.  Unlike for fluctuations around flat space, however, a tachyonic mass in anti-de Sitter space does not necessarily indicate an instability.  Instead, there's a negative lower bound on mass squared, called the Breitenlohner-Freedman (BF) bound~\cite{Breitenlohner:1982bm}, above which no perturbative instability occurs. For the Freund-Rubin solutions obtained in~\cite{Gubser:2001zr, Mourad:2016xbk}, encompassing both the heterotic and orientifold cases in ten dimensions, this analysis was performed in~\cite{Basile:2018irz}, finding some tachyons that could be removed by an orbifold projection or another modification of the internal manifold. For the $AdS_3\times S^3\times S^3\times S^1$ heterotic solution, in the supersymmetric case some modes saturate the bound, but none violate it.  The perturbative spectrum of the $O(16)\times O(16)$ string around this solution was analyzed in~\cite{Baykara:2022cwj}, and it was confirmed that the modes coming from the nine-dimensional background do not violate the BF bound.  Some of the classical moduli are also tachyonic; when the electric flux is large (so the solution is a perturbation to the tree-level result), these modes remain above the BF bound, but in the intrinsically quantum limit some of them drop below the bound and there is an instability~\cite{Fraiman:2023cpa}.

In~\cite{Robbins:2025wlm}, an analysis was performed for the $AdS_3\times S^3\times T^4$ background, with similar qualitative results.  As long as the magnetic flux was large, a perturbatively stable extremum existed.  If the electric flux was also large, then the solution could be interpreted as a correction to the tree-level result, matching the supersymmetric case.  Taking the electric flux to zero leads to intrinsically quantum extrema, with no tree-level analogs, but in this limit, some of the classical moduli from the $T^4$ violate the BF bound, leading to a linear instability.

In both of these examples, the surprising existence of intrinsically quantum extrema is somewhat spoiled by the fact that the classical moduli coming from the torus part of the background dip below the BF bound.  The existence of the intrinsically quantum extrema shows that electric flux is not necessary to solve the background equations, and it suggests that we should take a look at backgrounds without either electric flux or classical moduli. Indeed, such solutions were already considered in~\cite{Basile:2020xwi}.  Specifically, an $AdS_4\times S^3\times S^3$ backgrounds of the $O(16)\times O(16)$ heterotic string was constructed, generalizing the $AdS_7 \times S^3$ solution of \cite{Mourad:2016xbk}. For the latter, there was a scalar mode below the BF bound found in \cite{Basile:2018irz}, which could be removed by replacing $S^3$ with $\mathbb{RP}^3$. In addition, channels for non-perturbative instabilities were studied in \cite{Antonelli:2019nar}, particularly the nucleation of heterotic NS5-branes that could discharge the flux on the sphere. A holographic interpretation for these processes was put forth in \cite{Antonelli:2018qwz, Ghosh:2021lua}, and analyzed in more detail in \cite{Basile:2022zee}. 

The current article will focus on the case of the $AdS_4\times S^3\times S^3$ backgrounds constructed in~\cite{Basile:2020xwi} and also studied in~\cite{Raucci:2025bev}.  All such backgrounds are intrinsically quantum, in the sense of~\cite{Baykara:2022cwj}, meaning that they are not solutions of the tree-level equations of motion, and this is one motivation for studying them.  Moreover, unlike the intrinsically quantum solutions $AdS_3\times S^3\times S^3\times S^1$ and $AdS_3\times S^3\times T^4$ without electric flux, the $AdS_4\times S^3\times S^3$ backgrounds have no classical moduli.  These fields were the only source of perturbative instability in the $AdS_3$ examples\footnote{Here we restrict to intrinsically quantum vacua, and thus do not discuss the instabilities of the orientifold $AdS_3 \times S^7$ solutions supported by tree-level (annulus) potentials \cite{Mourad:2016xbk, Basile:2018irz}.}, so one may hope that there is a chance to find better stability in these $AdS_4$ models.  In the best case, one might find a stable (or more plausibly, given swampland conjectures~\cite{Ooguri:2016pdq}, a long-lived metastable) $AdS_4$ background which might be amenable to some sort of holographic description, such as those outlined in \cite{Antonelli:2018qwz, Ghosh:2021lua, Basile:2022zee}.

We will thus undertake a stability analysis of the $AdS_4\times S^3\times S^3$ backgrounds of the $O(16)\times O(16)$ heterotic string.  The perturbative analysis follows the same procedure as~\cite{Basile:2018irz,Baykara:2022cwj,Robbins:2025wlm} (following similar analyses done in the supersymmetric context~\cite{Eberhardt:2017fsi,Deger:1998nm}).  We find that there is one dangerous mode which can potentially lead to an instability, but it only violates the BF bound if the fluxes are of comparable order (specifically, if the ratio of the larger flux to the smaller flux is less than approximately 4.674); if one of the fluxes is much larger than the other, then there is no linear instability; in addition, one of the two three-sphere factors becomes much larger than both the other factor and the $AdS_4$, leading to an exotic inverse kind of scale separation \cite{Basile:2020xwi}. Intriguingly, it may be possible to eliminate this instability by replacing one of the spheres with a free $\Z_2$ quotient, i.e.~taking $\mathbb{RP}^3$ instead.

Looking at non-perturbative instability, we compute the decay rate for nucleation of NS5 branes wrapping arbitrary three-cycles inside $S^3\times S^3$, which can discharge some of the fluxes on the spheres. We find that such nucleation can always happen, but it happens much more rapidly when there is a large hierarchy between the two fluxes, in which case this channel tends to equalize the magnitudes of fluxes on the two spheres.  We are thus led to a surprisingly rich structure for the instabilities of these solutions.  If the two fluxes are of comparable magnitude, then there is a perturbative instability.  If there is a large hierarchy between the fluxes, in which case there is an inverted hierarchy, with one of the three-spheres much larger than the other three-sphere and the $AdS_3$ space whose sizes will be comparable, then the perturbative instability is absent but there is a non-perturbative instability to nucleating heterotic NS5-branes.  These can rapidly discharge the larger flux, eventually leading us back to the regime where the fluxes are of the same magnitude and the perturbative instability again takes over.

In section~\ref{sec:TreeLevel}, we write down the tree-level action and show that there are no $AdS_4\times S^3\times S^3$ solutions.  In section~\ref{sec:one-loop_solution} we add in the one-loop contribution to the effective action and construct solutions.  Section~\ref{sec:PerturbativeStability} computes the spectrum of fluctuations around the background by expanding the low-energy fields in harmonics on each of the three-spheres.  Section~\ref{sec:non-perturbative_stability} computes the viability and decay rate for brane nucleations.  Finally, section~\ref{sec:Discussion} concludes with some remarks and future directions, and an appendix sketches a minimal volume cycle argument that is needed in the brane nucleation calculation.

\section{Tree-level setup}
\label{sec:TreeLevel}

We adopt the conventions of \cite{Baykara:2022cwj} and write down the tree-level action of $O(16)\times O(16)$ heterotic theory as
\be\label{O(16) action}
S=\frac{1}{2\k^{2}_{10}}\int d^{10}x\;e^{-2\phi}\sqrt{-g}\lp R+4(\p\phi)^{2}-\frac{1}{12}|H_{3}|^{2}-\frac{1}{4}|F_{2}|^{2}\rp,
\ee
with
\be
\k_{10}^{2}=\hlf (2\pi)^{7}\al'^{4}e^{2\phi_{0}},
\ee
where $\phi_{0}$ is the VEV of dilaton, $H_{3}$ is the usual NS-NS field strength and $F_{2}$ is the field strength for the heterotic gauge field. We want to study the $AdS_{4}\times S^{3}\times \widetilde{S}^{3}$ compactifications of this action. The metric on this spacetime is taken as
\be
ds_{10}^{2}=ds_{AdS_{4}}^{2}+e^{2\chi}L^{2}d\Om^{2}_{3}+e^{2\chih}\widetilde{L}^{2}d\widetilde{\Om}^{2}_{3},
\ee
where $\chi(\widetilde{\chi})$ and $L(\widetilde{L})$ are radius moduli and radii for the two spheres respectively. Moreover, $d\Om_{3}(d\widetilde{\Om}_{3})$ are the metrics on the two spheres with unit radius. Compactifying \eqref{O(16) action} on $S^{3}\times \widetilde{S}^{3}$, we get a four-dimensional theory. We want to write down the potential of this four dimensional theory in the Einstein frame. The Ricci scalar term in the ten dimensional action dictates the appropriate transformation ($g_{4}=\mathcal{V}^{-1}\hat{g}_{4}$) to go the Einstein frame metric $\hat{g}_{4}$ in the four-dimensional action. This term transforms as shown below
$$
\frac{1}{2\k^{2}_{10}}\int d^{10}x\;e^{-2\phi}\;\sqrt{-g_{10}}\;R_{10}=\frac{1}{2\k^{2}_{10}}\int d^{10}x\;e^{-2\phi}\;\sqrt{-g_{10}}\;(R_{4}+R_{6})
$$
$$
=\frac{1}{2\k^{2}_{4}}\int d^{4}x\;\sqrt{-\hat{g}_{4}}(\hat{R}_{4}+\mathcal{V}^{-1}R_{6})
$$
where we have
$$
\mathcal{V}=\;e^{-2\phi+3(\chi+\chih)},\efill \k^{2}_{4}=\frac{\k^{2}_{10}}{4\pi^{4}L^{3}\widetilde{L}^{3}}.
$$
Going to the Einstein frame, the potential term in four dimensional becomes
\be
-\frac{1}{2\k^{2}_{4}}\int d^{4}x\;\sqrt{-\hat{g}_{4}}\;V_{\text{tree}}=\frac{1}{2\k^{2}_{4}}\int d^{4}x\;\mathcal{V}^{-1}\sqrt{-\hat{g}_{4}}\;\lp R_{6}-\frac{1}{12}|H^{m}_{3}|^{2}-\frac{1}{12}|H^{\widetilde{m}}_{3}|^{2}\rp,
\ee
which gives us
\begin{align}
V_{\text{tree}}=\ & -\mathcal{V}^{-1}\lp R_{6}-\frac{1}{12}|H^{m}_{3}|^{2}-\frac{1}{12}|H^{\tilde{m}}_{3}|^{2}\rp\non\\
=\ & -\mathcal{V}^{-1}\ls 6\lp \frac{e^{-2\chi}}{L^{2}}+\frac{e^{-2\chih}}{\widetilde{L}^{2}}\rp-2\al'^{2}\lp \frac{n^{2}}{L^{6}}e^{-6\chi}+\frac{\widetilde{n}^{2}}{\widetilde{L}^{6}}e^{-6\chih}\rp\rs.
\end{align}
Setting the derivatives of this potential with respect to~$\phi,\chi,\widetilde{\chi}$ to zero and solving for $L$ and $\widetilde{L}$ gives us inconsistent solutions and thus one doesn't have a tree level solution.

\section{One-loop correction}\label{sec:one-loop_solution}

\begin{figure}[ht]
    \centering
    \includegraphics[width=0.3\linewidth]{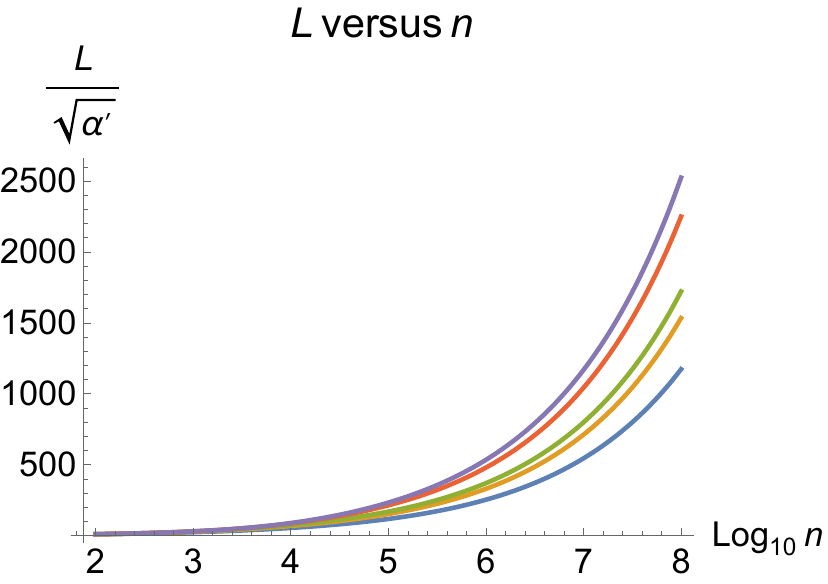}\efill \includegraphics[width=0.3\linewidth]{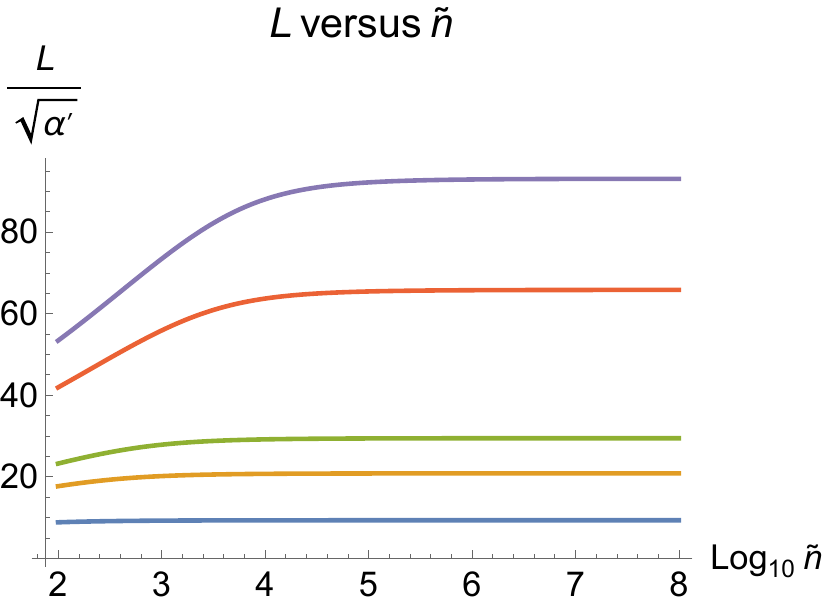} \efill\includegraphics[width=0.08\linewidth]{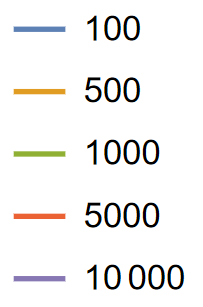}
    \caption{Plots of $L$ against $\log n$ for different values of $\widetilde{n}$ given in the legend and vice versa.}
    \label{fig: Lo plots}
\end{figure}

\begin{figure}[ht]
    \centering
    \includegraphics[width=0.3\linewidth]{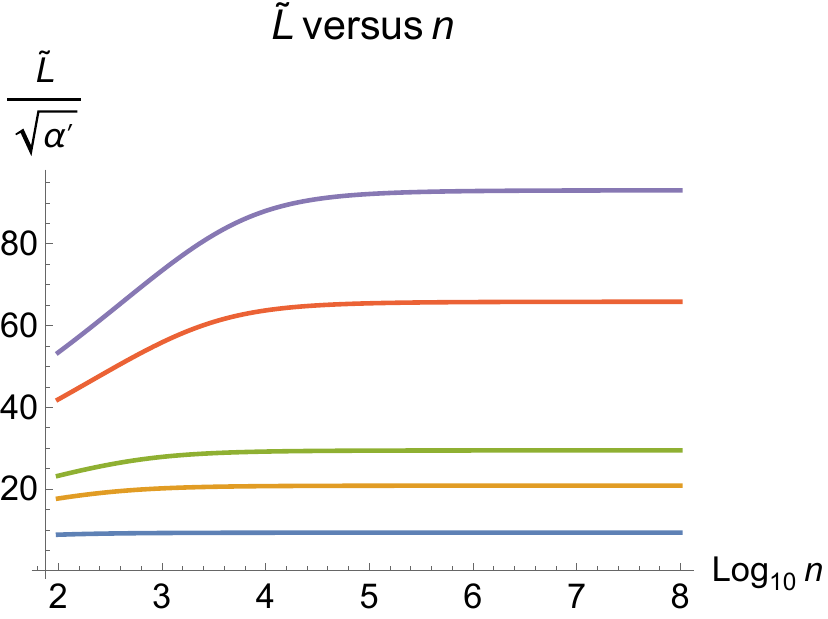}\efill\includegraphics[width=0.3\linewidth]{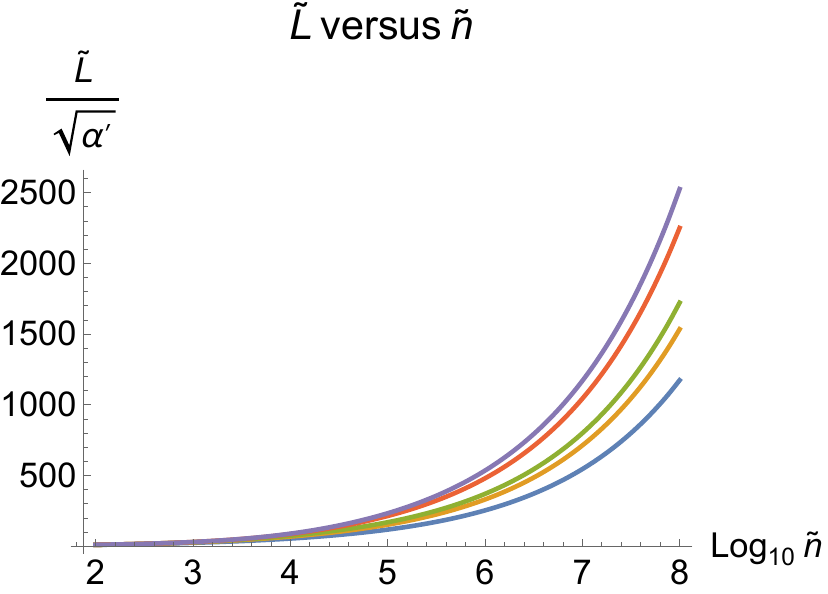}\efill\includegraphics[width=0.08\linewidth]{Figures/Legend.png}
    \caption{Plots of $\widetilde{L}$ against $\log n$ for different values of $\widetilde{n}$ given in the legend and vice versa.}
    \label{fig: Loh plots}
\end{figure}

\begin{figure}[ht]
    \centering
    \includegraphics[width=0.3\linewidth]{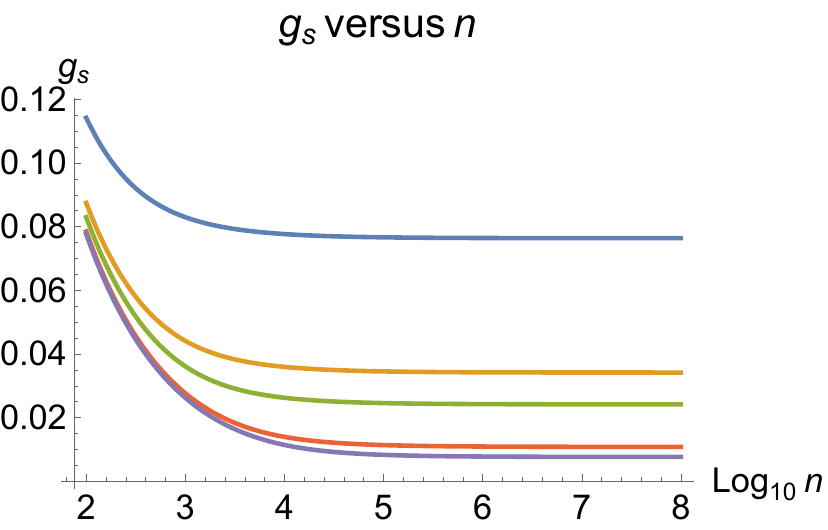}\efill \includegraphics[width=0.3\linewidth]{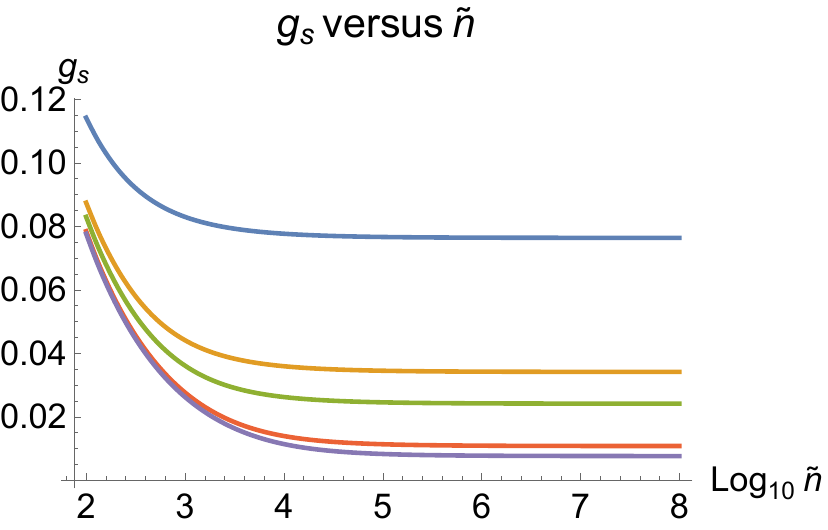}
    \efill\includegraphics[width=0.08\linewidth]{Figures/Legend.png}
    \caption{Plots of $g_{s}$ against $\log n$ for different values of $\widetilde{n}$ given in the legend and vice versa.}
    \label{fig: go plots}
\end{figure}

\begin{figure}[ht]
    \centering
    \includegraphics[width=0.3\linewidth]{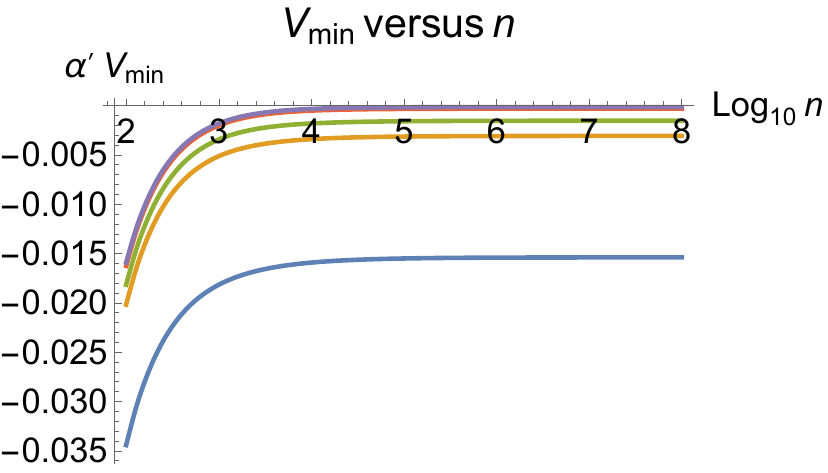}\efill \includegraphics[width=0.3\linewidth]{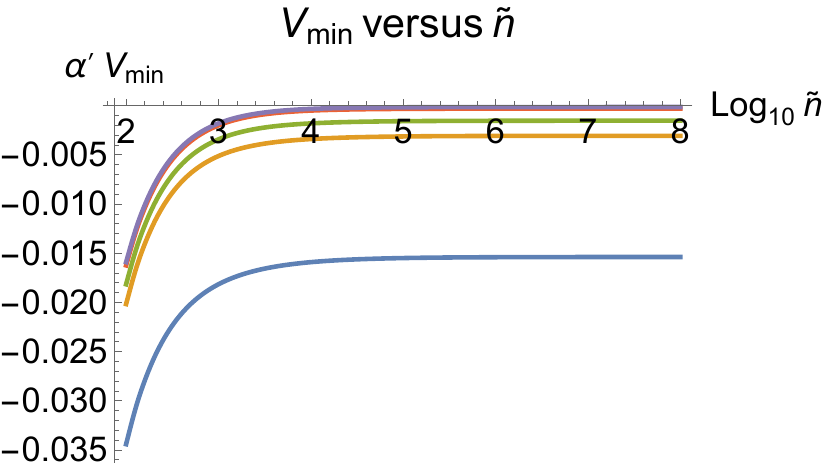}\efill  \includegraphics[width=0.08\linewidth]{Figures/Legend.png}
    \caption{Plots of $V_{\text{min}}$ against $\log n$ for different values of $\widetilde{n}$ given in the legend and vice versa.}
    \label{fig: Vmin plots}
\end{figure}

\begin{figure}[ht]
    \centering
    \includegraphics[width=0.3\linewidth]{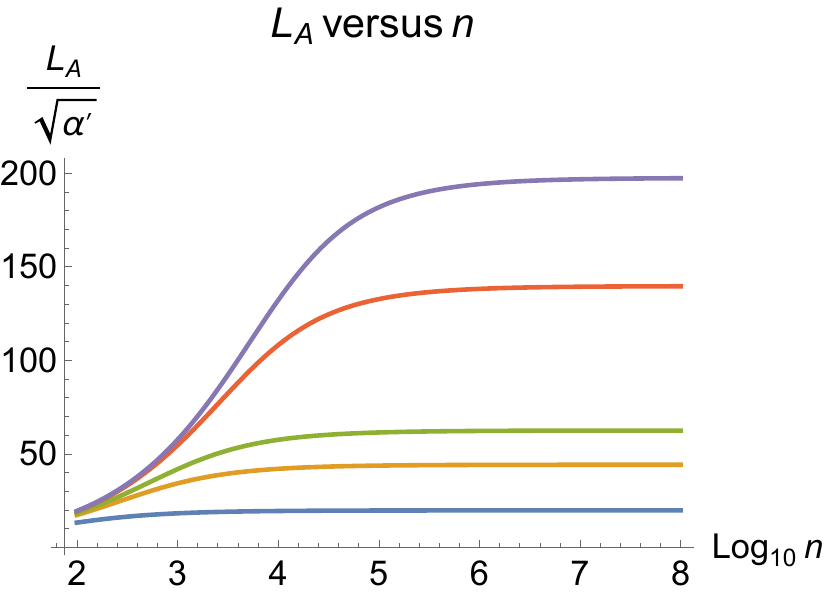}\efill \includegraphics[width=0.3\linewidth]{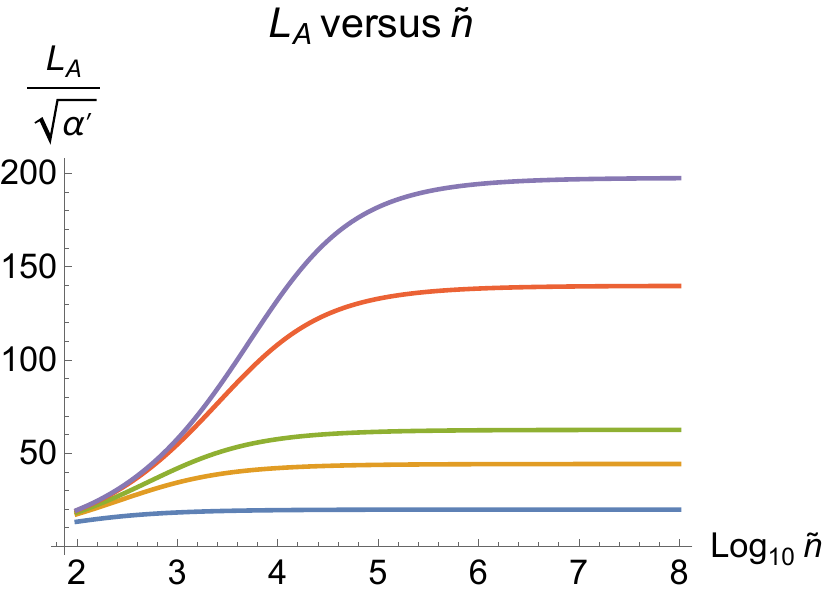}\efill  \includegraphics[width=0.08\linewidth]{Figures/Legend.png}
    \caption{Plots of $L_{A}$ against $\log n$ for different values of $\widetilde{n}$ given in the legend and vice versa.}
    \label{fig: LAdS plots}
\end{figure}

We will now add the one-loop correction to the tree level potential, and derive the one loop solutions for $L,\widetilde{L}$ and string coupling $g_{s}=e^{\phi_{0}}$. The ten dimensional one loop correction one gets was calculated in \cite{Alvarez-Gaume:1986ghj}. For our purposes, we will follow the conventions of \cite{Fraiman:2023cpa} and write the ten dimensional cosmological constant is as
\be
\Lambda_{10}=-\frac{1}{2(2\pi\sqrt{\al'})^{10}}\int_{\mathcal{F}}\frac{d^{2}\tau}{\tau^{2}_{2}}Z(\tau),
\ee
where $\mathcal{F}$ is the fundamental domain and $Z(\tau)$ is the partition function. We can interpret this cosmological constant as a contribution to the ten-dimensional scalar potential,
\begin{equation}
    -\int d^{10}x \sqrt{-g}\;\Lambda_{10}
    =-\frac{1}{2\k^{2}_{10}}\int d^{10}x  \sqrt{-g}\;V^{(10)},
\end{equation}
where
\begin{equation}
    V^{(10)}=2\k^{2}_{10}\;\Lambda_{10}.
\end{equation}
When we compactify on $S^{3}\times \widetilde{S}^{3}$, this contribution becomes
\begin{equation}
    -\frac{1}{2\k^{2}_{10}}\int d^{10}x\;\;\sqrt{-g}\;\;V^{(10)}=-\frac{1}{2\k^{2}_{4}}\int d^{4}x\;\;\sqrt{-g}\;e^{3(\chi+\chih)}\;V^{(10)}.
\end{equation}
Going to the four-dimensional Einstein frame, this term becomes
\begin{equation}
    -\frac{1}{2\k^{2}_{4}}\int d^{4}x\;\;\sqrt{-\hat{g}}\;\mathcal{V}^{-2}e^{3(\chi+\chih)}\;V^{(10)},
\end{equation}
resulting in the following one-loop potential in the four dimensional Einstein frame
\begin{align}
\label{one loop potential}
V_{1-\text{loop}}=\ & \mathcal{V}^{-2}e^{3(\chi+\chih)}V^{(10)}=2\underbrace{\lp-\frac{1}{4(2\pi)^{3}}\int_{\mathcal{F}}\frac{d^{2}\tau}{\tau^{2}_{2}}Z(\tau)\rp}_{\la}\mathcal{V}^{-2}e^{3(\chi+\chih)}\frac{g^{2}_{s}}{\al'}\non\\
=\ & 2\la\mathcal{V}^{-2}e^{3(\chi+\chih)}\frac{g^{2}_{s}}{\al'},
\end{align}
where the dimensionless quantity $\la$ is defined as
\be\label{lambda expression}
\la=-\frac{1}{4(2\pi)^{3}}\int_{\mathcal{F}}\frac{d^{2}\tau}{\tau^{2}_{2}}Z(\tau).
\ee
We can calculate $\la$ by using the $\R^{1,9}$ partition function (though this is only justified if all the three components of the background, i.e.~$AdS_{4}\times S^{3}\times \widetilde{S}^{3}$, are large in string units). Using the expression for $Z(\tau)$ \cite{Alvarez-Gaume:1986ghj} (also see \cite{Fraiman:2023cpa, Robbins:2025wlm} for notation and conventions that we follow), we can calculate $\la$ to get
\begin{align}
\la=-\frac{1}{8(2\pi)^{3}}\int \frac{d^{2}\tau}{\tau^{4}_{2}}\frac{1}{\eta^{24}\bar{\eta}^{8}}&\ls(\bar{V}_{8}+\bar{S}_{8})\sum_{\pi}(-1)^{2\d.\pi}q^{\hlf |\pi|^{2}}+(\bar{O}_{8}-\bar{C}_{8})\sum_{\pi}q^{\hlf |\pi+\d|^{2}}\right.\non\\
&\qquad\left.-(\bar{O}_{8}+\bar{C}_{8})\sum_{\pi}(-1)^{2\d.\pi}q^{\hlf |\pi+\d|^{2}}
\rs\approx 0.741,
\end{align}
where the sums over $\pi$ take place over the $E_{8}\times E_{8}$ Narain lattice and $\d=(0^{7},1;0^{7},1)$. To evaluate these sums, we used the lattice sums given in \cite{Robbins:2025wlm}.\newline\newline
Including the one-loop correction, we get the one-loop corrected potential as
\begin{multline}
V=V_{\text{tree}}+V_{1-\text{loop}}\non\\
=-2\mathcal{V}^{-1}\ls 3\lp \frac{e^{-2\chi}}{L^{2}}+\frac{e^{-2\chih}}{\widetilde{L}^{2}}\rp-\al'^{2}\lp \frac{n^{2}}{L^{6}}e^{-6\chi}+\frac{\widetilde{n}^{2}}{\widetilde{L}^{6}}e^{-6\chih}\rp\rs+2\la\mathcal{V}^{-2}e^{3(\chi+\chih)}\frac{g^{2}_{s}}{\al'},
\end{multline}
where $L,\widetilde{L}$ and $g_{s}$ are now one loop corrected quantities. Since we didn't have any tree level solution for $L,\widetilde{L}$ and $g_{s}$, we don't need to distinguish the tree level radii and coupling from their one loop counterparts. Setting the derivatives of the one loop potential to zero gives us
\begin{align}\label{one loop pot phi der}
\left.\p_{\phi}V\right|_{\phi,\chi,\widetilde{\chi}=0}=0\quad\Rightarrow\quad & 3\lp \frac{1}{L^{2}}+\frac{1}{\widetilde{L}^{2}}\rp-\al'^{2}\lp \frac{n^{2}}{L^{6}}+\frac{\widetilde{n}^{2}}{\widetilde{L}^{6}}\rp-\frac{2g^{2}_{s}\la}{\al'}=0,\\
\label{one loop pot chi der}
\left.\p_{\chi}V\right|_{\phi,\chi,\widetilde{\chi}=0}=0\quad\Rightarrow\quad & \frac{5}{L^{2}}+\frac{3}{\widetilde{L}^{2}}-\al'^{2}\lp \frac{3n^{2}}{L^{6}}+\frac{\widetilde{n}^{2}}{\widetilde{L}^{6}}\rp-\frac{g^{2}_{s}\la}{\al'}=0,\\
\label{one loop pot chih der}
\left.\p_{\widetilde{\chi}}V\right|_{\phi,\chi,\widetilde{\chi}=0}=0\quad\Rightarrow\quad & \frac{3}{L^{2}}+\frac{5}{\widetilde{L}^{2}}-\al'^{2}\lp \frac{n^{2}}{L^{6}}+\frac{3\widetilde{n}^{2}}{\widetilde{L}^{6}}\rp-\frac{g^{2}_{s}\la}{\al'}=0.
\end{align}
We note that \eqref{one loop pot phi der},\eqref{one loop pot chi der} and \eqref{one loop pot chih der} are invariant under the following scaling
\be\label{scalings}
\begin{pmatrix}n\\ \widetilde{n}
\end{pmatrix}
\rr \om\begin{pmatrix}n\\ \widetilde{n}
\end{pmatrix},\efill\efill\begin{pmatrix}L\\\widetilde{L}
\end{pmatrix}\rr \sqrt{\om}\begin{pmatrix}L\\\widetilde{L}
\end{pmatrix},\efill\efill g_{s}\rr \frac{g_{s}}{\sqrt{\om}},
\ee
for some positive $\om$, and if any expression is invariant under this scaling, it should only depend on the ratio of the fluxes $n$ and $\widetilde{n}$. Equations \eqref{one loop pot phi der},\eqref{one loop pot chi der}, and \eqref{one loop pot chih der} can be solved in closed form but the solution is cumbersome, and hence, we will plot the solutions numerically. $L,\widetilde{L},g_{s}$ and $V_{\text{min}}$ (the minimum value of the potential) have been plotted in Figures~\ref{fig: Lo plots}, \ref{fig: Loh plots}, \ref{fig: go plots} and \ref{fig: Vmin plots} respectively. The graphs in Figure~\ref{fig: Vmin plots} show that the uplift to de Sitter doesn't happen, which is consistent with the no-go theorem in \cite{Basile:2020mpt}. The expression for the AdS radius $L_{A}$ in terms of $L$ and $\widetilde{L}$ can be deduced from \eqref{one loop pot phi der}, \eqref{one loop pot chi der} and \eqref{one loop pot chih der} as
\be\label{AdS length scale}
\frac{1}{L^{2}_{A}}=\frac{2}{9}\lp \frac{1}{L^{2}}+\frac{1}{\widetilde{L}^{2}}\rp,
\ee
and note that $L_{A}\rr\sqrt{\om}L_{A}$ under the scaling mentioned in \eqref{scalings}. $L_{A}$ has been plotted for different fluxes in Figure~\ref{fig: LAdS plots}. The plots of $L$ and $\widetilde{L}$ clearly show that if $n\gg$ $\widetilde{n}$ then $L\gg \widetilde{L}$ and vice versa. Moreover, we see that $L_{A}$ peaks at a small value and is small as compared to the bigger $S^{3}$. Therefore, one can deduce the inverse scale separation that was mentioned in \cite{Basile:2020xwi}. If $n=\widetilde{n}$, then $L$ and $\widetilde{L}$ are equal, as they should be due to the exchange symmetry. In this case, we can solve \eqref{one loop pot phi der},\eqref{one loop pot chi der} and \eqref{one loop pot chih der} exactly to get
\be\label{equal flux Lo Loh}
L=\widetilde{L}=\lp\al' n\sqrt{\frac{3}{5}}\rp^{1/2},\efill g_{s}=\lp \frac{4}{3\la n}\sqrt{\frac{5}{3}}\rp^{1/2},\efill L_{A}=\frac{3}{2}L=\frac{3}{2}\widetilde{L},
\ee
which means that $L,\widetilde{L}$ and $L_{A}$ all increase with increasing $n$.

\section{Perturbative stability analysis}
\label{sec:PerturbativeStability}

\subsection{Background solution}

The ten dimensional effective action for $O(16)\times O(16)$ with one-loop corrected potential (in string frame) is
\be
S=\frac{1}{2\k^{2}_{10}}\int d^{10}x\;\sqrt{-g}\ls e^{-2\phi}\lp R+4(\p \phi)^{2}-\frac{1}{12}|H_{3}|^{2}\rp-\frac{2\la g^{2}_{s}}{\al'}\rs.
\ee
This action results in the following equations of motion for the dilaton, the $B$-field, and the metric respectively
\begin{align}
    0=\ & \nabla^M\nabla_M\phi+\frac{1}{4}R-\frac{1}{48}\left|H_3\right|^2-\nabla^M\phi\nabla_M\phi,\\
    0=\ & \nabla^P\left(e^{-2\phi}H_{MNP}\right),\\
    0=\ & \hlf g_{MN}\left(R+4\nabla^P\nabla_P\phi-4\nabla^P \phi\nabla_P\phi-\frac{1}{12}\left|H_3\right|^2-e^{2\phi}\Lambda\right)-R_{MN}\non\\
    &-2\nabla_M\nabla_N\phi+\frac{1}{4}H_M{}^{PQ}H_{NPQ}\non\\
    =\ & -\hlf g_{MN}e^{2\phi}\Lambda-R_{MN}-2\nabla_M\nabla_N\phi+\frac{1}{4}H_M{}^{PQ}H_{NPQ},
\end{align}
where $\Lambda=2\lambda g_s^2/\alpha'$.  For the metric equation, we used the dilaton equation of motion to simplify the result on the second line.

We make an ansatz that the solution will have the form of $AdS_4\times S^3\times S^3$ with constant dilaton $\phi=0$ (since we absorbed its VEV into the coupling constant $g_s$) and $H_3$ wrapping each of the two three-spheres. We will use Greek indices $\mu,\nu,\rho$ for $AdS_4$, Roman letters $a,b,c$ from the beginning of the alphabet for the first $S^3$, and Roman letters $i,j,k$ for the second $S^3$.  Our ansatz means that the non-vanishing components of the ten-dimensional Riemann tensor are
\begin{align}
    R_{\mu\nu\rho\sigma}=\ & -L_A^{-2}\left(g_{\mu\rho}g_{\nu\sigma}-g_{\mu\sigma}g_{\nu\rho}\right),\quad R_{\mu\nu}=-3L_A^{-2}g_{\mu\nu},\\
    R_{abcd}=\ & L^{2}\left(g_{ac}g_{bd}-g_{ad}g_{bc}\right),\quad R_{ab}=2L^{-2}g_{ab},\\
    R_{ijk\ell}=\ & \wtL^{-2}\left(g_{ik}g_{j\ell}-g_{i\ell}g_{jk}\right),\quad R_{ij}=2\wtL^{-2}g_{ij},
\end{align}
and $H_3$ fluxes are
\begin{align}
    H_{abc}=\ & 2\mcL^{-1}\e_{abc},\\
    H_{ijk}=\ & 2\wtmcL^{-1}\e_{ijk},
\end{align}
where $\mathcal{L}$ and $\widetilde{\mathcal{L}}$ are some length scales. Here $\e_{abc}$ and $\e_{ijk}$ are the volume forms for the three-spheres, satisfying
\begin{equation}
    \e_{abc}\e^{def}=6\d^{[d}_a\d^e_b\d^{f]}_c,
\end{equation}
where indices are raised with the background metric, $\e^{def}=g^{dd'}g^{ee'}g^{ff'}\e_{d'e'f'}$, and our conventions for symmetrizing and anti-symmetrizing are~
\begin{equation}
    T_{(a_1\cdots a_k)}=\frac{1}{k!}\sum_{\sigma\in S_k}T_{a_{\sigma_1}\cdots a_{\sigma_k}},\efill\efill T_{[a_1\cdots a_k]}=\frac{1}{k!}\sum_{\sigma\in S_k}(\text{sign}(\sigma))T_{a_{\sigma_1}\cdots a_{\sigma_k}}.
\end{equation}
Plugging this ansatz into the equations of motion, we find that we have a solution provided that
\begin{equation}
    L_A^{-2}=\frac{1}{6}\Lambda,\qquad\mcL^{-2}=L^{-2}+\frac{1}{4}\Lambda,\qquad\wtmcL^{-2}=\wtL^{-2}+\frac{1}{4}\Lambda,\qquad L^{-2}+\wtL^{-2}=\frac{3}{4}\Lambda.
\end{equation}
which is consistent with \eqref{one loop pot phi der},\eqref{one loop pot chi der},\eqref{one loop pot chih der} and \eqref{AdS length scale}. Additionally, the fluxes need to be properly quantized, leading to
\begin{equation}\label{expressions for curly Ls}
    \mcL=\frac{L^3}{\alpha'n},\qquad\wtmcL=\frac{\wtL^3}{\alpha'\wtn},\qquad n,\wtn\in\Z.
\end{equation}
Note that this means that the ten-dimensional Ricci scalar and $|H_3|^2$ are given by
\begin{equation}
    R^M{}_M=-12L_A^{-2}+6L^{-2}+6\wtL^{-2}=\frac{5}{2}\Lambda,\qquad|H_3|^2=H^{MNP}H_{MNP}=24\mcL^{-2}+24\wtmcL^{-2}=30\Lambda.
\end{equation}
There is an overall scaling symmetry in the equations that determine the radii and the string coupling.  To see this, let us write
\begin{equation}
    n^2=x(1+y),\qquad\wtn^2=x(1-y),
\end{equation}
where we have introduced two new variables,
\begin{equation}
    x=\frac{n^2+\wtn^2}{2},\qquad y=\frac{n^2-\wtn^2}{n^2+\wtn^2}=\frac{1-\frac{\wtn^2}{n^2}}{1+\frac{\wtn^2}{n^2}}.
\end{equation}
In other words, $x$ is the average of $n^2$ and $\wtn^2$, while $y$ is a function of the ratio $\wtn/n$.  Then the $x$ dependence of $L$, $\wtL$, and $g_s$ (via $\Lambda$) is fixed, and we can write
\begin{equation}
    L^{-2}=\frac{f(y)}{\alpha'x^{1/2}},\qquad\wtL^{-2}=\frac{f(-y)}{\alpha'x^{1/2}},\qquad\Lambda=\frac{4}{3\alpha'x^{1/2}}\left(f(y)+f(-y)\right),
\end{equation}
and the function $f(y)$ is determined by the following equation (together with the fact that $f(y)>0$)
\begin{equation}\label{f equation}
    (1+y)f(y)^3=\frac{4}{3}f(y)+\frac{1}{3}f(-y).
\end{equation}
To see this, we can eliminate $f(-y)$ from this equation to get an equation for non-trivial $f(y)$ alone, which turns out to be
$$
3(1-y)f(y)^2(3(1+y)f(y)^2-4)^3=4(3(1+y)f(y)^2-4)+1.
$$
This equation gives us eight different branches, but only one of these branches gives $f(0)=\sqrt{5/3}$, which is consistent with \eqref{f equation} and has a positive $f(y)$. The string coupling is related to $\Lambda$ by
\begin{equation}
    g^{2}_s=\frac{\alpha'}{2\lambda}\Lambda=\frac{2}{3\lambda x^{1/2}}\left(f(y)+f(-y)\right).
\end{equation}
For small $y$ (fluxes nearly equal in magnitude), we have
\begin{equation}
    f(y)\simeq\sqrt{\frac{5}{3}}\left(1-\frac{5}{12}y+\cdots\right),
\end{equation}
which translates at leading order to
\begin{equation}
    L\simeq\wtL\simeq \left(\frac{3}{5}\right)^{1/4}\sqrt{\alpha'}|n|^{1/2},\quad L_A\simeq\frac{3}{2}\left(\frac{3}{5}\right)^{1/4}\sqrt{\al'}|n|^{1/2},\quad g_s\simeq \frac{2}{3}\left( 15\right)^{1/4}\la^{-1/2}|n|^{-1/2}.
\end{equation}
In this case all three radii are of the same order, and all can be taken to be large by taking the flux large, which also makes the string coupling small. For $y$ approaching 1 (i.e. when $|n|\gg|\wtn|$) we have
\begin{align}
    f(y)\simeq\frac{1}{\sqrt{3}}\left(1-y\right)^{-1/6}\left(1+\frac{11}{16}\left(1-y\right)^{1/3}\cdots\right)\\
    \qquad f(-y)\simeq\sqrt{\frac{4}{3}}\frac{1}{\sqrt{1-y}}\left(1+\frac{1}{16}\left(1-y\right)^{1/3}\cdots\right),
\end{align}
translating to
\begin{align}
    L\simeq 2^{-1/6}\cdot 3^{1/4}\cdot\sqrt{\alpha'}|n|^{1/3}|\wtn|^{1/6},&\qquad\qquad\wtL\simeq 2^{-1/2}\cdot 3^{1/4}\cdot\sqrt{\al'}|\wtn|^{1/2},\non\\
    \qquad L_A\simeq 2^{-1}\cdot 3^{5/4}\cdot\sqrt{\al'}|\wtn|^{1/2},&\qquad\qquad g_s\simeq 2\cdot 3^{-3/4}\cdot\la^{-1/2}|\wtn|^{-1/2}.
\end{align}
Here the sphere with the larger flux through it has a radius that is much larger than the other sphere's or than the AdS radius, which are of the same order.  By taking the smaller flux to be large as well, all radii can be made large and the string coupling can be made small.

\subsection{Linearized perturbations}

We are interested in studying linearized fluctuations around this background solution.  To this end we expand each of the component fields into a background and a fluctuation piece, defining new fluctuating fields $\phi$, $H_{\mu\nu}$, $M$, $S_{\mu a}$, $\wtS_{\mu i}$, $K_{ab}$, $N$, $W_{ai}$, $\wtK_{ij}$, $\wtN$, $X_{\mu\nu}$, $Z_{\mu a}$, $\wtZ_{\mu i}$, $V_a$, $P_{ai}$, and $\wtV_i$, according to
\begin{align}
    \phi\rr\ & 0+\phi,\\
    g_{\mu\nu}\rr\ & g_{\mu\nu}+H_{\mu\nu}+g_{\mu\nu}M,\\
    g_{\mu a}\rr\ & 0+S_{\mu a},\\
    g_{\mu i}\rr\ & 0+\wtS_{\mu i},\\
    g_{ab}\rr\ & g_{ab}+K_{ab}+g_{ab}N,\\
    g_{ai}\rr\ & 0+W_{ai},\\
    g_{ij}\rr\ & g_{ij}+\wtK_{ij}+g_{ij}\wtN,\\
    B_{\mu\nu}\rr\ & 0+X_{\mu\nu},\\
    B_{\mu a}\rr\ & 0+Z_{\mu a},\\
    B_{\mu i}\rr\ & 0+\wtZ_{\mu i},\\
    B_{ab}\rr\ & B_{ab}+\e_{abc}V^c,\\
    B_{ai}\rr\ & 0+P_{ai},\\
    B_{ij}\rr\ & B_{ij}+\e_{ijk}\wtV^k.
\end{align}
The symmetric tensors $H_{\mu\nu}$, $K_{ab}$, and $\wtK_{ij}$ are chosen to be traceless. Next, we will impose some gauge conditions. Following \cite{Eberhardt:2017fsi, Baykara:2022cwj}, we choose to impose Lorentz-like gauge conditions
$$
\nabla^{a}\d g_{a\m}=\nabla^{a}\d g_{\{ab\}}=\nabla^{a}\d g_{ai}=\nabla^{a}\d B_{a M}=0
$$
where $\{ab\}$ denotes the traceless part. These conditions translate to the following
\begin{equation}
    \nabla^aS_{\mu a}=0,\qquad \nabla^bK_{ab}=0,\qquad\nabla^aW_{ai}=0,
\end{equation}
\begin{equation}
    \nabla^aZ_{\mu a}=0,\qquad\nabla_{[a}V_{b]}=0,\qquad\nabla^aP_{ai}=0.
\end{equation}
These gauge conditions will not fix all of the gauge freedom from diffeomorphisms and $B$-field gauge transformations; we will discuss the residual gauge transformations in section~\ref{subsec:ResidualGaugeTransformations} below.\newline\newline
Plugging these expansions into the equations of motion and expanding to first order in fluctuations leads to the following, starting with the dilaton,
\begin{align}
    0=\ & \frac{1}{4}\nabla^\mu\nabla^\nu H_{\mu\nu}+\hlf\nabla^\mu\nabla^i\wtS_{\mu i}+\frac{1}{4}\nabla^i\nabla^j\wtK_{ij}-\hlf\mcL^{-1}\nabla^aV_a-\hlf\wtmcL^{-1}\nabla^i\wtV_i\non\\
    & +\square_A\left(\phi-\frac{3}{4}M-\frac{3}{4}N-\frac{3}{4}\wtN\right)+\square\left(\phi-M-\hlf N-\frac{3}{4}\wtN\right)+\wtsquare\left(\phi-M-\frac{3}{4}N-\hlf\wtN\right)\non\\
    &+\Lambda\left(\hlf M+\frac{3}{8}N+\frac{3}{8}\wtN\right).
\end{align}
We have introduced abbreviations
\begin{equation}
    \square_A=\nabla^\mu\nabla_\mu,\qquad\square=\nabla^a\nabla_a,\qquad\wtsquare=\nabla^i\nabla_i,
\end{equation}
for the Laplacian operators acting on $AdS_4$ and each of the $S^3$s.\newline\newline
For the metric equations of motion (including the shift by the dilaton equation of motion, and splitting the $\mu\nu$, $ab$, and $ij$ equations into trace and traceless parts), we have
\begin{align}
    0=\ & -\nabla^\mu\nabla^\nu H_{\mu\nu}-\nabla^\mu\nabla^i\wtS_{\mu i}+\square_A\left(-2\phi+3M+\frac{3}{2}N+\frac{3}{2}\wtN\right)+2\left(\square+\wtsquare\right)M\non\\
    &+\Lambda\left(-4\phi-2M\right),\\
    0=\ & -\nabla^\rho\nabla_{\{\mu}H_{\n\}\rho}+\hlf\left(\square_A+\square+\wtsquare\right)H_{\mu\nu}-\hlf\Lambda H_{\mu\nu}-\nabla^i\nabla_{\{\mu}\wtS_{\n\}i}\non\\
    &+\nabla_{\{\mu}\nabla_{\n\}}\left(-2\phi+M+\frac{3}{2}N+\frac{3}{2}\wtN\right),\\
    0=\ & -\hlf\nabla_a\nabla^\nu H_{\mu\nu}+\hlf\left(\square_A+\square+\wtsquare\right)S_{\mu a}-\hlf\nabla_\mu\nabla^\nu S_{\nu a}-L^{-2}S_{\mu a}-\frac{1}{4}\Lambda S_{\mu a}-\hlf\nabla_a\nabla^i\wtS_{\mu i}\non\\
    & \quad -\hlf\nabla_\mu\nabla^iW_{ai}-\mcL^{-1}\e_a{}^{bc}\nabla_bZ_{\mu c}+\mcL^{-1}\nabla_\mu V_a+\nabla_\mu\nabla_a\left(-2\phi+\frac{3}{2}M+N+\frac{3}{2}\wtN\right),\\
    0=\ & -\hlf\nabla_i\nabla^\nu H_{\mu\nu}+\hlf\left(\square_A+\square+\wtsquare\right)\wtS_{\mu i}-\hlf\nabla_\mu\nabla^\nu\wtS_{\nu i}-\hlf\nabla_i\nabla^j\wtS_{\mu j}-\wtL^{-2}\wtS_{\mu i}-\frac{1}{4}\Lambda\wtS_{\mu i}\non\\
    & \quad -\hlf\nabla_\mu\nabla^j\wtK_{ij}-\wtmcL^{-1}\e_i{}^{jk}\nabla_j\wtZ_{\mu k}+\wtmcL^{-1}\nabla_\mu\wtV_i+\nabla_\mu\nabla_i\left(-2\phi+\frac{3}{2}M+\frac{3}{2}N+\wtN\right),\\
    0=\ & 6\mcL^{-1}\nabla^aV_a+\frac{3}{2}\left(\square_A+\wtsquare\right)N-12L^{-2}N+\square\left(-2\phi+2M+2N+\frac{3}{2}\wtN\right)\non\\
    &+\Lambda\left(-3\phi-\frac{9}{2}N\right),\\
    0=\ & -\nabla^\mu\nabla_{(a}S_{|\mu|b)}+\hlf\left(\square_A+\square+\wtsquare\right)K_{ab}-L^{-2}K_{ab}-\nabla^i\nabla_{(a}W_{b)i}\non\\
    &+\nabla_{\{a}\nabla_{b\}}\left(-2\phi+2M+\hlf N+\frac{3}{2}\wtN\right),\\
    0=\ & -\hlf\nabla_i\nabla^\mu S_{\mu a}-\hlf\nabla_a\nabla^\mu\wtS_{\mu i}+\hlf\left(\square_A+\square+\wtsquare\right)W_{ai}-\hlf\nabla_i\nabla^jW_{aj}-\frac{5}{4}\Lambda W_{ai}\non\\
    &-\hlf\nabla_a\nabla^j\wtK_{ij}+\mcL^{-1}\nabla_iV_a+\mcL^{-1}\e_a{}^{bc}\nabla_bP_{ci}-\wtmcL^{-1}\e_i{}^{jk}\nabla_jP_{ak}+\wtmcL^{-1}\nabla_a\wtV_i\non\\
    &+\nabla_a\nabla_i\left(-2\phi+2M+N+\wtN\right),\\
    0=\ & -\nabla^\mu\nabla^i\wtS_{\mu i}-\nabla^i\nabla^j\wtK_{ij}+6\wtmcL^{-1}\nabla^i\wtV_i+\frac{3}{2}\left(\square_A+\square\right)\wtN-12\wtL^{-2}\wtN\non\\
    &+\wtsquare\left(-2\phi+2M+\frac{3}{2}N+2\wtN\right)+\Lambda\left(-3\phi-\frac{9}{2}\wtN\right),\\
    0=\ & -\nabla^\mu\nabla_{\{i}\wtS_{|\mu|j\}}+\hlf\left(\square_A+\square+\wtsquare\right)\wtK_{ij}-\nabla^k\nabla_{\{i}\wtK_{j\}k}+2\wtL^{-2}\wtK_{ij}\non\\
    &+\nabla_{\{i}\nabla_{j\}}\left(-2\phi+2M+\frac{3}{2}N+\hlf\wtN\right).
\end{align}
We have also introduced notation for the traceless symmetric combination of indices, i.e.~
\begin{equation}
    T_{\{ab\}}:=\hlf T_{ab}+\hlf T_{ba}-\frac{1}{3}g_{ab}T^c{}_c.
\end{equation}
Finally the equations from the $B$-field are (and for $B_{ab}$ and $B_{ij}$ components we contract with the $\e^{abc}$ and $\e^{ijk}$ tensors respectively)
\begin{align}
    0=\ & \left(\square_A+\square+\wtsquare\right)X_{\mu\nu}+2\nabla^\rho\nabla_{[\mu}X_{\nu]\rho}+2\nabla^i\nabla_{[\mu}\wtZ_{\nu]i},\\
    0=\ & -2\mcL^{-1}\e_a{}^{bc}\nabla_bS_{\mu c}-\nabla_a\nabla^\nu X_{\mu\nu}+\left(\square_A+\square+\wtsquare\right)Z_{\mu a}-\nabla_\mu\nabla^\nu Z_{\nu a}-2L^{-2}Z_{\mu a}+\hlf\Lambda Z_{\mu a}\non\\
    &-\nabla_a\nabla^i\wtZ_{\mu i}+\nabla_\mu\nabla^iP_{ai},\\
    0=\ & -2\wtmcL^{-1}\e_i{}^{jk}\nabla_j\wtS_{\mu k}-\nabla_i\nabla^\nu X_{\mu\nu}+\left(\square_A+\square+\wtsquare\right)\wtZ_{\mu i}-\nabla_\mu\nabla^\nu\wtZ_{\nu i}-\nabla_i\nabla^j\wtZ_{\mu j}-2\wtL^{-2}\wtZ_{\mu i}\non\\
    &+\hlf\Lambda\wtZ_{\mu i}+\e_i{}^{jk}\nabla_\mu\nabla_j\wtV_k,\\
    0=\ & -4\mcL^{-1}\nabla^\mu S_{\mu a}-4\mcL^{-1}\nabla^iW_{ai}-2\e_a{}^{bc}\nabla^\mu\nabla_bZ_{\mu c}+2\left(\square_A+\wtsquare\right)V_a+2\nabla_a\nabla^bV_b\non\\
    & \quad +2\e_a{}^{bc}\nabla_b\nabla^iP_{ci}+\mcL^{-1}\nabla_a\left(-8\phi+8M-6N+6\wtN\right),\\
    0=\ & 2\mcL^{-1}\e_a{}^{bc}\nabla_bW_{ci}-2\wtmcL^{-1}\e_i{}^{jk}\nabla_jW_{ak}+\nabla_i\nabla^\mu Z_{\mu a}-\nabla_a\nabla^\mu\wtZ_{\mu i}+\left(\square_A+\square+\wtsquare\right)P_{ai}\non\\
    & \quad -\nabla_i\nabla^jP_{aj}-\frac{3}{2}\Lambda P_{ai}+\e_i{}^{jk}\nabla_a\nabla_j\wtV_k,\\
    0=\ & -4\wtmcL^{-1}\nabla^\mu\wtS_{\mu i}-2\e_i{}^{jk}\nabla^\mu\nabla_j\wtZ_{\mu k}+2\left(\square_A+\square\right)\wtV_i+2\nabla_i\nabla^j\wtV_j\non\\
    &+\wtmcL^{-1}\nabla_i\left(-8\phi+8M+6N-6\wtN\right).
\end{align}

\subsection{Spherical harmonic expansion}

We now wish to perform a harmonic expansion on each of the $S^3$s, in order to get the effective four-dimensional theory.  We do this using $S^3$ spherical harmonics \cite{rubin1984eigenvalues, Salam:1981xd, d1984fermion, VanNieuwenhuizen:1985be, Duff:1986hr, castellani1991supergravity, vanNieuwenhuizen:2012zk}.  In particular we will use the spherical harmonic functions $Y^{(\ell,0)}$, $Y^{(\ell,{\pm} 1)}_a$, and $Y^{(\ell,{\pm} 2)}_{ab}$, defined for $\ell\ge 0$, $\ell\ge 1$, and $\ell\ge 2$ respectively.  These obey several useful identities,
\begin{align}
    \square Y^{(\ell,0)}=\ & -L^{-2}\ell(\ell+2)Y^{(\ell,0)},\\
    \nabla^aY^{(\ell,{\pm} 1)}_a=\ & 0,\\
    \square Y^{(\ell,{\pm} 1)}_a=\ & -L^{-2}(\ell^2+2\ell-1)Y^{(\ell,{\pm} 1)}_a,\\
    \e_a{}^{bc}\nabla_bY^{(\ell,{\pm} 1)}_c=\ & \pm L^{-1}(\ell+1)Y^{(\ell,{\pm} 1)}_a,\\
    Y^{(\ell,\pm 2)}_{ba}=\ & Y^{(\ell,{\pm} 2)}_{ab},\\
    g^{ab}Y^{(\ell,{\pm} 2)}_{ab}=\ & 0,\\
    \nabla^bY^{(\ell,{\pm} 2)}_{ab}=\ & 0,\\
    \square Y^{(\ell,{\pm} 2)}_{ab}=\ & -L^{-2}(\ell^2+2\ell-2)Y^{(\ell,{\pm} 2)}_{ab}.
\end{align}
Additionally, for low values of $\ell$ we have special relations
\begin{equation}
    \nabla_aY^{(0,0)}=0,\qquad\nabla_{\{a}\nabla_{b\}}Y^{(1,0)}=0,\qquad\nabla_{(a}Y^{(1,{\pm} 1)}_{b)}=0.
\end{equation}
Now, every scalar on $S^3$ can be expanded in a linear combination of the $Y^{(\ell,0)}$, $\ell\ge 0$, every vector on $S^3$ can be expanded as a linear combination of $Y^{(\ell,1)}_a$, $Y^{(\ell,-1)}_a$, and $\nabla_aY^{(\ell,0)}$ for $\ell\ge 1$, and every traceless symmetric tensor on $S^3$ can be expanded in a linear combination of $Y^{(\ell,2)}_{ab}$, $Y^{(\ell,-2)}_{ab}$, $\nabla_{(a}Y^{(\ell,1)}_{b)}$, $\nabla_{(a}Y^{(\ell,-1)}_{b)}$, and $\nabla_{\{a}\nabla_{b\}}Y^{(\ell,0)}$, for $\ell\ge 2$.  Then every one of our fluctuating fields can be expanded in this way, for example,
\begin{align}
    \phi=\ & \sum_{\ell=0}^\infty\sum_{\wtell=0}^\infty\phi^{(\ell,0)(\wtell,0)}Y^{(\ell,0)}\widetilde{Y}^{(\wtell,0)},\\
    S_{\mu a}=\ & \sum_{\ell=1}^\infty\sum_{\wtell=0}^\infty\left(S_\mu^{(\ell,1)(\wtell,0)}Y^{(\ell,1)}_a+S_\mu^{(\ell,-1)(\wtell,0)}Y^{(\ell,-1)}_a+S_\mu^{(\ell,0)(\wtell,0)}\nabla_aY^{(\ell,0)}\right)\widetilde{Y}^{(\wtell,0)},\\
    K_{ab}=\ & \sum_{\ell=2}^\infty\sum_{\wtell=0}^\infty\left(K^{(\ell,2)(\wtell,0)}Y^{(\ell,2)}_{ab}+K^{(\ell,-2)(\wtell,0)}Y^{(\ell,-2)}_{ab}+K^{(\ell,1)(\wtell,0)}\nabla_{(a}Y^{(\ell,1)}_{b)}+K^{(\ell,-1)(\wtell,0)}\nabla_{(a}Y^{(\ell,-1)}_{b)}\right.\non\\
    & \qquad\left.+K^{(\ell,0)(\wtell,0)}\nabla_{\{a}\nabla_{b\}}Y^{(\ell,0)}\right)\widetilde{Y}^{(\wtell,0)},\\
    P_{ai}=\ & \sum_{\ell=1}^\infty\sum_{\wtell=1}^\infty\left(P^{(\ell,1)(\wtell,1)}Y^{(\ell,1)}_a\widetilde{Y}^{(\wtell,1)}_i+P^{(\ell,1)(\wtell,-1)}Y^{(\ell,1)}_a\widetilde{Y}^{(\wtell,-1)}_i+P^{(\ell,-1)(\wtell,1)}Y^{(\ell,-1)}_a\widetilde{Y}^{(\wtell,1)}_i\right.\non\\
    &\qquad\left. +P^{(\ell,-1)(\wtell,-1)}Y^{(\ell,-1)}_a\widetilde{Y}^{(\wtell,-1)}_i+P^{(\ell,1)(\wtell,0)}Y^{(\ell,1)}_a\nabla_i\widetilde{Y}^{(\wtell,0)}+P^{(\ell,-1)(\wtell,0)}Y^{(\ell,-1)}_a\nabla_i\widetilde{Y}^{(\wtell,0)}\right.\non\\
    &\qquad\left. +P^{(\ell,0)(\wtell,1)}\nabla_aY^{(\ell,0)}\widetilde{Y}^{(\wtell,1)}_i+P^{(\ell,0)(\wtell,-1)}\nabla_aY^{(\ell,0)}\widetilde{Y}^{(\wtell,-1)}_i+P^{(\ell,0)(\wtell,0)}\nabla_aY^{(\ell,0)}\nabla_i\widetilde{Y}^{(\wtell,0)}\right).
\end{align}
Our gauge condition now becomes simply the statement that certain components in this expansion must vanish.  For instance, since
\begin{equation}
    \nabla^aS_{\mu a}=\sum_{\ell=1}^\infty\sum_{\wtell=0}^\infty\left( -L^{-2}\ell(\ell+2)S_\mu^{(\ell,0)(\wtell,0)}Y^{(\ell,0)}\right)\widetilde{Y}^{(\wtell,0)},
\end{equation}
the gauge condition $\nabla^aS_{\mu a}=0$ is simply equivalent to the condition that $S_\mu^{(\ell,0)(\wtell,0)}=0$, $\ell\ge 1,\wtell\ge 0$.  All together, the gauge conditions imply
\begin{equation}
\label{eq:SHMetricGaugeConditions}
    S_\mu^{(\ell,0)(\wtell,0)}=K^{(\ell,{\pm} 1)(\wtell,0)}=K^{(\ell,0)(\wtell,0)}=W^{(\ell,0)(\wtell,{\pm} 1)}=W^{(\ell,0)(\wtell,0)}=0,
\end{equation}
\begin{equation}
\label{eq:SHBFieldGaugeConditions}
    Z_\mu^{(\ell,0)(\wtell,0)}=V^{(\ell,{\pm} 1)(\wtell,0)}=P^{(\ell,0)(\wtell,{\pm} 1)}=P^{(\ell,0)(\wtell,0)}=0.
\end{equation}

The surviving fluctuating fields are then
\begin{align}
    \ell\ge 0,\wtell\ge 0:\ & \phi^{(\ell,0)(\wtell,0)},\ H_{\mu\nu}^{(\ell,0)(\wtell,0)},\ M^{(\ell,0)(\wtell,0)},\ N^{(\ell,0)(\wtell,0)},\ \wtN^{(\ell,0)(\wtell,0)},\ X_{\mu\nu}^{(\ell,0)(\wtell,0)},\\
    \ell\ge 1,\wtell\ge 0:\ & S_\mu^{(\ell,{\pm} 1)(\wtell,0)},\ Z_\mu^{(\ell,{\pm} 1)(\wtell,0)},\ V^{(\ell,0)(\wtell,0)},\\
    \ell\ge 0,\wtell\ge 1:\ & \wtS_\mu^{(\ell,0)(\wtell,{\pm} 1)},\ \wtS_\mu^{(\ell,0)(\wtell,0)},\ \wtZ_\mu^{(\ell,0)(\wtell,{\pm} 1)},\ \wtZ_\mu^{(\ell,0)(\wtell,0)},\ \wtV^{(\ell,0)(\wtell,{\pm} 1)},\ \wtV^{(\ell,0)(\wtell,0)},\\
    \ell\ge 2,\wtell\ge 0:\ & K^{(\ell,{\pm} 2)(\wtell,0)},\\
    \ell\ge 1,\wtell\ge 1:\ & W^{(\ell,{\pm} 1)(\wtell,{\pm'}1)},\ W^{(\ell,{\pm} 1)(\wtell,0)},\ P^{(\ell,{\pm} 1)(\wtell,{\pm'}1)},\ P^{(\ell,{\pm} 1)(\wtell,0)},\\
    \ell\ge 0,\wtell\ge 2:\ & \wtK^{(\ell,0)(\wtell,{\pm} 2)},\ \wtK^{(\ell,0)(\wtell,{\pm} 1)},\ \wtK^{(\ell,0)(\wtell,0)},
\end{align}
where the prime on $\pm'$ means that this sign is independent of the sign without the prime i.e. $\pm$. We can also take the equations of motion and decompose them into spherical harmonic components.  Each equation will only contain fluctuating fields with the same spherical harmonic structure, so we don't need to bother writing the $(\ell,m)(\wtell,\widetilde{m})$ superscripts.  For each sector $(\ell,m)(\wtell,\widetilde{m})$ we list the equations that appear, grouped by the range of $\ell$ and $\wtell$ for which the equations apply.\newline\newline
$(\ell,{\pm} 2)(\wtell,0)\text{ sectors with }\ell\ge 2,\wtell\ge 0:$
\be\label{eq:l2hl0EoM}
0=\hlf\square_AK+\left[-\hlf L^{-2}\ell(\ell+2)-\hlf\wtL^{-2}\wtell(\wtell+2)\right]K,
\ee
$(\ell\,{\pm} 1)(\wtell\,{\pm'}1)\text{ sectors with }\ \ell\ge 1,\wtell\ge 1:$
\begin{align}\label{eq:l1hl1WEoM}
0&=\hlf\square_AW+\left[-\hlf L^{-2}(\ell+1)^2-\hlf\wtL^{-2}(\wtell+1)^2-\hlf\Lambda\right]W+\left[\pm\frac{\ell+1}{L\mcL}\mp'\frac{\wtell+1}{\wtL\wtmcL}\right]P\\
\label{eq:l1hl1PEoM}
0&=\square_AP+\left[\pm 2\mcL^{-1}L^{-1}(\ell+1)\mp'2\wtmcL^{-1}\wtL^{-1}(\wtell+1)\right]W+\left[ -L^{-2}(\ell+1)^2-\wtL^{-2}(\wtell+1)^2\right]P
\end{align}
$(\ell,0)(\wtell,{\pm 2})\text{ sectors with }\ell\ge 0,\wtell\ge 2:$
\be\label{eq:l0hl2EoM}
0=\hlf\square_A\wtK+\left[-\hlf L^{-2}\ell(\ell+2)-\hlf\wtL^{-2}\wtell(\wtell+2)\right]\wtK,
\ee
$(\ell,{\pm}1)(\wtell,0)\text{ sectors with }\ell\ge 2,\wtell\ge 0:$
\be\label{eq:l1hl0SDiv}
0=-\nabla^\mu S_\mu+\wtL^{-2}\wtell(\wtell+2)W,
\ee
$(\ell,{\pm}1)(\wtell,0)\text{ sectors with }\ell\ge 1,\wtell\ge 1:$
\begin{align}\label{eq:l1hl0WEoM}
0&=-\hlf\nabla^\mu S_\mu+\hlf\square_AW+\left[ -\hlf L^{-2}(\ell+1)^2-\hlf\Lambda\right]W\pm\mcL^{-1}L^{-1}(\ell+1)P,\\
\label{eq:l1hl0PEoM}
0&=\square_AP+\nabla^\mu Z_\mu\pm 2\mcL^{-1}L^{-1}(\ell+1)W-L^{-2}(\ell+1)^2P
\end{align}
$(\ell,{\pm}1)(\wtell,0)\text{ sectors with }\ell\ge 1,\wtell\ge 0:$
\begin{align}
\label{eq:l1hl0SEoM}
0&=\hlf\square_AS_\mu-\hlf\nabla_\mu\nabla^\nu S_\nu+\left[ -\hlf L^{-2}(\ell+1)^2-\hlf\wtL^{-2}\wtell(\wtell+2)-\frac{1}{4}\Lambda\right]S_\mu+\hlf\wtL^{-2}\wtell(\wtell+2)\nabla_\mu W\non\\
&\qquad\mp\mcL^{-1}L^{-1}(\ell+1)Z_\mu,\\
\label{eq:l1hl0ZEoM}
0&=\square_AZ_\mu-\nabla_\mu\nabla^\nu Z_\nu\mp 2\mcL^{-1}L^{-1}(\ell+1)S_\mu+\left[ -L^{-2}(\ell+1)^2-\wtL^{-2}\wtell(\wtell+2)+\hlf\Lambda\right]Z_\mu\non\\
&\qquad-\wtL^{-2}\wtell(\wtell+2)\nabla_\mu P,\\
\label{eq:l1hl0ZDiv}
0&=-4\mcL^{-1}\nabla^\mu S_\mu+4\mcL^{-1}\wtL^{-2}\wtell(\wtell+2)W\mp 2L^{-1}(\ell+1)\nabla^\mu Z_\mu\mp 2L^{-1}\wtL^{-2}(\ell+1)\wtell(\wtell+2)P,
\end{align}
$(\ell,0)(\wtell,\pm 1)\text{ sectors with }\ell\ge 1,\wtell\ge 1:$
\begin{align}\label{eq:l0hl1SDiv}
0&=-\hlf\nabla^\mu\wtS_\mu+\frac{1}{4}\wtL^{-2}(\wtell^2+2\wtell-3)\wtK+\wtmcL^{-1}\wtV,\\
\label{eq:l0hl1ZDiv}
0&=-\nabla^\mu\wtZ_\mu\pm\wtL^{-1}(\wtell+1)\wtV,
\end{align}
$(\ell,0)(\wtell,\pm 1)\text{ sectors with } \ell\ge 0,\wtell\ge 2:$
\be\label{eq:l0hl1KEoM}
0=\hlf\square_A\wtK-\nabla^\mu\wtS_\mu-\hlf L^{-2}\ell(\ell+2)\wtK,
\ee
$(\ell,0)(\wtell,\pm 1)\text{ sectors with } \ell\ge 0,\wtell\ge 1:$
\begin{align}\label{eq:l0hl1SEoM}
0&=\hlf\square_A\wtS_\mu-\hlf\nabla_\mu\nabla^\nu\wtS_\nu+\left[ -\hlf L^{-2}\ell(\ell+2)-\hlf\wtL^{-2}(\wtell+1)^2-\frac{1}{4}\Lambda\right]\wtS_\mu\non\\
&\qquad+\frac{1}{4}\wtL^{-2}(\wtell^2+2\wtell-3)\nabla_\mu\wtK\mp\wtmcL^{-1}\wtL^{-1}(\wtell+1)\wtZ_\mu+\wtmcL^{-1}\nabla_\mu\wtV,\\
\label{eq:l0hl1ZEoM}
0&=\square_A\wtZ_\mu-\nabla_\mu\nabla^\nu\wtZ_\nu\mp 2\wtmcL^{-1}\wtL^{-1}(\wtell+1)\wtS_\mu+\left[ -L^{-2}\ell(\ell+2)-\wtL^{-2}(\wtell+1)^2+\hlf\Lambda\right]\wtZ_\mu\non\\
&\qquad\pm\wtL^{-1}(\wtell+1)\nabla_\mu\wtV,\\
\label{eq:l0hl1VEoM}
0&=2\square_A\wtV-4\wtmcL^{-1}\nabla^\mu\wtS_\mu\mp 2\wtL^{-1}(\wtell+1)\nabla^\mu\wtZ_\mu-2L^{-2}\ell(\ell+2)\wtV,
\end{align}
$(\ell,0)(\wtell,0)\text{ sectors with }\ell\ge 2,\wtell\ge 0:$
\be\label{eq:l0hl0Algebraic}
0=-2\phi+2M+\hlf N+\frac{3}{2}\wtN,
\ee
$(\ell,0)(\wtell,0)\text{ sectors with }\ell\ge 1,\wtell\ge 1:$
\begin{align}\label{eq:l0hl0hSDiv}
0&=-\hlf\nabla^\mu\wtS_\mu+\frac{1}{3}\wtL^{-2}(\wtell^2+2\wtell-3)\wtK+\mcL^{-1}V+\wtmcL^{-1}\wtV-2\phi+2M+N+\wtN,\\
\label{eq:l0hl0hZDiv}
0&=-\nabla^\mu\wtZ_\mu,
\end{align}
$(\ell,0)(\wtell,0)\text{ sectors with }\ell\ge 0,\wtell\ge 2:$
\be\label{eq:l0hl0hKEoM}
0=\hlf\square_A\wtK-\nabla^\mu\wtS_\mu+\left[ -\hlf L^{-2}\ell(\ell+2)+\frac{1}{6}\wtL^{-2}\wtell(\wtell+2)\right]\wtK-2\phi+2M+\frac{3}{2}N+\hlf\wtN,
\ee
$(\ell,0)(\wtell,0)\text{ sectors with }\ell\ge 1,\wtell\ge 0:$
\begin{align}\label{eq:l0hl0HDiv}
0&=-\hlf\nabla^\nu H_{\mu\nu}+\hlf\wtL^{-2}\wtell(\wtell+2)\wtS_\mu+\mcL^{-1}\nabla_\mu V+\nabla_\mu\left( -2\phi+\frac{3}{2}M+N+\frac{3}{2}\wtN\right),\\
\label{eq:l0hl0XDiv}
0&=-\nabla^\nu X_{\mu\nu}+\wtL^{-2}\wtell(\wtell+2)\wtZ_\mu,\\
\label{eq:l0hl0VEoM}
0&=2\square_AV+\left[-2L^{-2}\ell(\ell+2)-2\wtL^{-2}\wtell(\wtell+2)\right]V+\mcL^{-1}\left(-8\phi+8M-6N+6\wtN\right),
\end{align}
$(\ell,0)(\wtell,0)\text{ sectors with }\ell\ge 0,\wtell\ge 1:$
\begin{align}\label{eq:l0hl0hSEoM}
0&=\hlf\square_A\wtS_\mu-\hlf\nabla_\mu\nabla^\nu\wtS_\nu-\hlf\nabla^\nu H_{\mu\nu}+\left[ -\hlf L^{-2}\ell(\ell+2)-\frac{1}{4}\Lambda\right]\wtS_\mu\non\\
&\qquad +\frac{1}{3}\wtL^{-2}(\wtell^2+2\wtell-3)\nabla_\mu\wtK+\wtmcL^{-1}\nabla_\mu\wtV+\nabla_\mu\left(-2\phi+\frac{3}{2}M+\frac{3}{2}N+\wtN\right),\\
\label{eq:l0hl0hZEoM}
0&=\square_A\wtZ_\mu-\nabla_\mu\nabla^\nu\wtZ_\nu-\nabla^\nu X_{\mu\nu}+\left[ -L^{-2}\ell(\ell+2)+\hlf\Lambda\right]\wtZ_\mu,\\
\label{eq:l0hl0hVEoM}
0&=2\square_A\wtV-4\wtmcL^{-1}\nabla^\mu\wtS_\mu+\left[ -2L^{-2}\ell(\ell+2)-2\wtL^{-2}\wtell(\wtell+2)\right]\wtV\non\\
&\qquad+\wtmcL^{-1}\left( -8\phi+8M+6N-6\wtN\right),
\end{align}
$(\ell,0)(\wtell,0)\text{ sectors with }\ell\ge 0,\wtell\ge 0:$
\begin{align}\label{eq:l0hl0phiEoM}
0&=\square_A\left(\phi-\frac{3}{4}M-\frac{3}{4}N-\frac{3}{4}\wtN\right)+\frac{1}{4}\nabla^\mu\nabla^\nu H_{\mu\nu}-\hlf\wtL^{-2}\wtell(\wtell+2)\nabla^\mu\wtS_\mu+\hlf\wtmcL^{-1}\wtL^{-2}\wtell(\wtell+2)\wtV\non\\
&\quad +\frac{1}{6}\wtL^{-4}\wtell(\wtell-1)(\wtell+2)(\wtell+3)\wtK+\hlf\mcL^{-1}L^{-2}\ell(\ell+2)V-\frac{\ell(\ell+2)}{L^2}\left(\phi-M-\frac{N}{2} -\frac{3\wtN}{4}\right)\non\\
& \quad -\wtL^{-2}\wtell(\wtell+2)\left(\phi-M-\frac{3}{4}N-\hlf\wtN\right)+\Lambda\left(\hlf M+\frac{3}{8}N+\frac{3}{8}\wtN\right),\\
\label{eq:l0hl0MEoM}
0&=\square_A\left(-2\phi+3M+\frac{3}{2}N+\frac{3}{2}\wtN\right)-\nabla^\mu\nabla^\nu H_{\mu\nu}+\wtL^{-2}\wtell(\wtell+2)\nabla^\mu\wtS_\mu\non\\
& \qquad +\left[ -2L^{-2}\ell(\ell+2)-2\wtL^{-2}\wtell(\wtell+2)\right]M+\Lambda\left(-4\phi-2M\right),\\
\label{eq:l0hl0HEoM}
0&=\hlf\square_AH_{\mu\nu}-\nabla^\rho\nabla_{\{\mu}H_{\nu\}\rho}+\left[ -\hlf L^{-2}\ell(\ell+2)-\hlf\wtL^{-2}\wtell(\wtell+2)-\hlf\Lambda\right]H_{\mu\nu}\non\\
& \qquad +\wtL^{-2}\wtell(\wtell+2)\nabla_{\{\mu}\wtS_{\nu\}}+\nabla_{\{\mu}\nabla_{\nu\}}\left(-2\phi+M+\frac{3}{2}N+\frac{3}{2}\wtN\right),\\
\label{eq:l0hl0NEoM}
0&=\frac{3}{2}\square_AN-6\mcL^{-1}L^{-2}\ell(\ell+2)V-12L^{-2}N-\frac{3}{2}\wtL^{-2}\wtell(\wtell+2)N\non\\
& \qquad -L^{-2}\ell(\ell+2)\left(-2\phi+2M+2N+\frac{3}{2}\wtN\right)+\Lambda\left(-3\phi-\frac{9}{2}N\right),\\
\label{eq:l0hl0hNEoM}
0&=\frac{3}{2}\square_A\wtN+\wtL^{-2}\wtell(\wtell+2)\nabla^\mu\wtS_\mu-\frac{2}{3}\wtL^{-4}\wtell(\wtell-1)(\wtell+2)(\wtell+3)\wtK-6\wtmcL^{-1}\wtL^{-2}\wtell(\wtell+2)\wtV-12\wtL^{-2}\wtN\non\\
& \qquad -\frac{3}{2}L^{-2}\ell(\ell+2)\wtN-\wtL^{-2}\wtell(\wtell+2)\left(-2\phi+2M+\frac{3}{2}N+2\wtN\right)+\Lambda\left(-3\phi-\frac{9}{2}\wtN\right),\\
\label{eq:l0hl0XEoM}
0&=\square_AX_{\mu\nu}+2\nabla^\rho\nabla_{[\mu}X_{\nu]\rho}+\left[ -L^{-2}\ell(\ell+2)-\wtL^{-2}\wtell(\wtell+2)\right]X_{\mu\nu}-2\wtL^{-2}\wtell(\wtell+2)\nabla_{[\mu}\wtZ_{\nu]}.
\end{align}
\subsection{Residual gauge transformations}
\label{subsec:ResidualGaugeTransformations}
Next we will take a closer look at the gauge transformations, starting with the ten-dimensional diffeomorphisms.  Taking the diffeomorphism parameter as $\xi_M$, we can decompose it as
\begin{align}
    \xi_\mu=\ & \sum_{\ell=0}^\infty\sum_{\wtell=0}^\infty\al_\mu^{(\ell,0)(\wtell,0)}Y^{(\ell,0)}\widetilde{Y}^{(\wtell,0)},\\
    \xi_a=\ & \sum_{\ell=1}^\infty\sum_{\wtell=0}^\infty\left(\beta^{(\ell,1)(\wtell,0)}Y^{(\ell,1)}_a+\beta^{(\ell,-1)(\wtell,0)}Y^{(\ell,-1)}_a+\beta^{(\ell,0)(\wtell,0)}\nabla_aY^{(\ell,0)}\right)\widetilde{Y}^{(\wtell,0)},\\
    \xi_i=\ & \sum_{\ell=0}^\infty\sum_{\wtell=1}^\infty\left(\g^{(\ell,0)(\wtell,1)}\widetilde{Y}^{(\wtell,1)}_i+\g^{(\ell,0)(\wtell,-1)}\widetilde{Y}^{(\wtell,-1)}_i+\g^{(\ell,0)(\wtell,0)}\nabla_i\widetilde{Y}^{(\wtell,0)}\right)Y^{(\ell,0)}.
\end{align}
Then for example from $\d S_{\mu a}=\nabla_\mu\xi_a+\nabla_a\xi_\mu$, we would have
\begin{equation}
    \d S_\mu^{(\ell,0)(\wtell,0)}=\nabla_\mu\beta^{(\ell,0)(\wtell,0)}+\al_\mu^{(\ell,0)(\wtell,0)},
\end{equation}
thus preserving the gauge condition $S_\mu^{(\ell,0)(\wtell,0)}=0$ requires $\al_\mu^{(\ell,0)(\wtell,0)}=-\nabla_\mu\beta^{(\ell,0)(\wtell,0)}$ for ($\ell\ge 1$ and $\wtell\ge 0$).  Similarly, the other conditions \eqref{eq:SHMetricGaugeConditions} lead to the following four conditions
$$
\beta^{(\ell,{\pm}1)(\wtell,0)}=0\qquad (\ell\ge 2,\wtell\ge 0),
$$
$$
\beta^{(\ell,0)(\wtell,0)}=0\qquad (\ell\ge 2,\wtell\ge 0),
$$ 
$$
\g^{(\ell,0)(\wtell,{\pm}1)}=0\qquad (\ell\ge 1,\wtell\ge 1),
$$
$$
\g^{(\ell,0)(\wtell,0)}=-\beta^{(\ell,0)(\wtell,0)}\qquad (\ell\ge 1,\wtell\ge 1).
$$
The residual unfixed gauge transformations are
\begin{equation}
    \al_\mu^{(0,0)(\wtell,0)},\ \beta^{(1,{\pm}1)(\wtell,0)},\ \beta^{(1,0)(\wtell,0)},\quad(\wtell\ge 0),\qquad\qquad \g^{(0,0)(\wtell,{\pm}1)},\ \g^{(0,0)(\wtell,0)},\quad(\wtell\ge 1),
\end{equation}
with
\begin{equation}
    \g^{(1,0)(\wtell,0)}=-\beta^{(1,0)(\wtell,0)},\quad(\wtell\ge 1),\qquad\qquad\al_\mu^{(1,0)(\wtell,0)}=-\nabla_\mu\beta^{(1,0)(\wtell,0)},\quad(\wtell\ge 0).
\end{equation}
The $B$-field components will also transform under these residual diffeomorphisms, and by combining with compensating $B$-field gauge transformations, we can preserve the gauge conditions.  In practice the easiest way to determine the transformations of the $B$-field components under these residual diffeomorphisms is simply to demand that all the equations of motion are invariant.\newline\newline
For the $B$-field gauge transformations, we similarly decompose the gauge parameter,
\begin{align}
    \lambda_\mu=\ & \sum_{\ell=0}^\infty\sum_{\wtell=0}^\infty\la_\mu^{(\ell,0)(\wtell,0)}Y^{(\ell,0)}\widetilde{Y}^{(\wtell,0)},\\
    \la_a=\ & \sum_{\ell=1}^\infty\sum_{\wtell=0}^\infty\left(\mu^{(\ell,1)(\wtell,0)}Y^{(\ell,1)}_a+\mu^{(\ell,-1)(\wtell,0)}Y^{(\ell,-1)}_a+\mu^{(\ell,0)(\wtell,0)}\nabla_aY^{(\ell,0)}\right)\widetilde{Y}^{(\wtell,0)},\\
    \la_i=\ & \sum_{\ell=0}^\infty\sum_{\wtell=1}^\infty\left(\nu^{(\ell,0)(\wtell,1)}\widetilde{Y}^{(\wtell,1)}_i+\nu^{(\ell,0)(\wtell,{-}1)}\widetilde{Y}^{(\ell,{-}1)}+\nu^{(\ell,0)(\wtell,0)}\nabla_i\widetilde{Y}^{(\wtell,0)}\right)Y^{(\ell,0)}.
\end{align}
Preserving the gauge conditions \eqref{eq:SHBFieldGaugeConditions} then forces the following
\begin{align}
\mu^{(\ell,{\pm}1)(\wtell,0)}&=0\qquad (\ell\ge 1,\wtell\ge 0),\\
\nu^{(\ell,0)(\wtell,{\pm}1)}&=0 \qquad (\ell\ge 1,\wtell\ge 1),\\
\la_\mu^{(\ell,0)(\wtell,0)}&=\nabla_\mu\mu^{(\ell,0)(\wtell,0)} \qquad(\ell\ge 1,\wtell\ge 0),\\
\nu^{(\ell,0)(\wtell,0)}&=\mu^{(\ell,0)(\wtell,0)} \qquad (\ell\ge 1,\wtell\ge 1).
\end{align}
Moreover, transformations satisfying any of the following three conditions
\be
\nu^{(\ell,0)(\wtell,0)}=\mu^{(\ell,0)(\wtell,0)}\quad\text{and}\quad\la_\mu^{(\ell,0)(\wtell,0)}=\nabla_\mu\mu^{(\ell,0)(\wtell,0)}\quad\text{for}\quad\ell\ge 1,\wtell\ge 1,
\ee
\be
\la_\mu^{(\ell,0)(0,0)}=\nabla_\mu\mu^{(\ell,0)(0,0)}\quad\text{for}\quad\ell\ge 1,
\ee
\be
\la_\mu^{(0,0)(\wtell,0)}=\nabla_\mu\nu^{(0,0)(\wtell,0)}\quad\text{for}\quad\wtell\ge 1,
\ee
correspond to exact one-form gauge parameters under which no field transforms, so we can discard those cases.  The remaining residual gauge transformations then correspond to
\begin{equation}
    \la_\mu^{(0,0)(\wtell,0)}\quad(\wtell\ge 0),\qquad\qquad\nu^{(0,0)(\wtell,{\pm}1)},\ \nu^{(0,0)(\wtell,0)}\quad(\wtell\ge 1).
\end{equation}
Altogether, the nonvanishing transformations of our fluctuating fields under these residual gauge transformations for $\ell=1$ are given as
\begin{align}
    \d S_\mu^{(1\,{\pm}1)(\wtell,0)}=\ & \nabla_\mu\beta^{(1,{\pm}1)(\wtell,0)}\qquad(\wtell\ge 0),\\
    \d W^{(1,{\pm}1)(\wtell,0)}=\ & \beta^{(1,{\pm}1)(\wtell,0)}\qquad(\wtell\ge 1),\\
    \d Z_\mu^{(1,{\pm}1)(\wtell,0)}=\ & \mp\mcL^{-1}L\nabla_\mu\beta^{(1,{\pm}1)(\wtell,0)}\qquad(\wtell\ge 0),\\
    \d P^{(1,{\pm}1)(\wtell,0)}=\ & \pm\mcL^{-1}L\beta^{(1\,{\pm}1)(\wtell,0)}\qquad(\wtell\ge 1),\\
    \d H_{\mu\nu}^{(1,0)(\wtell,0)}=\ & -2\nabla_{\{\mu}\nabla_{\nu\}}\beta^{(1,0)(\wtell,0)}\qquad(\wtell\ge 0),\\
    \d M^{(1,0)(\wtell,0)}=\ & -\hlf\square_A\beta^{(1,0)(\wtell,0)}\qquad(\wtell\ge 0),\\
    \d\wtS_\mu^{(1,0)(\wtell,0)}=\ & -2\nabla_\mu\beta^{(1,0)(\wtell,0)}\qquad(\wtell\ge 1),\\
    \d N^{(1,0)(\wtell,0)}=\ & -2L^{-2}\beta^{(1,0)(\wtell,0)}\qquad(\wtell\ge 0),\\
    \d\wtK^{(1,0)(\wtell,0)}=\ & -2\beta^{(1,0)(\wtell,0)}\qquad(\wtell\ge 2),\\
    \d\wtN^{(1,0)(\wtell,0)}=\ & \frac{2}{3}\wtL^{-2}\wtell(\wtell+2)\beta^{(1,0)(\wtell,0)}\qquad(\wtell\ge 1),\\
    \d V^{(1,0)(\wtell,0)}=\ & 2\mcL^{-1}\beta^{(1,0)(\wtell,0)}\qquad(\wtell\ge 0),\\
    \d\wtV^{(1,0)(\wtell,0)}=\ & -2\wtmcL^{-1}\beta^{(1,0)(\wtell,0)}\qquad(\wtell\ge 1),
\end{align}
and the nonvanishing transformations for $\ell=0$ are given as
\begin{align}
    \d\wtS_\mu^{(0,0)(\wtell,{\pm}1)}=\ & \nabla_\mu\g^{(0,0)(\wtell,{\pm}1)}\qquad(\wtell\ge 1),\\
    \d\wtK^{(0,0)(\wtell,{\pm}1)}=\ & 2\g^{(0,0)(\wtell,{\pm}1)}\qquad(\wtell\ge 2),\\
    \d\wtZ_\mu^{(0,0)(\wtell,{\pm}1)}=\ & \nabla_\mu\nu^{(0,0)(\wtell,{\pm}1)}\qquad(\wtell\ge 1),\\
    \d\wtV^{(0,0)(\wtell,{\pm}1)}=\ & 2\wtmcL^{-1}\g^{(0,0)(\wtell,{\pm}1)}\pm\wtL^{-1}(\wtell+1)\nu^{(0,0)(\wtell,{\pm}1)}\qquad(\wtell\ge 1),\\
    \d H_{\mu\nu}^{(0,0)(\wtell,0)}=\ & 2\nabla_{\{\m}\al_{\n\}}^{(0,0)(\wtell,0)}\qquad(\wtell\ge 0),\\
    \d M^{(0,0)(\wtell,0)}=\ & \hlf\nabla^\mu\al_\mu^{(0,0)(\wtell,0)}\qquad(\wtell\ge 0),\\
    \d\wtS_\mu^{(0,0)(\wtell,0)}=\ & \al_\mu^{(0,0)(\wtell,0)}+\nabla_\mu\g^{(0,0)(\wtell,0)}\qquad(\wtell\ge 1),\\
    \d\wtK^{(0,0)(\wtell,0)}=\ & 2\g^{(0,0)(\wtell,0)}\qquad(\wtell\ge 2),\\
    \d\wtN^{(0,0)(\wtell,0)}=\ & -\frac{2}{3}\wtL^{-2}\wtell(\wtell+2)\g^{(0,0)(\wtell,0)}\qquad(\wtell\ge 1),\\
    \d X_{\mu\nu}^{(0,0)(\wtell,0)}=\ & 2\nabla_{[\mu}\la_{\nu]}^{(0,0)(\wtell,0)}\qquad(\wtell\ge 0),\\
    \d\wtZ_\mu^{(0,0)(\wtell,0)}=\ & \nabla_\mu\nu^{(0,0)(\wtell,0)}-\la_\mu^{(0,0)(\wtell,0)}\qquad(\wtell\ge 1),\\
    \d\wtV^{(0,0)(\wtell,0)}=\ & 2\wtmcL^{-1}\g^{(0,0)(\wtell,0)}\qquad(\wtell\ge 1).
\end{align}

\subsection{Spectrum of fluctuations}
\begin{figure}
    \centering
    \includegraphics[width=0.65\linewidth]{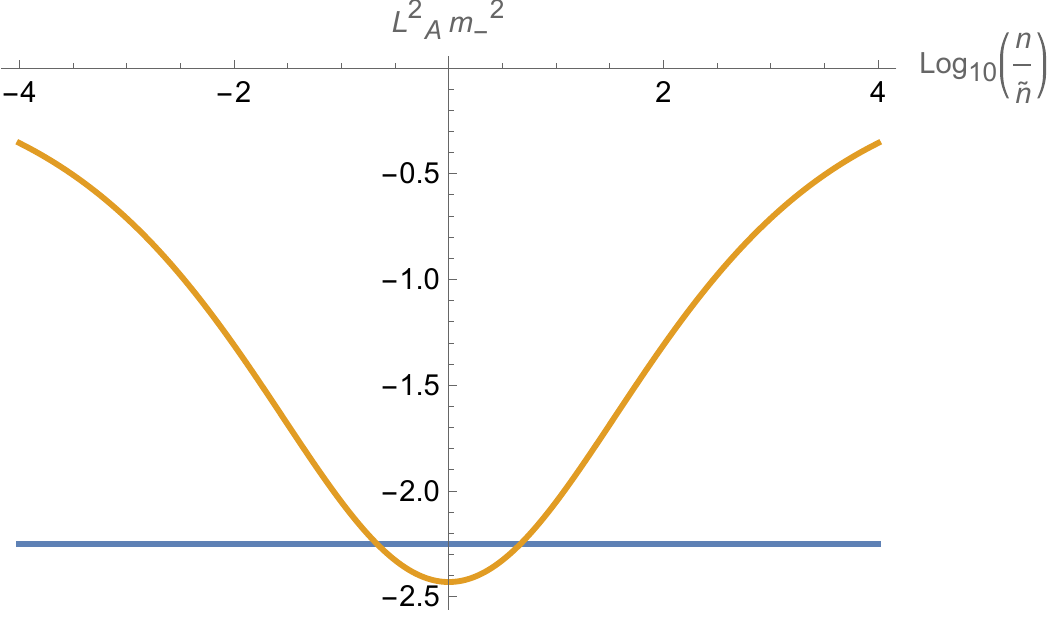}
    \caption{The plot of $L^{2}_{A}m^{2}_{-}$ with $m^{2}_{-}$ coming from \eqref{(l,pm1)(lh,pm'1), l geq 1, lh geq 1 scalar} for $\ell=\ellh=1$ and $\pm'=\mp$ (orange) that violates the BF bound (blue) against $\log(n/\widetilde{n})$.}
    \label{fig: BF bound violating eigenvalues}
\end{figure}

Now we are finally ready to take a look at the spectrum.  Let's start with $(\ell,{\pm}2)(\wtell,0)$, for $\ell\ge 2,\wtell\ge 0$.  There is only one scalar field, $K^{(\ell,{\pm}2)(\wtell,0)}$, in each of these sectors, and it is always gauge invariant.  Its equation of motion~\eqref{eq:l2hl0EoM} is
\begin{equation}
    \square_AK=\left[L^{-2}\ell(\ell+2)+\wtL^{-2}\wtell(\wtell+2)\right]K,
\end{equation}
from which we read off the mass,
\begin{equation}\label{(l,pm2)(lh,0), l geq 2, lh geq 0 scalar}
    m^2=L^{-2}\ell(\ell+2)+\wtL^{-2}\wtell(\wtell+2),
\end{equation}
which is always positive.\newline\newline
Similarly, for the $(\ell,0)(\wtell,{\pm}2)$ sector with $\ell\ge 0,\wtell\ge 2$, we have only the gauge invariant scalar $\wtK$ and from~\eqref{eq:l0hl2EoM} we find the same positive mass as above.  Note that taking the $(\ell,{\pm}2)(\wtell,0)$ and $(\ell,0)(\wtell,{\pm}2)$ sectors combined gives a symmetry under exchange of the two $S^3$s, as should be the case. In the sector $(\ell,{\pm}1)(\wtell,{\pm'}1)$ for $\ell\ge 1,\wtell\ge 1$, there are two gauge invariant scalars, $W$ and $P$, whose equations~\eqref{eq:l1hl1WEoM} and~\eqref{eq:l1hl1PEoM} can be summarized as
\begin{align}
    \square_AW=\ & \left[L^{-2}(\ell+1)^2+\wtL^{-2}(\wtell+1)^2+\Lambda\right]W+\left[\mp 2\mcL^{-1}L^{-1}(\ell+1)\pm' 2\wtmcL^{-1}\wtL^{-1}(\wtell+1)\right]P,\\
    \square_AP=\ & \left[\mp 2\mcL^{-1}L^{-1}(\ell+1)\pm' 2\wtmcL^{-1}\wtL^{-1}(\wtell+1)\right]W+\left[L^{-2}(\ell+1)^2+\wtL^{-2}(\wtell+1)^2\right]P.
\end{align}
This leads to masses
\begin{equation}\label{(l,pm1)(lh,pm'1), l geq 1, lh geq 1 scalar}
    m^2_{\pm''}=L^{-2}(\ell+1)^2+\wtL^{-2}(\wtell+1)^2+\frac{\Lambda}{2}\pm''\sqrt{\left(2\mcL^{-1}L^{-1}(\ell+1)\mp\pm'2\wtmcL^{-1}\wtL^{-1}(\wtell+1)\right)^2+\frac{\Lambda^2}{4}}.
\end{equation}
These masses respect the symmetry of interchanging the two spheres. It can be seen that $m^{2}_{-}$ violate the BF bound for $\ell=\wtell=1$, $\pm'=\mp$ and equal fluxes $n=\widetilde{n}$. In this scenario, $m^{2}_{-}$ becomes
\be
m_-^2=\frac{7-\sqrt{61}}{2}\Lambda\simeq -0.405\,\Lambda\qquad (\ell=\ellh=1,\;\pm'=\mp,\;n=\widetilde{n}),
\ee
and it is below the scalar BF bound which can be written as
\be
m_{BF}^2=-\frac{9}{4}L_A^{-2}=-\frac{3}{8}\Lambda=-0.375\,\Lambda.
\ee
Numerical analysis indicates that the BF bound is violated by this mode if $0.214\lesssim|n/\tilde{n}|\lesssim 4.674$.  The mass squared with $\ell=\ellh=1$ and $\pm=\mp'$ is plotted in Figure~\ref{fig: BF bound violating eigenvalues}, and one can clearly see that the BF bound violation happens for these modes only when the fluxes are of comparable magnitude.
\newline\newline
Next we turn to the $(\ell,{\pm}1)(\wtell,0)$ sector, first in the case where $\ell\ge 2$ and $\wtell\ge 1$.  The relevant fields are then two vectors and two scalars, $S_\mu$, $Z_\mu$, $W$, and $P$.  Two of the equations, \eqref{eq:l1hl0SDiv} and~\eqref{eq:l1hl0ZDiv}, can be used to fix the divergences of the vectors,
\begin{align}
    \nabla^\mu S_\mu=\ & \wtL^{-2}\wtell(\wtell+2)W,\\
    \nabla^\mu Z_\mu=\ & -\wtL^{-2}\wtell(\wtell+2)P.
\end{align}
Substituting these into the remaining two scalar valued equations~\eqref{eq:l1hl0WEoM} and~\eqref{eq:l1hl0PEoM} gives us the mass matrix for $W$ and $P$,
\begin{align}
    \square_AW=\ & \left[L^{-2}(\ell+1)^2+\wtL^{-2}\wtell(\wtell+2)+\Lambda\right]W\mp 2\mcL^{-1}L^{-1}(\ell+1)P,\\
    \square_AP=\ & \mp 2\mcL^{-1}L^{-1}(\ell+1)W+\left[L^{-2}(\ell+1)^2+\wtL^{-2}\wtell(\wtell+2)\right]P,
\end{align}
corresponding to masses
\begin{equation}\label{(l,pm1)(lh,0), l geq 1, lh geq 1 scalar}
    m^2_{\pm'}=L^{-2}(\ell+1)^2+\wtL^{-2}\wtell(\wtell+2)+\frac{\Lambda}{2}\pm'\sqrt{4\mcL^{-2}L^{-2}(\ell+1)^2+\frac{\Lambda^2}{4}},
\end{equation}
which are all positive.  To deal with the vectors, we can first do a field redefinition to get divergence free vectors $S_\mu'$ and $Z_\mu'$.  The details of how this is done won't matter much, but if we write the vectors and scalars as $v_{X\,\mu}$ and $\varphi_I$ with $X,I=1,2$, and read off the mass matrix $M_{IJ}$ from $\square_A\varphi_I=M_{IJ}\varphi_J$ and a matrix $D_{XI}$ from $\nabla^\mu v_{X\,\mu}=D_{XI}\varphi_I$, then we can define new divergence-free vectors $v_{X\,\mu}'=v_{X\,\mu}-D_{XI}(M^{-1})_{IJ}\nabla_\mu\varphi_J$.  Explicitly for this case,
\begin{align*}
    S_\mu'=\ & S_\mu-\frac{\wtL^{-2}\wtell(\wtell+2)\left(\left(L^{-2}(\ell+1)^2+\wtL^{-2}\wtell(\wtell+2)\right)\nabla_\mu W\pm 2\mcL^{-1}L^{-1}(\ell+1)\nabla_\mu P\right)}{\left(L^{-2}(\ell+1)^2+\wtL^{-2}\wtell(\wtell+2)+\Lambda\right)\left(L^{-2}(\ell+1)^2+\wtL^{-2}\wtell(\wtell+2)\right)-4\mcL^{-2}L^{-2}(\ell+1)^2},\\
    Z_\mu'=\ & Z_\mu+\frac{\wtL^{-2}\wtell(\wtell+2)\left(\pm 2\mcL^{-1}L^{-1}(\ell+1)\nabla_\mu W+\left(L^{-2}(\ell+1)^2+\wtL^{-2}\wtell(\wtell+2)+\Lambda\right)\nabla_\mu P\right)}{\left(L^{-2}(\ell+1)^2+\wtL^{-2}\wtell(\wtell+2)+\Lambda\right)\left(L^{-2}(\ell+1)^2+\wtL^{-2}\wtell(\wtell+2)\right)-4\mcL^{-2}L^{-2}(\ell+1)^2}.
\end{align*}
These new vectors have simplified equations of motion from~\eqref{eq:l1hl0SEoM}, \eqref{eq:l1hl0ZEoM}, decoupled from the scalars,
\begin{align}
    \square_AS_\mu'=\ & \left[L^{-2}(\ell+1)^2+\wtL^{-2}\wtell(\wtell+2)+\frac{\Lambda}{2}\right]S_\mu'\pm 2\mcL^{-1}L^{-1}(\ell+1)Z_\mu',\\
    \square_AZ_\mu'=\ & \pm 2\mcL^{-1}L^{-1}(\ell+1)S_\mu'+\left[L^{-2}(\ell+1)^2+\wtL^{-2}\wtell(\wtell+2)-\frac{\Lambda}{2}\right]Z_\mu'.
\end{align}
To read off the masses of the vectors, we note that a free vector should correspond to an effective action (as usual $F_{\mu\nu}=2\nabla_{[\mu}A_{\nu]}$)
\begin{equation}
    S=-\int d^4x\sqrt{-g}\left(\frac{1}{4}F^{\mu\nu}F_{\mu\nu}+\hlf m^2A^\mu A_\mu\right),
\end{equation}
where $m^2$ here corresponds to what we want to identify as the mass (so $m=0$ is the case with additional gauge symmetry, for instance).  The corresponding equation of motion would be
\begin{equation}
    0=\nabla^\nu F_{\mu\nu}+m^2A_\mu,
\end{equation}
and if we are working on AdS$_4$ of radius $L_A$ with a vector that is divergence free, $\nabla^\mu A_\mu=0$, this equation becomes
\begin{equation}
    \square_AA_\mu=\left(m^2-3L_A^{-2}\right)A_\mu=\left(m^2-\frac{\Lambda}{2}\right)A_\mu.
\end{equation}
So we need to take the shift of $3L_A^{-2}=\frac{1}{2}\Lambda$ into account when reading the mass $m^2$ off from the vector equations of motion.  Then the Breitenlohner-Freedman bound for vectors~\cite{Ishibashi:2004wx} is usually quoted in terms of this $m^2$,
\begin{equation}
    m_{BF,vec}^2=-\frac{(D-3)^2}{4L_A^2}=-\frac{1}{4L_A^2}=-\frac{1}{24}\Lambda.
\end{equation}
Consulting the equation of motion for the vectors, we find the masses
\begin{equation}\label{(l, pm1)(lh,0), l geq 1, lh geq 1 vector}
    m^2_{\pm'}=L^{-2}(\ell+1)^2+\wtL^{-2}\wtell(\wtell+2)+\frac{\Lambda}{2}\pm'\sqrt{4\mcL^{-2}L^{-2}(\ell+1)^2+\frac{\Lambda^2}{4}},
\end{equation}
the same as the scalars, and in particular, they are all positive.\newline\newline
If we consider the $(\ell,{\pm}1)(0,0)$ sector with $\ell\ge 2$, then the scalar fields are no longer present and the vectors are already divergence free, and their equations of motion are
\begin{align}
    \label{4.161(1)}
    \square_AS_\mu=\ & \left[L^{-2}(\ell+1)^2+\frac{\Lambda}{2}\right]S_\mu\pm 2\mcL^{-1}L^{-1}(\ell+1)Z_\mu,\\\label{4.161(2)}
    \square_AZ_\mu=\ & \pm 2\mcL^{-1}L^{-1}(\ell+1)S_\mu+\left[L^{-2}(\ell+1)^2-\frac{\Lambda}{2}\right]Z_\mu,
\end{align}
with masses given by~\eqref{(l, pm1)(lh,0), l geq 1, lh geq 1 vector} with $\wtell=0$. 
\newline\newline
For the $(1,{\pm}1)(\wtell,0)$ sector with $\wtell\ge 1$, we have scalars and vectors.  The equations are as before except that we don't have the equation~\eqref{eq:l1hl0SDiv} determining the divergence of $S_\mu$, and in addition we have residual gauge freedom parameterized by $\beta^{(1,{\pm}1)(\wtell,0)}$.  We can completely use up the gauge freedom to set $W=0$, and then the equation~\eqref{eq:l1hl0WEoM} becomes an equation fixing $\nabla^\mu S_\mu$,
\begin{equation}
    \nabla^\mu S_\mu=\pm 4\mcL^{-1}L^{-1}P.
\end{equation}
The divergence of $Z_\mu$ from~\eqref{eq:l1hl0ZDiv} then changes as well,
\begin{equation}
    \nabla^\mu Z_\mu=-\wtL^{-2}\wtell(\wtell+2)P-4\mcL^{-2}P.
\end{equation}
Substituting into the remaining scalar equation~\eqref{eq:l1hl0PEoM} gives us
\begin{equation}
    \square_AP=\left[\wtL^{-2}\wtell(\wtell+2)+4L^{-2}+4\mcL^{-2}\right]P,
\end{equation}
which corresponds to a mass
\begin{equation}\label{(l,pm1)(lh,0), l= 1, lh geq 1 scalar}
    m_P^2=\wtL^{-2}\wtell(\wtell+2)+8L^{-2}+\Lambda.
\end{equation}
We can then define divergence free vectors
\begin{align}
    S_\mu'=\ & S_\mu\mp\frac{4\mcL^{-1}L^{-1}}{m_P^2}\nabla_\mu P,\\
    Z_\mu'=\ & Z_\mu+\frac{\wtL^{-2}\wtell(\wtell+2)+4L^{-2}+\Lambda}{m_P^2}\nabla_\mu P.
\end{align}
These obey equations
\begin{align}
    \square_AS_\mu'=\ & \left[\wtL^{-2}\wtell(\wtell+2)+4L^{-2}+\frac{\Lambda}{2}\right]S_\mu'\pm 4\mcL^{-1}L^{-1}Z_\mu',\\
    \square_AZ_\mu'=\ & \pm 4\mcL^{-1}L^{-1}S_\mu'+\left[\wtL^{-2}\wtell(\wtell+2)+4L^{-2}-\frac{\Lambda}{2}\right]Z_\mu'
\end{align}
with manifestly positive masses
\begin{equation}\label{(l,pm1)(lh,0), l= 1, lh geq 1 vector}
    m_+^2=\wtL^{-2}\wtell(\wtell+2)+8L^{-2}+\Lambda,\qquad m_-^2=\wtL^{-2}\wtell(\wtell+2).
\end{equation}
For the $(1,{\pm}1)(0,0)$ sector, we have only the vector fields.  We can use the residual gauge transformation parameterized by $\beta^{(1,{\pm}1)(0,0)}$ to set $\nabla^\mu S_\mu=0$, which still leaves transformations satisfying $\square_A\beta=0$.  One of the surviving equations, \eqref{eq:l1hl0ZDiv}, forces $\nabla^\mu Z_\mu=0$ as well.  The remaining equations~\eqref{eq:l1hl0SEoM} and~\eqref{eq:l1hl0ZEoM} are
\begin{align}
    \square_AS_\mu=\ & \left[ 4L^{-2}+\frac{\Lambda}{2}\right]S_\mu\pm 4\mcL^{-1}L^{-1}Z_\mu,\\
    \square_AZ_\mu=\ & \pm 4\mcL^{-1}L^{-1}S_\mu+\left[4L^{-2}-\frac{\Lambda}{2}\right]Z_\mu,
\end{align}
resulting in masses
\begin{equation}\label{(l, pm1)(lh,0), l=1, lh=0 vector}
    m_+^2=8L^{-2}+\Lambda,\qquad m_-^2=0.
\end{equation}
The massless mode can be removed by the residual gauge freedom (note that for $\beta$ satisfying $\square_A\beta=0$, we have $\square_A(\nabla_\mu\beta)=-\frac{1}{2}\Lambda\nabla_\mu\beta$, so the massless vector mode satisfying $\square_AA_\mu=-\frac{1}{2}\Lambda A_\mu$ can be gauged away by shifting it by something proportional to $\nabla_\mu\beta$) and thus only the massive vector mode with $m_+^2=8L^{-2}+\Lambda$ survives.
\newline\newline
Proceeding next to the $(\ell,0)(\wtell,{\pm}1)$ sector, we start with the generic case $\ell\ge 1$, $\wtell\ge 2$, with two vectors $\wtS_\mu$ and $\wtZ_\mu$ and two scalars $\wtK$ and $\wtV$.  The vector divergences from~\eqref{eq:l0hl1SDiv} and~\eqref{eq:l0hl1ZDiv} are
\begin{align}
    \nabla^\mu\wtS_\mu=\ & \hlf\wtL^{-2}(\wtell^2+2\wtell-3)\wtK+2\wtmcL^{-1}\wtV,\\
    \nabla^\mu\wtZ_\mu=\ & \pm\wtL^{-1}(\wtell+1)\wtV.
\end{align}
Substituting these into the other scalar equations~\eqref{eq:l0hl1KEoM} and~\eqref{eq:l0hl1VEoM} results in
\begin{align}
    \square_A\wtK=\ & \left[L^{-2}\ell(\ell+2)+\wtL^{-2}(\wtell^2+2\wtell-3)\right]\wtK+4\wtmcL^{-1}\wtV,\\
    \square_A\wtV=\ & \wtmcL^{-1}\wtL^{-2}(\wtell^2+2\wtell-3)\wtK+\left[L^{-2}\ell(\ell+2)+\wtL^{-2}(\wtell+1)^2+4\wtmcL^{-2}\right]\wtV.
\end{align}
The scalar masses are
\be\label{(l,0)(lh,pm1), l geq 1, lh geq 2 scalar}
m^2_{\pm'}=L^{-2}\ell(\ell+2)+\wtL^{-2}(\wtell+1)^2+\frac{\Lambda}{2}\pm'\sqrt{4\wtmcL^{-2}\wtL^{-2}(\wtell+1)^2+\frac{\Lambda^2}{4}}.
\ee

This agrees with the $(\ell,{\pm}1)(\wtell,0)$ sector results upon swapping the two spheres.  Using the same trick as before, we can define divergence-free versions of the vectors, which then satisfy equations
\begin{align}
    \square_A\wtS_\mu'=\ & \left[L^{-2}\ell(\ell+2)+\wtL^{-2}(\wtell+1)^2+\frac{\Lambda}{2}\right]\wtS_\mu'\pm 2\wtmcL^{-1}\wtL^{-1}(\wtell+1)\wtZ_\mu',\\
    \square_A\wtZ_\mu'=\ & \pm 2\wtmcL^{-1}\wtL^{-1}(\wtell+1)\wtS_\mu'+\left[L^{-2}\ell(\ell+2)+\wtL^{-2}(\wtell+1)^2-\frac{\Lambda}{2}\right]\wtZ_\mu',
\end{align}
and give rise to the following masses
\begin{equation}\label{(l,0)(lh,pm1), l geq 1, lh geq 2 vector}
    m^2_{\pm'}=L^{-2}\ell(\ell+2)+\wtL^{-2}(\wtell+1)^2+\frac{\Lambda}{2}\pm'\sqrt{4\wtmcL^{-2}\wtL^{-2}(\wtell+1)^2+\frac{\Lambda^2}{4}},
\end{equation}
again in agreement with swapping the spheres.
\newline\newline
For $(\ell,0)(1,{\pm}1)$ sector with $\ell\ge 1$, we lose the scalar $\wtK$ and one of the equations.  The remaining scalar equations~\eqref{eq:l0hl1SDiv} and~\eqref{eq:l0hl1ZDiv} fix
\begin{align}
    \nabla^\mu\wtS_\mu=\ & 2\wtmcL^{-1}\wtV,\\
    \nabla^\mu\wtZ_\mu=\ & \pm 2\wtL^{-1}\wtV,
\end{align}
and~\eqref{eq:l0hl1VEoM} gives the following
\begin{equation}\label{(l,0)(lh,pm1), l geq 1, lh=1 scalar}
    \square_A\wtV=\left[L^{-2}\ell(\ell+2)+4\wtL^{-2}+4\wtmcL^{-2}\right]\wtV\Rightarrow m_{\wtV}^2=L^{-2}\ell(\ell+2)+8\wtL^{-2}+\Lambda.
\end{equation}
Then shifting the vectors to get divergence free combinations, the vector equations are
\begin{align}
    \square_A\wtS_\mu'=\ & \left[L^{-2}\ell(\ell+2)+4\wtL^{-2}+\frac{\Lambda}{2}\right]\wtS_\mu'\pm 4\wtmcL^{-1}\wtL^{-1}\wtZ_\mu',\\
    \square_A\wtZ_\mu'=\ & \pm 4\wtmcL^{-1}\wtL^{-1}\wtS_\mu'+\left[L^{-2}\ell(\ell+2)+4\wtL^{-2}-\frac{\Lambda}{2}\right]\wtZ_\mu',
\end{align}
giving
\begin{equation}\label{(l,0)(lh,pm1), l geq 1, lh= 1 vector}
    m_+^2=L^{-2}\ell(\ell+2)+8\wtL^{-2}+\Lambda,\qquad m_-^2=L^{-2}\ell(\ell+2).
\end{equation}
For the $(0,0)(\wtell,{\pm}1)$ sector with $\wtell\ge 2$, we have all four fields but we lose the two equations which gave us the vector divergences.  On the other hand, we have a two parameter family of gauge transformations given by $\g^{(0,0)(\wtell,{\pm}1)}$ and $\nu^{(0,0)(\wtell,{\pm}1)}$.  These can be completely used up by setting the scalars $\wtK$ and $\wtV$ to zero.  Upon doing so, the equations~\eqref{eq:l0hl1KEoM} and~\eqref{eq:l0hl1VEoM} force $\nabla^\mu\wtS_\mu=\nabla^\mu\wtZ_\mu=0$, and from~\eqref{eq:l0hl1SEoM}, \eqref{eq:l0hl1ZEoM} the vectors obey
\begin{align}
    \square_A\wtS_\mu=\ & \left[\wtL^{-2}(\wtell+1)^2+\frac{\Lambda}{2}\right]\wtS_\mu\pm 2\wtmcL^{-1}\wtL^{-1}(\wtell+1)\wtZ_\mu,\\
    \square_A\wtZ_\mu=\ & \pm 2\wtmcL^{-1}\wtL^{-1}(\wtell+1)\wtS_\mu+\left[\wtL^{-2}(\wtell+1)^2-\frac{\Lambda}{2}\right]\wtZ_\mu,
\end{align}
so
\begin{equation}\label{(l,0)(lh,pm1), l=0, lh geq 2 vector}
    m^2_{\pm'}=\wtL^{-2}(\wtell+1)^2+\frac{\Lambda}{2}\pm'\sqrt{4\wtmcL^{-2}\wtL^{-2}(\wtell+1)^2+\frac{\Lambda^2}{4}}.
\end{equation}
Turning next to the $(0,0)(1,{\pm}1)$ sector, we have fields $\wtS_\mu$, $\wtZ_\mu$, and $\wtV$, and we have the two gauge transformations described above.  We can use the $\n^{(0,0)(1,{\pm}1)}$ gauge transformations to set $\wtV=0$, and we can use $\gamma^{(0,0)(1,{\pm}1)}$, with a compensating transformation 
$$
\nu^{(0,0)(1,{\pm}1)}=\mp\wtmcL^{-1}\wtL\gamma^{(0,0)(1,{\pm}1)}
$$
(to preserve $\wtV=0$) to enforce $\nabla^\mu\wtS_\mu=0$, leaving a residual gauge transformation satisfying $\square_A\g=0$.  The equations then also fix $\nabla^\mu\wtZ_\mu=0$, and the two vectors satisfy
\begin{align}
    \square_A\wtS_\mu=\ & \left[4\wtL^{-2}+\frac{\Lambda}{2}\right]\wtS_\mu\pm 4\wtmcL^{-1}\wtL^{-1}\wtZ_\mu,\\
    \square_A\wtZ_\mu=\ & \pm 4\wtmcL^{-1}\wtL^{-1}\wtS_\mu+\left[4\wtL^{-2}-\frac{\Lambda}{2}\right]\wtZ_\mu,
\end{align}
corresponding to masses
\begin{equation}\label{(l,0)(lh,pm1), l=0, lh= 1 vector}
    m_-^2=0,\qquad m_+^2=8\wtL^{-2}+\Lambda.
\end{equation}
As before, the residual gauge transformation is exactly what is required to remove the $m_-^2$ massless mode, leaving a single vector with $m_+^2$.\newline\newline
Finally, we turn to the $(\ell,0)(\wtell,0)$ sectors, starting with the generic case $\ell\ge 2$, $\wtell\ge 2$.  A priori we have seven scalar fields, $\phi$, $M$, $N$, $\wtN$, $V$, $\wtV$, and $\wtK$, two vector fields $\wtS_\mu$ and $\wtZ_\mu$, an antisymmetric tensor $X_{\mu\nu}$, and a symmetric traceless tensor $H_{\mu\nu}$.  As with the vectors, we should establish how to identify the masses for the two tensor fields, and what the corresponding perturbative stability bounds are.  For the antisymmetric tensor, we identify the mass by considering an action
\begin{equation}
    S=-\int d^4x\sqrt{-g}\left(\frac{1}{12}H^{\mu\nu\rho}H_{\mu\nu\rho}+\frac{1}{4}m^2B^{\mu\nu}B_{\mu\nu}\right),
\end{equation}
where $H_{\mu\nu\rho}=3\nabla_{[\mu}B_{\nu\rho]}$, so that $m^2=0$ is the expected massless case.  The corresponding Breitenlohener-Freedman bound is~\cite{Basile:2018irz}
\begin{equation}
    m_{BF,2-form}^2=-\frac{(D-5)^2}{4L_A^2}=-\frac{1}{4L_A^2}=-\frac{1}{24}\Lambda.
\end{equation}
If $B_{\mu\nu}$ is divergence free, $\nabla^\nu B_{\mu\nu}=0$, then the action leads to an equation of motion
\begin{equation}
    \square_AB_{\mu\nu}=\left(m^2-4L_A^{-2}\right)B_{\mu\nu}=\left(m^2-\frac{2}{3}\Lambda\right)B_{\mu\nu}.
\end{equation}
For the traceless symmetric tensor, the massless case should correspond to the linearized Einstein-Hilbert action.  This leads us to identify the mass from the equation of motion
\begin{equation}
    \square_Ah_{\mu\nu}=\left(m^2-2L_A^{-2}\right)h_{\mu\nu}=\left(m^2-\frac{1}{3}\Lambda\right)h_{\mu\nu},
\end{equation}
where $h_{\mu\nu}$ is traceless symmetric and satisfies $\nabla^\nu h_{\mu\nu}=0$.  With this convention, the stability bound is~\cite{Ishibashi:2004wx,Basile:2018irz}
\begin{equation}
    m_{BF,spin 2}^2=-\frac{(D-1)^2}{4L_A^2}=-\frac{9}{4L_A^2}=-\frac{3}{8}\Lambda.
\end{equation}

\begin{figure}
    \centering
    \includegraphics[width=0.28\linewidth]{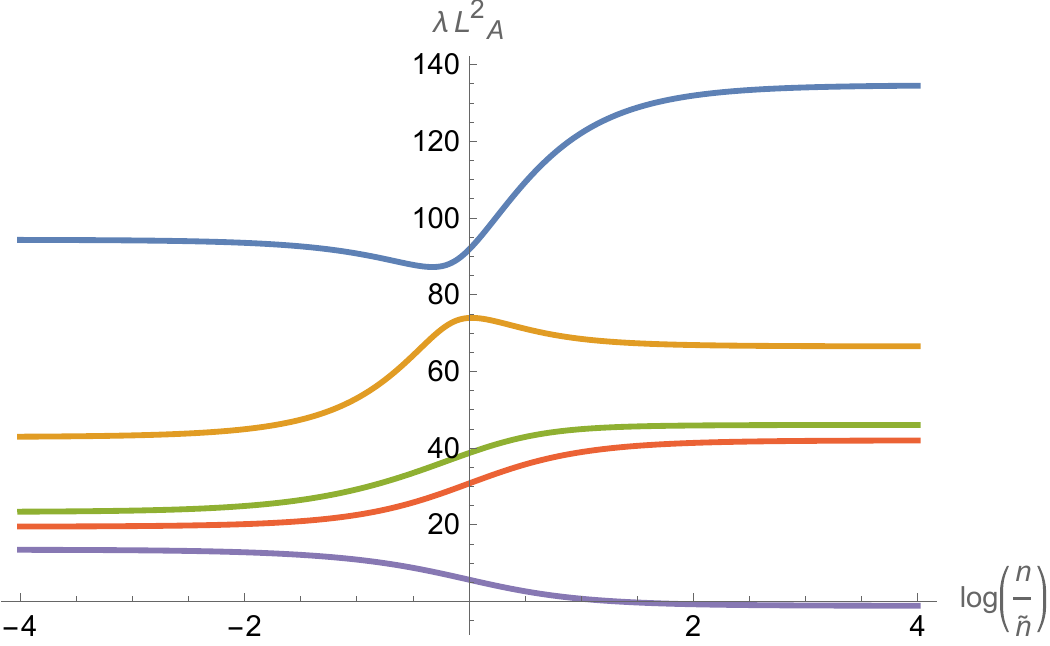}\efill\includegraphics[width=0.28\linewidth]{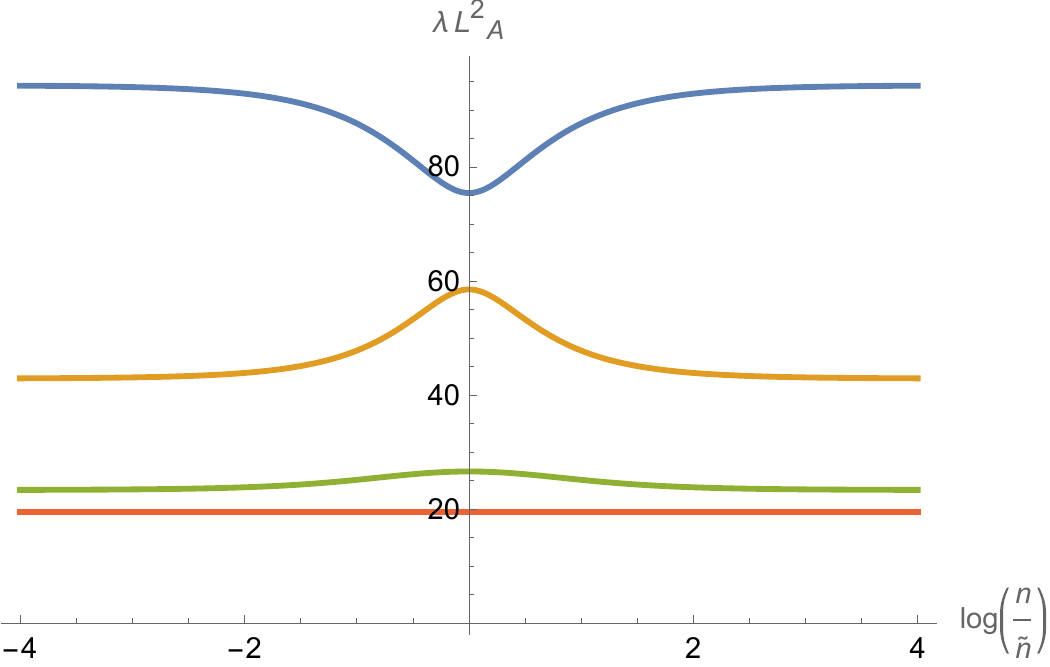}\efill \includegraphics[width=0.28\linewidth]{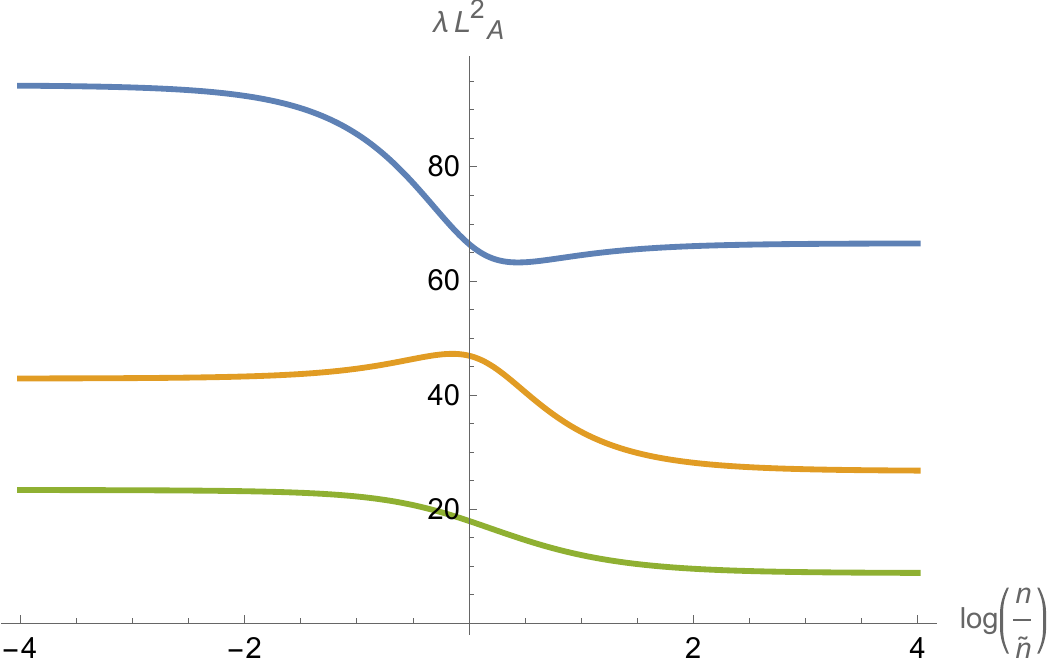}
    \caption{The plots of the eigenvalues for the mass matrices for the $(\ell,0)(\widetilde{\ell},0)$ sectors with (left to right) $(\ell,\ellh)=(1,2), (1,1)$ and $(1,0)$ against $\log\lp n/\wtn\rp$. None of these eigenvalues violates the BF bound.}
    \label{fig:l=1 eigenvalues}
\end{figure}

\begin{figure}
    \centering
    \includegraphics[width=0.28\linewidth]{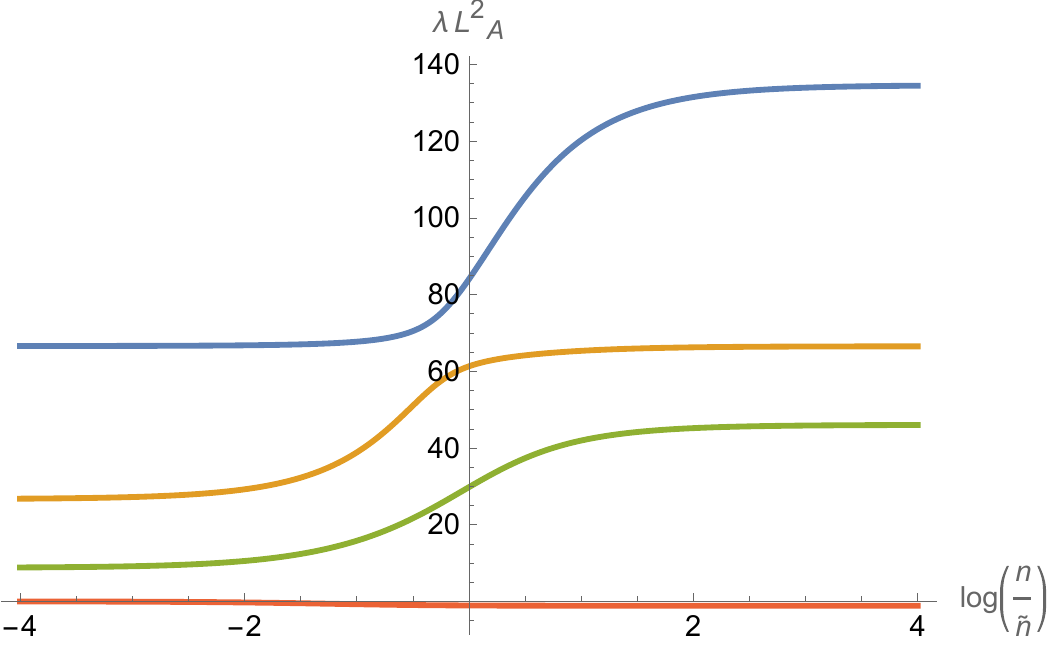}\efill \includegraphics[width=0.28\linewidth]{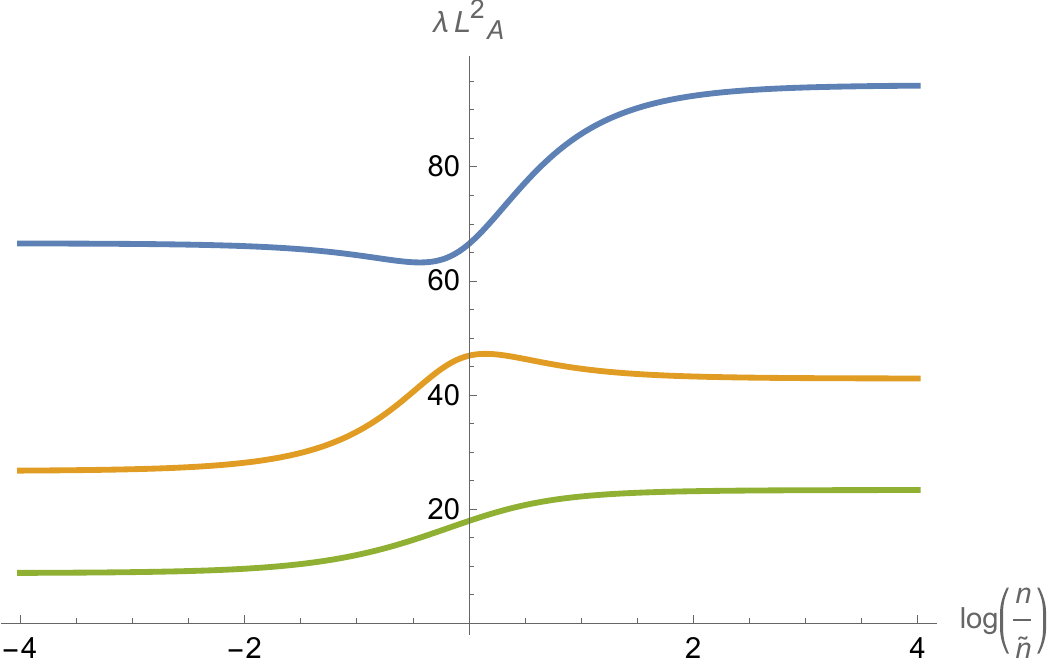}\efill \includegraphics[width=0.28\linewidth]{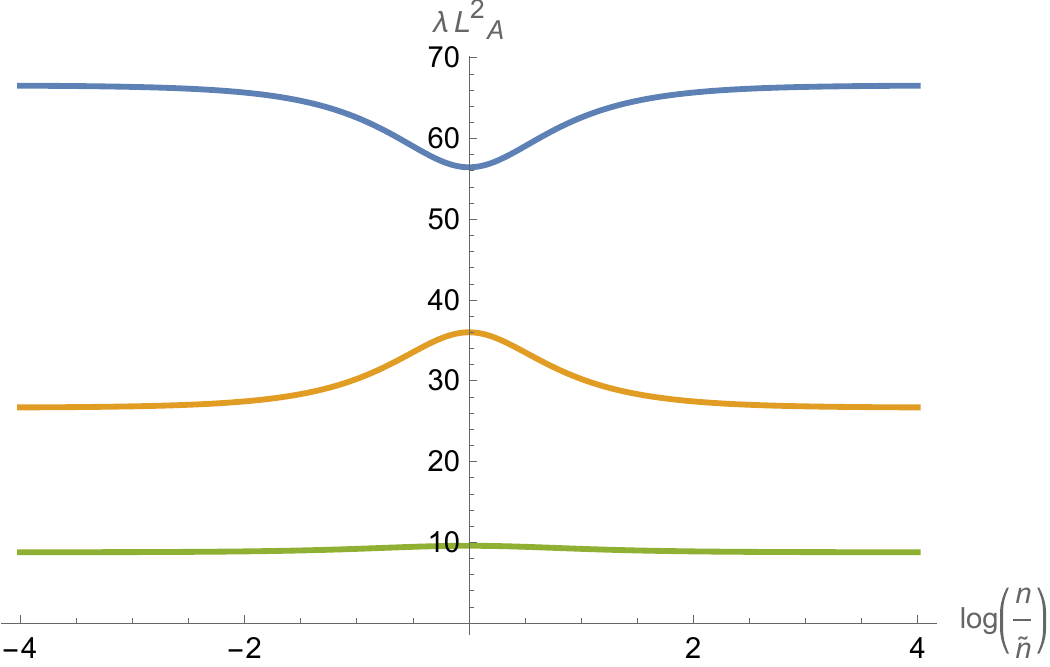}
    \caption{The plots of the eigenvalues for the mass matrices for the $(\ell,0)(\widetilde{\ell},0)$ sectors with (left to right) $(\ell,\ellh)=(0,2), (0,1)$ and $(0,0)$ against $\log \lp n/\wtn\rp$. None of these eigenvalues violates the BF bound.}
    \label{fig:l=0 eigenvalues}
\end{figure}

Returning to the analysis, one equation, \eqref{eq:l0hl0Algebraic}, is completely algebraic and can be used to eliminate one of the scalar fields,
\begin{equation}
    \phi=M+\frac{1}{4}N+\frac{3}{4}\wtN.
\end{equation}
The vector $\wtZ_\mu$ is divergence free~\eqref{eq:l0hl0hZDiv}, while the other vector satisfies~\eqref{eq:l0hl0hSDiv}
\begin{equation}
    \nabla^\mu\wtS_\mu=N-\wtN+2\mcL^{-1}V+2\wtmcL^{-1}\wtV+\frac{2}{3}\wtL^{-2}(\wtell^2+2\wtell-3)\wtK.
\end{equation}
We can use another scalar equation~\eqref{eq:l0hl0MEoM} to eliminate $\nabla^\mu\nabla^\nu H_{\mu\nu}$,
\begin{align}
    \nabla^\mu\nabla^\nu H_{\mu\nu}=\ & \square_A\left(M+N\right)-\Lambda\left(6M+N+3\wtN\right)-2L^{-2}\ell(\ell+2)M\non\\
    &+\wtL^{-2}\wtell(\wtell+2)\left( -2M+N-\wtN+2\mcL^{-1}V+2\wtmcL^{-1}\wtV+\frac{2}{3\wtL^{2}}(\wtell^2+2\wtell-3)\wtK\right).
\end{align}
Substituting these expressions into the remaining six scalar equations, \eqref{eq:l0hl0hKEoM}, \eqref{eq:l0hl0VEoM}, \eqref{eq:l0hl0hVEoM}, \eqref{eq:l0hl0phiEoM}, \eqref{eq:l0hl0NEoM}, and \eqref{eq:l0hl0hNEoM}, gives us our mass matrix,
\begin{align}\label{(l,0)(lh,0), l geq 2, lh geq 2 scalar 1}
    \square_AM=\ & \left[L^{-2}\ell(\ell+2)+\wtL^{-2}\wtell(\wtell+2)+3\Lambda\right]M+\left[4L^{-2}+\frac{3}{2}\Lambda\right]N+\frac{3}{2}\Lambda\wtN\non\\
    &\quad+\mcL^{-1}\left[L^{-2}\ell(\ell+2)+\wtL^{-2}\wtell(\wtell+2)\right]V,\\
    \label{(l,0)(lh,0), l geq 2, lh geq 2 scalar 2}
    \square_AN=\ & 2\Lambda M+\left[L^{-2}\ell(\ell+2)+\wtL^{-2}\wtell(\wtell+2)+8L^{-2}+\frac{7}{2}\Lambda\right]N+\frac{3}{2}\Lambda\wtN\non\\
    &\quad+4\mcL^{-1}L^{-2}\ell(\ell+2)V,\\
    \label{(l,0)(lh,0), l geq 2, lh geq 2 scalar 3}
    \square_A\wtN=\ & 2\Lambda M+\hlf\Lambda N+\left[L^{-2}\ell(\ell+2)+\wtL^{-2}\wtell(\wtell+2)+8\wtL^{-2}+\frac{9}{2}\Lambda\right]\wtN\non\\
    &\quad -\frac{4}{3}\wtL^{-2}\wtell(\wtell+2)\lp\mcL^{-1}V-2\wtmcL^{-1}\wtV\rp,\\
    \label{(l,0)(lh,0), l geq 2, lh geq 2 scalar 5}
    \square_A\wtV=\ & 4\wtmcL^{-1}\wtN+4\mcL^{-1}\wtmcL^{-1}V+\left[L^{-2}\ell(\ell+2)+\wtL^{-2}\wtell(\wtell+2)+4\wtmcL^{-2}\right]\wtV\non\\
    &\quad+\frac{4}{3}\wtmcL^{-1}\wtL^{-2}(\wtell^2+2\wtell-3)\wtK,\\
    \label{(l,0)(lh,0), l geq 2, lh geq 2 scalar 4}
    \square_AV=\ & 4\mcL^{-1}N+\left[L^{-2}\ell(\ell+2)+\wtL^{-2}\wtell(\wtell+2)\right]V,\\
    \label{(l,0)(lh,0), l geq 2, lh geq 2 scalar 6}
    \square_A\wtK=\ & 4\mcL^{-1}V+4\wtmcL^{-1}\wtV+\left[L^{-2}\ell(\ell+2)+\wtL^{-2}\wtell(\wtell+2)-4\wtL^{-2}\right]\wtK.
\end{align}
Though it is not obvious from this form of the mass matrix, all six eigenvalues are above the BF bound for all $\ell,\wtell\ge 2$, and there is a symmetry in the spectrum upon exchanging the two $S^3$s (i.e.~swapping $\ell$, $L$, $\mcL$ with $\wtell$, $\wtL$, $\wtmcL$). See Figure \ref{fig:l,lh>2 eigenvalues} for some relevant plots.
\begin{figure}
    \centering
    \includegraphics[width=0.3\linewidth]{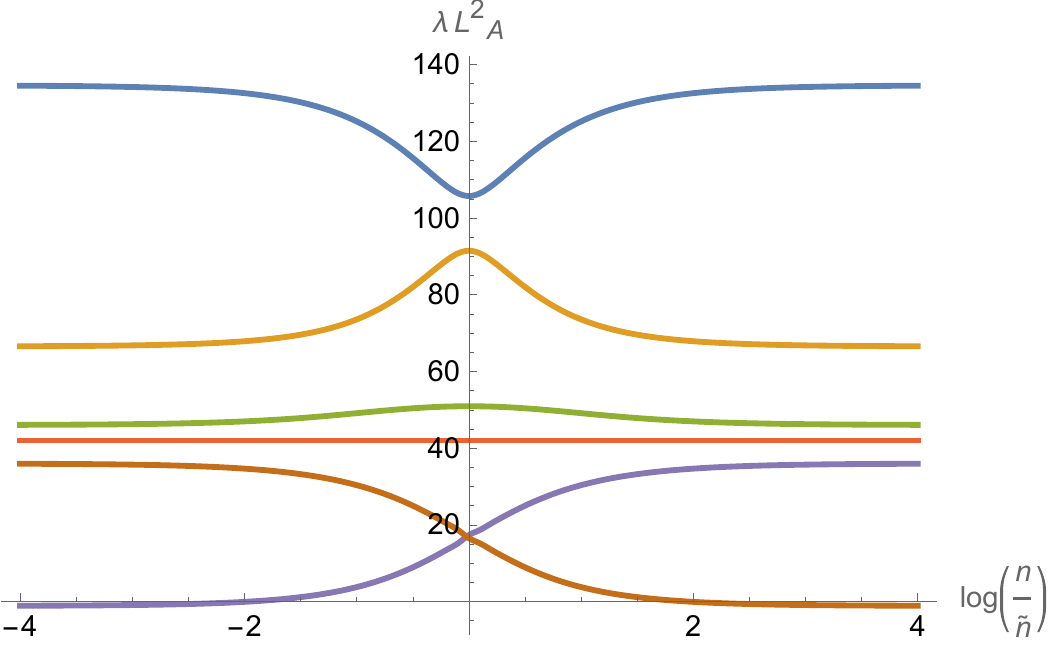}\hfill\includegraphics[width=0.3\linewidth]{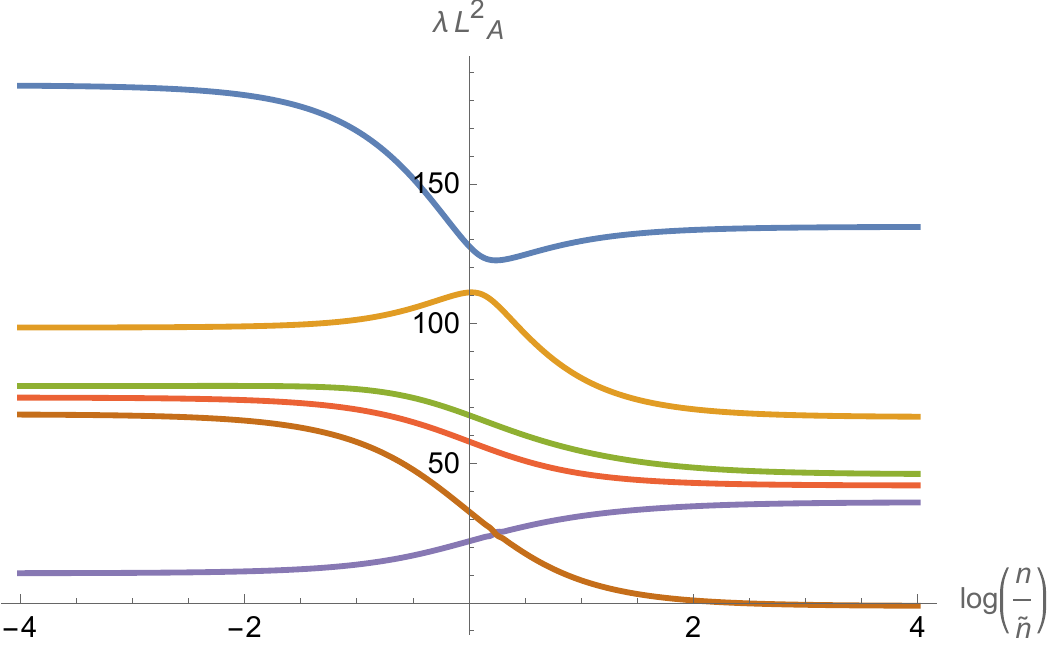}\hfill\includegraphics[width=0.3\linewidth]{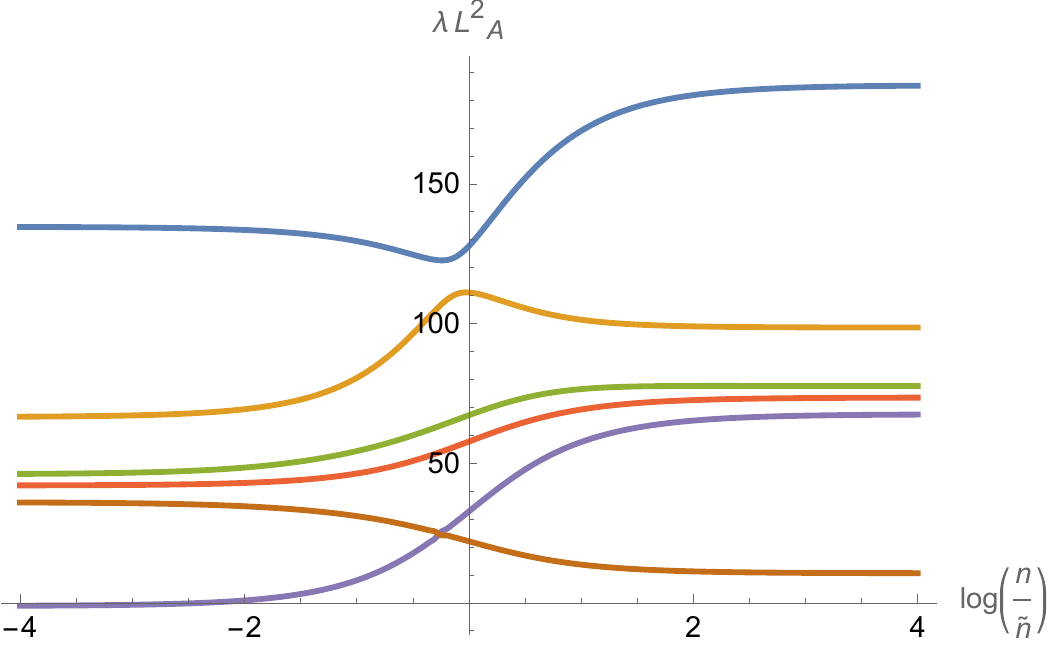}
    \caption{The plots of six eigenvalues of the mass matrices for the $(\ell,0)(\widetilde{\ell},0)$ sectors with (left to right) $(\ell,\ellh)=(2,2), (3,2)$ and $(2,3)$ against $\log\lp n/\widetilde{n}\rp$. None of these eigenvalues violates the BF bound. The symmetry between the two $S^{3}$ factors is demonstrated by the middle and the right plot.}
    \label{fig:l,lh>2 eigenvalues}
\end{figure}
\newline\newline
We can play the same trick as before, shifting the vectors by gradients of linear combinations of the scalars to produce a new divergence free vectors $\wtS_\mu'$ and equations (from~\eqref{eq:l0hl0hSEoM}, \eqref{eq:l0hl0hZEoM})
\begin{align}
    \square_A\wtS_\mu'=\ & \left[L^{-2}\ell(\ell+2)+\wtL^{-2}\wtell(\wtell+2)+\frac{\Lambda}{2}\right]\wtS_\mu',\\
    \square_A\wtZ_\mu=\ & \left[L^{-2}\ell(\ell+2)+\wtL^{-2}\wtell(\wtell+2)-\frac{\Lambda}{2}\right]\wtZ_\mu,
\end{align}
which are accommodating enough to decouple, giving masses
\begin{equation}\label{(l,0)(lh,0), l geq 2, lh geq 2 vector}
    m_{\wtS}^2=L^{-2}\ell(\ell+2)+\wtL^{-2}\wtell(\wtell+2)+\Lambda,\qquad m_{\wtZ}^2=L^{-2}\ell(\ell+2)+\wtL^{-2}\wtell(\wtell+2).
\end{equation}
A similar shifting trick serves to redefine the tensor fields.  For instance,
\begin{equation}
    X_{\mu\nu}':=X_{\mu\nu}+\frac{2\wtL^{-2}\wtell(\wtell+2)}{L^{-2}\ell(\ell+2)+\wtL^{-2}\wtell(\wtell+2)}\nabla_{[\mu}\wtZ_{\nu]}\qquad\Rightarrow\qquad\nabla^\nu X_{\mu\nu}'=0,
\end{equation}
and
\begin{equation}\label{(l,0)(lh,0), l geq 2, lh geq 2 tensor 1}
    \square_AX_{\mu\nu}'=\left[L^{-2}\ell(\ell+2)+\wtL^{-2}\wtell(\wtell+2)-\frac{2}{3}\Lambda\right]X_{\mu\nu}',
\end{equation}
and therefore the mass of the $X'_{\m\n}$ field is,
\begin{equation}
    m_X^2=L^{-2}\ell(\ell+2)+\wtL^{-2}\wtell(\wtell+2)>0.
\end{equation}
Similarly we can shift $H_{\mu\nu}$ by derivatives of scalars and vectors to produce $H_{\mu\nu}'$ satisfying $\nabla^\nu H_{\mu\nu}'=0$ and
\begin{equation}
    \square_AH_{\mu\nu}'=\left[L^{-2}\ell(\ell+2)+\wtL^{-2}\wtell(\wtell+2)-\frac{1}{3}\Lambda\right]H_{\mu\nu}',
\end{equation}
thus, we can read off the mass of the $H_{\m\n}$ field as
\begin{equation}\label{(l,0)(lh,0), l geq 2, lh geq 2 tensor 2}
    m_H^2=L^{-2}\ell(\ell+2)+\wtL^{-2}\wtell(\wtell+2)>0.
\end{equation}
We still need to work through the special cases with low $\ell$ or $\wtell$.  Start with $\ell\ge 2$, $\wtell=1$.  We lose one scalar, $\wtK$, and one equation, \eqref{eq:l0hl0hKEoM}.  Everything else works the same way as before, so we just need to set $\wtell=1$ and remove $\wtK$ from the results above. The eigenvalues for the $(\ell,\ellh)=(2,1)$ modes are plotted in Figure \ref{fig:l>=2,lh=1 eigenvalues}. None of the fields violates the BF bounds. Similarly, for $\ell\ge 2$, $\wtell=0$ we drop the two vectors $\wtS_\mu$ and $\wtZ_\mu$ and the scalars $\wtV$ and $\wtK$ and lose several equations, \eqref{eq:l0hl0hSDiv}, \eqref{eq:l0hl0hZDiv}, \eqref{eq:l0hl0hKEoM}, \eqref{eq:l0hl0hSEoM}, \eqref{eq:l0hl0hZEoM}, and~\eqref{eq:l0hl0hVEoM}, involving them.  However, everything else remains the same, and we can take the previous results, set $\wtell=0$ and drop those fields, and get our answers. The scalar masses for the $(\ell,\ellh)=(2,0)$ modes are plotted in Figure \ref{fig:l>=2,lh=0 eigenvalues}.\newline\newline
\begin{figure}
    \centering
    \includegraphics[width=0.3\linewidth]{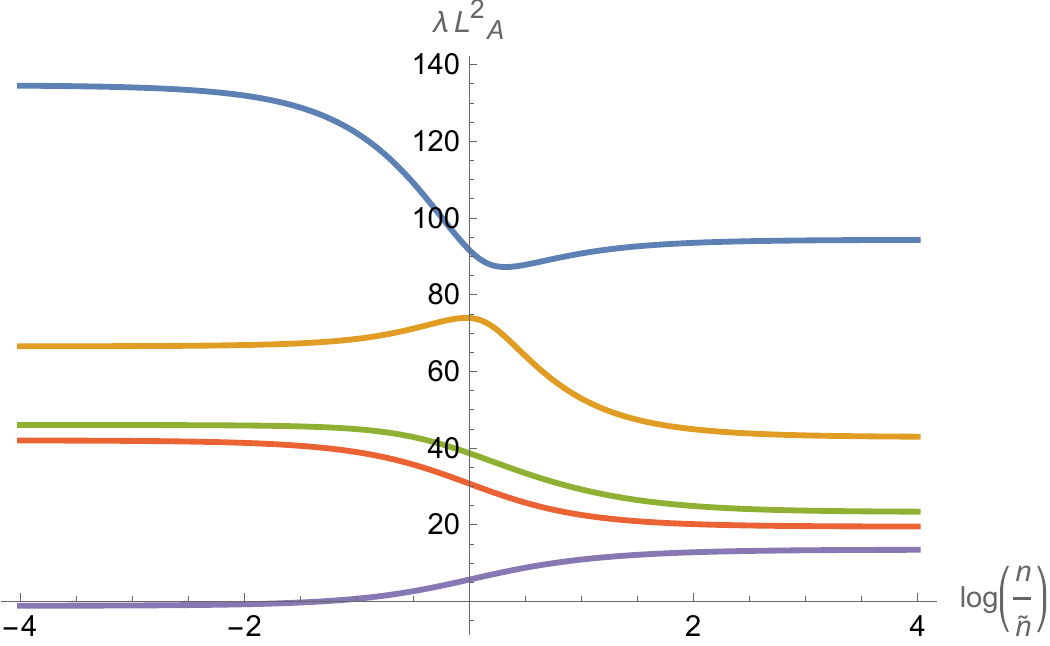}
    \caption{The plots of the five eigenvalues for the mass matrix for the $(2,0)(1,0)$ sector against $\log\lp n/\wtn\rp$. None of these eigenvalues violates the BF bound.}
    \label{fig:l>=2,lh=1 eigenvalues}
\end{figure}
\begin{figure}
    \centering
    \includegraphics[width=0.3\linewidth]{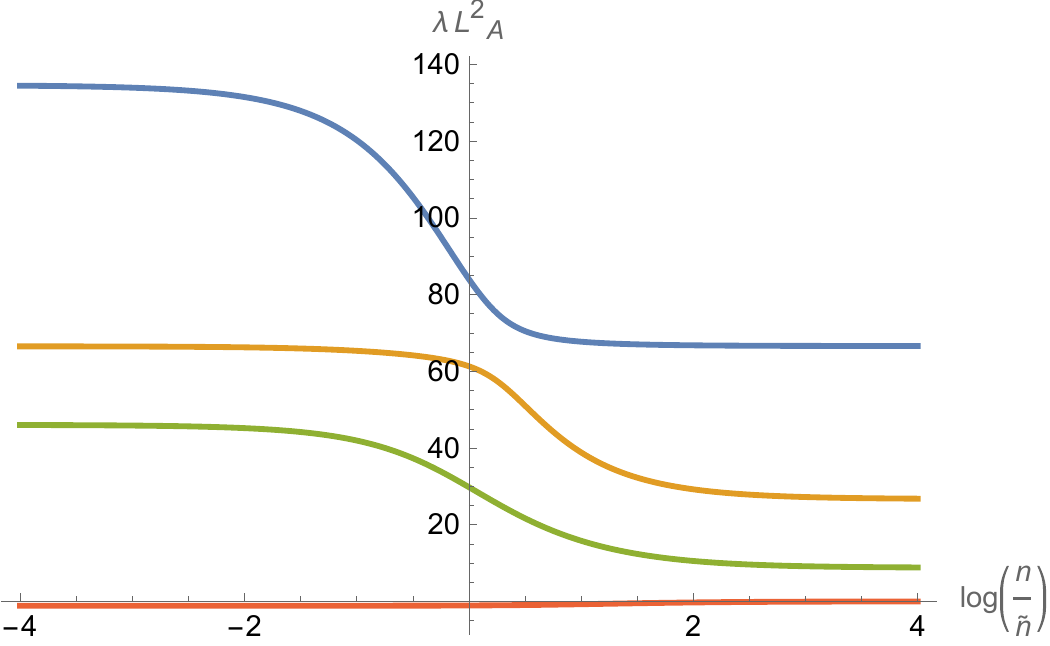}
    \caption{The plots of the eigenvalues for the mass matrix for the $(2,0)(0,0)$ sector against $\log\lp n/\wtn\rp$. None of these eigenvalues violates the BF bound.}
    \label{fig:l>=2,lh=0 eigenvalues}
\end{figure}
For $\ell<2$ we have to deal with residual gauge transformations.  In the case $(1,0)(\wtell,0)$ sector with $\wtell\ge 2$, we don't lose any fields but we do lose the one algebraic equation, \eqref{eq:l0hl0Algebraic}, that we used to eliminate $\phi$.  On the other hand we have gauge transformations parameterized by $\beta^{(1\,0)(\wtell\,0)}$.  There are two paths we can follow to proceed, and it is instructive to see how they give the same result.  On the one hand, we can use the gauge freedom to impose the earlier algebraic equation $\phi=M+\frac{1}{4}N+\frac{3}{4}\wtN$, since in this case the rest of the equations will be identical.  If we do this there is still some left over gauge freedom, since a $\beta$ satisfying $\square_A\beta=[-L^{-2}+\wtL^{-2}\wtell(\wtell+2)]\beta$ will not disturb this gauge choice.  Indeed, the six-by-six mass matrix has, for $\ell=1$, one eigenvalue that is exactly $-L^{-2}+\wtL^{-2}\wtell(\wtell+2)$, and we can use the residual gauge transformation to precisely remove this mode.  This leaves five scalars along with the vectors and tensors whose masses are given by the formulas above with $\ell=1$.  The alternative path is to instead use the full gauge freedom to set $\wtK=0$, which leaves the six scalars $\phi$, $M$, $N$, $\wtN$, $V$, and $\wtV$.  In this case, however, a different equation, \eqref{eq:l0hl0hKEoM}, substituting $\nabla^\mu\wtS_\mu$ from~\eqref{eq:l0hl0hSDiv}, becomes algebraic,
\begin{equation}
    0=2\phi-2M-\hlf N-\frac{3}{2}\wtN-2\mcL^{-1}V-2\wtmcL^{-1}\wtV,
\end{equation}
allowing us again to eliminate $\phi$,
\begin{equation}
    \phi=M+\frac{1}{4}N+\frac{3}{4}\wtN+\mcL^{-1}V+\wtmcL^{-1}\wtV.
\end{equation}
The scalar equations with $\wtK$ removed then yield a five-by-five mass matrix whose eigenvalues precisely match the previous approach. The scalar masses for the $(\ell,\ellh)=(1,2)$ modes are plotted in Figure \ref{fig:l=1 eigenvalues}. The vector and tensor equations come out exactly the same as well.
\newline\newline
For $(1,0)(1,0)$ sector, the scalar $\wtK$ is absent from the start and we still have the gauge transformations parameterized by $\beta$.  The gauge freedom can be completely used to set $\wtV=0$, and then~\eqref{eq:l0hl0hVEoM} becomes algebraic and can be used to eliminate $\phi$,
\begin{equation}
    \phi=M+\frac{1}{4}N+\frac{7}{4}\wtN+\mcL^{-1}V.
\end{equation}
We are left with a four-by-four scalar mass matrix,
\begin{align}\label{(l,0)(lh,0), l =1, lh =1 scalar 1}
    \square_AM=\ & \frac{13}{4}\Lambda M-\left[12\wtL^{-2}+\frac{21}{4}\Lambda\right]\wtN,\\
    \label{(l,0)(lh,0), l =1, lh =1 scalar 2}
    \square_AN=\ & 2\Lambda M+\left[8L^{-2}+\frac{23}{4}\Lambda\right]N+\left[4\wtL^{-2}+\hlf\Lambda\right]\wtN+\left[8L^{-2}+2\Lambda\right]\mcL^{-1}V,\\
    \label{(l,0)(lh,0), l =1, lh =1 scalar 3}
    \square_A\wtN=\ & 2\Lambda M+\hlf\Lambda N+\left[12\wtL^{-2}+\frac{35}{4}\Lambda\right]\wtN+2\Lambda\mcL^{-1}V,\\
    \label{(l,0)(lh,0), l =1, lh =1 scalar 4}
    \square_AV=\ & 4\mcL^{-1}N+4\mcL^{-1}\wtN+\left[4L^{-2}+\frac{13}{4}\Lambda\right]V.
\end{align}
and the scalar masses are plotted in Figure \ref{fig:l=1 eigenvalues}.  The vectors decouple from each other.  The vector $\wtZ_\mu$ is divergence-free and has mass
\begin{equation}\label{(l,0)(lh,0), l =1, lh =1 vector 1}
    m_{\wtZ}^2=\frac{9}{4}\Lambda,
\end{equation}
while we can shift $\wtS_\mu$ by gradients of scalars to get a divergence-free vector $\wtS_\mu'$ with mass
\begin{equation}\label{(l,0)(lh,0), l =1, lh =1 vector 2}
    m_{\wtS}^2=\frac{13}{4}\Lambda.
\end{equation}
We can similarly shift the antisymmetric and symmetric tensors to get divergence-free versions with masses
\begin{equation}\label{(l,0)(lh,0), l =1, lh =1 tensor}
    m_X^2=m_H^2=\frac{9}{4}\Lambda.
\end{equation}
For the $(1,0)(0,0)$ sector, we lose both vectors as well as the scalars $\wtK$ and $\wtV$.  We can also use residual gauge freedom to set $V=0$, and then one of the equations, \eqref{eq:l0hl0VEoM}, becomes algebraic allowing us to fix
\begin{equation}
\phi=M-\frac{3}{4}N+\frac{3}{4}\wtN.
\end{equation}
We're left with a three-by-three mass matrix for the scalars,
\begin{align}\label{(l,0)(lh,0), l =1, lh =0 scalar 1}
    \square_AM=\ & \left[3L^{-2}+5\Lambda\right]M+15L^{-2}N+3\Lambda\wtN,\\
    \label{(l,0)(lh,0), l =1, lh =0 scalar 2}
    \square_AN=\ & 2\Lambda M+\left[15L^{-2}+\frac{3}{2}\Lambda\right]N+\frac{3}{2}\Lambda\wtN,\\
    \label{(l,0)(lh,0), l =1, lh =0 scalar 3}
    \square_A\wtN=\ & 2\Lambda M-\frac{3}{2}\Lambda N+\left[3L^{-2}+8\wtL^{-2}+\frac{9}{2}\Lambda\right]\wtN.
\end{align}
and the scalar masses are plotted in Figure \ref{fig:l=1 eigenvalues}.  The tensors have masses
\begin{equation}\label{(l,0)(lh,0), l =1, lh =0 tensor}
    m_H^2=m_X^2=3L^{-2}.
\end{equation}
Finally, dropping to $\ell=0$, we lose the scalar $V$, and we have gauge transformations parameterized by $\al_\m^{(0,0)(\wtell,0)}$ and $\la_\mu^{(0,0)(\wtell,0)}$ with $\wtell\ge 0$, and $\g^{(0,0)(\wtell,0)}$ and $\nu^{(0,0)(\wtell,0)}$ with $\wtell\ge 1$.  For the case $\wtell\ge 2$, we can use the $\g$ transformation to set $\wtK=0$, the $\al_\mu$ transformation to set $\wtS_\mu=0$, and the $\la_\mu$ transformation to set $\wtZ_\mu=0$.  There is no residual gauge freedom (any surviving gauge freedom from $\nu$ corresponds to an exact one-form).  Equation~\eqref{eq:l0hl0hKEoM} then implies that
\begin{equation}
\label{eq:00tl0phiEqn}
    \phi=M+\frac{3}{4}N+\frac{1}{4}\wtN.
\end{equation}
This leaves a four-by-four mass matrix for the scalars, 
\begin{align}\label{(l,0)(lh,0), l =0, lh geq 2 scalar 1}
    \square_AM=\ & \left[\wtL^{-2}\wtell(\wtell+2)+3\Lambda\right]M+\frac{3}{2}\Lambda N+\left[4\wtL^{-2}+\frac{3}{2}\Lambda\right]\wtN+\wtmcL^{-1}\wtL^{-2}\wtell(\wtell+2)\wtV,\\
    \label{(l,0)(lh,0), l =0, lh geq 2 scalar 2}
    \square_AN=\ & 2\Lambda M+\left[8L^{-2}+\wtL^{-2}\wtell(\wtell+2)+\frac{9}{2}\Lambda\right]N+\hlf\Lambda\wtN,\\
    \label{(l,0)(lh,0), l =0, lh geq 2 scalar 3}
    \square_A\wtN=\ & 2\Lambda M+\frac{3}{2}\Lambda N+\left[8\wtL^{-2}+\wtL^{-2}\wtell(\wtell+2)+\frac{7}{2}\Lambda\right]\wtN+4\wtmcL^{-1}\wtL^{-2}\wtell(\wtell+2)\wtV,\\
    \label{(l,0)(lh,0), l =0, lh geq 2 scalar 4}
    \square_A\wtV=\ & 4\wtmcL^{-1}\wtN+\wtL^{-2}\wtell(\wtell+2)\wtV,
\end{align}
with the masses plotted in Figure \ref{fig:l=0 eigenvalues}. The tensors have masses
\begin{equation}\label{(l,0)(lh,0), l =0, lh geq 2 tensor}
    m_H^2=m_X^2=\wtL^{-2}\wtell(\wtell+2).
\end{equation}
The $(0,0)(1,0)$ sector is almost identical.  The $\wtK$ scalar is gone from the outset and we lose the algebraic equation that fixed $\phi$, but we can now use the $\g$ gauge freedom to impose the equation \eqref{eq:00tl0phiEqn} instead.  This still leaves a bit of gauge freedom, namely parameters $\g^{(0,0)(1,0)}$ satisfying $\square_A\g=-\wtL^{-2}\g$ (note that preserving the gauge $\wtS_\mu=0$ requires the $\g$ gauge transformation to be accompanied by a compensating transformation $\al_\m=-\nabla_\mu\g$, and hence $M$ transforms as $\d M=-\hlf\square_A\g$).  Having imposed~\eqref{eq:00tl0phiEqn}, everything proceeds as above with $\wtell=1$.  One of the eigenvalues turns out to be precisely $-\wtL^{-2}$, and can be removed by the residual gauge transformation. The remaining three scalar masses are plotted in Figure \ref{fig:l=0 eigenvalues}.  The tensor masses are given by setting $\wtell=1$ in~\eqref{(l,0)(lh,0), l =0, lh geq 2 tensor},
\begin{equation}\label{(l,0)(lh,0), l =0, lh = 1 tensor}
    m_H^2=m_X^2=3\wtL^{-2}.
\end{equation}
Finally, for the $(0,0)(0,0)$ sector we start with only the fields $\phi$, $M$, $N$, $\wtN$, $H_{\mu\nu}$, and $X_{\mu\nu}$.  The only gauge freedom is from $\al_\mu^{(0,0)(0,0)}$ and $\la_\mu^{(0,0)(0,0)}$.  We can use some of the $\alpha_\mu$ transformation to again impose~\eqref{eq:00tl0phiEqn}, leaving residual gauge transformations satisfying $\nabla^\mu\alpha_\mu=0$ and a three-by-three mass matrix
\begin{align}\label{(l,0)(lh,0), l =0, lh = 0 scalar 1}
    \square_AM=\ & 3\Lambda M+\frac{3}{2}\Lambda N+\left[4\wtL^{-2}+\frac{3}{2}\Lambda\right]\wtN,\\
    \label{(l,0)(lh,0), l =0, lh = 0 scalar 2}
    \square_AN=\ & 2\Lambda M+\left[8L^{-2}+\frac{9}{2}\Lambda\right]+\hlf\Lambda\wtN,\\
    \label{(l,0)(lh,0), l =0, lh = 0 scalar 3}
    \square_A\wtN=\ & 2\Lambda M+\frac{3}{2}\Lambda N+\left[8\wtL^{-2}+\frac{7}{2}\Lambda\right]\wtN.
\end{align}
The scalar masses are plotted in Figure \ref{fig:l=0 eigenvalues}.  The tensors are both massless
\be\label{(l,0)(lh,0), l =0, lh = 0 tensor}
m^{2}_{H}=m^{2}_{X}=0
\ee
and the residual gauge transformations from $\alpha_\mu$ and $\lambda_\mu$ are the appropriate ones for such fields. In table \ref{tab: masses for different sectors}, we give the equations that contain the expressions for the squared masses for scalars, vectors, and tensors for different sectors.
\begin{table}[ht]
    \centering
    \begin{tabular}{c|c|c|c|c}
    \textbf{Sectors}&\textbf{Subcases}& \textbf{Scalar masses}&\textbf{Vector masses}&\textbf{Tensor masses}\\
    \hline
     $(\ell,\pm 2)(\ellh,0)$& $\ell\geq 2,\;\ellh\geq 0$ & \eqref{(l,pm2)(lh,0), l geq 2, lh geq 0 scalar}  & - & -  \\
     \hline
     $(\ell,\pm 1)(\ellh,0)$& $\ell\geq 1,\;\ellh\geq 1$ & \eqref{(l,pm1)(lh,0), l geq 1, lh geq 1 scalar}  & \eqref{(l, pm1)(lh,0), l geq 1, lh geq 1 vector} & -  \\
     & $\ell\geq 2,\;\ellh=0$ & - & \eqref{(l, pm1)(lh,0), l geq 1, lh geq 1 vector} & -  \\
     & $\ell= 1,\;\ellh\geq 1$ & \eqref{(l,pm1)(lh,0), l= 1, lh geq 1 scalar}  & \eqref{(l,pm1)(lh,0), l= 1, lh geq 1 vector} & -  \\
     & $\ell= 1,\;\ellh= 0$ & - & \eqref{(l, pm1)(lh,0), l=1, lh=0 vector} & -  \\
     \hline
     $(\ell,\pm 1)(\ellh,\pm'1)$& $\ell\geq 1,\;\ellh\geq 1$ & \eqref{(l,pm1)(lh,pm'1), l geq 1, lh geq 1 scalar}* & - & -  \\
     \hline
     $(\ell,0)(\ellh,\pm2)$& $\ell\geq 0,\;\ellh\geq 2$ & \eqref{(l,pm2)(lh,0), l geq 2, lh geq 0 scalar}  & - & -  \\
     \hline
      $(\ell,0)(\ellh,\pm 1)$& $\ell\geq 1,\;\ellh\geq 2$ & \eqref{(l,0)(lh,pm1), l geq 1, lh geq 2 scalar} & \eqref{(l,0)(lh,pm1), l geq 1, lh geq 2 vector} & -  \\
     & $\ell\geq 1,\;\ellh=1$ & \eqref{(l,0)(lh,pm1), l geq 1, lh=1 scalar} & \eqref{(l,0)(lh,pm1), l geq 1, lh= 1 vector} & -  \\
     & $\ell= 0,\;\ellh\geq 2$ & - & \eqref{(l,0)(lh,pm1), l=0, lh geq 2 vector} & -  \\
     & $\ell= 0,\;\ellh= 1$ & - & \eqref{(l,0)(lh,pm1), l=0, lh= 1 vector}& -  \\
     \hline
      $(\ell,0)(\ellh,0)$& $\ell\geq 2,\;\ellh\geq 2$ & EVs \eqref{(l,0)(lh,0), l geq 2, lh geq 2 scalar 1}-\eqref{(l,0)(lh,0), l geq 2, lh geq 2 scalar 6}& \eqref{(l,0)(lh,0), l geq 2, lh geq 2 vector} & \eqref{(l,0)(lh,0), l geq 2, lh geq 2 tensor 1}, \eqref{(l,0)(lh,0), l geq 2, lh geq 2 tensor 2}  \\
      & $\ell\geq 2,\;\ellh=1$ & 5 EVs from \eqref{(l,0)(lh,0), l geq 2, lh geq 2 scalar 1}-\eqref{(l,0)(lh,0), l geq 2, lh geq 2 scalar 6}& \eqref{(l,0)(lh,0), l geq 2, lh geq 2 vector} & \eqref{(l,0)(lh,0), l geq 2, lh geq 2 tensor 1}, \eqref{(l,0)(lh,0), l geq 2, lh geq 2 tensor 2}  \\
      & $\ell\geq 2,\;\ellh=0$ & 4 EVs from \eqref{(l,0)(lh,0), l geq 2, lh geq 2 scalar 1}-\eqref{(l,0)(lh,0), l geq 2, lh geq 2 scalar 6}& \eqref{(l,0)(lh,0), l geq 2, lh geq 2 vector} & \eqref{(l,0)(lh,0), l geq 2, lh geq 2 tensor 1}, \eqref{(l,0)(lh,0), l geq 2, lh geq 2 tensor 2}  \\
     & $\ell= 1,\;\ellh\ge 2$ & 5 EVs from \eqref{(l,0)(lh,0), l geq 2, lh geq 2 scalar 1}-\eqref{(l,0)(lh,0), l geq 2, lh geq 2 scalar 6}& \eqref{(l,0)(lh,0), l geq 2, lh geq 2 vector} & \eqref{(l,0)(lh,0), l geq 2, lh geq 2 tensor 1}, \eqref{(l,0)(lh,0), l geq 2, lh geq 2 tensor 2}   \\
     & $\ell= 1,\;\ellh=1$ & EVs from \eqref{(l,0)(lh,0), l =1, lh =1 scalar 1}-\eqref{(l,0)(lh,0), l =1, lh =1 scalar 4}& \eqref{(l,0)(lh,0), l =1, lh =1 vector 1}, \eqref{(l,0)(lh,0), l =1, lh =1 vector 2} & \eqref{(l,0)(lh,0), l =1, lh =1 tensor} \\
     & $\ell= 1,\;\ellh= 0$ & EVs from \eqref{(l,0)(lh,0), l =1, lh =0 scalar 1}-\eqref{(l,0)(lh,0), l =1, lh =0 scalar 3}  & - & \eqref{(l,0)(lh,0), l =1, lh =0 tensor}  \\
     & $\ell= 0,\;\ellh\geq 2$ & EVs from \eqref{(l,0)(lh,0), l =0, lh geq 2 scalar 1}-\eqref{(l,0)(lh,0), l =0, lh geq 2 scalar 4} & - & \eqref{(l,0)(lh,0), l =0, lh geq 2 tensor}  \\
     & $\ell= 0,\;\ellh=1$ & 3 EVs from \eqref{(l,0)(lh,0), l =0, lh geq 2 scalar 1}-\eqref{(l,0)(lh,0), l =0, lh geq 2 scalar 4} & - & \eqref{(l,0)(lh,0), l =0, lh = 1 tensor}  \\ 
     & $\ell= 0,\;\ellh=0$ & EVs from \eqref{(l,0)(lh,0), l =0, lh = 0 scalar 1}-\eqref{(l,0)(lh,0), l =0, lh = 0 scalar 3} & - &  \eqref{(l,0)(lh,0), l =0, lh = 0 tensor}
    \end{tabular}
    \caption{The equations for the expressions of the squared masses for the scalars, vectors, and tensors for different sectors. All the squared masses satisfy the BF bound, except for the one with an asterisk, which violates the BF bound for $\ell,\ellh=1$ and $\pm'=\mp$. EVs is a shorthand for Eigenvalues}
    \label{tab: masses for different sectors}
\end{table}
\section{Non-perturbative stability analysis}\label{sec:non-perturbative_stability}
Having studied perturbative instabilities in detail, we now turn to non-perturbative instabilities which are typically mediated by tunneling processes. Among the possible decay channels, including Coleman-de Luccia \cite{Coleman:1980dl} vacuum bubbles and bubbles of nothing \cite{Witten:1981gj, Dibitetto:2020csn} (see also \cite{GarciaEtxebarria:2020xsr}), non-supersymmetric flux vacua are typically plagued by instabilities mediated by brane nucleation, leading to vacuum bubbles of the Brown-Teitelboim type \cite{Brown:1987dd}. These processes involve a gradual discharge of flux from the internal manifold into charged branes, which subsequently expand carrying the flux away to infinity. As such, these flux jumps interpolate between ``thin-wall'' bubbles and ``giant leaps'' of flux \cite{Brown:2010bc, Brown:2010mg}; in the limit in which all of the flux is removed, they can be interpreted as bubbles of nothing \cite{Brown:2010mf, Brown:2011gt} attached to a singular flux configuration \cite{Bomans:2021ara}.\newline\newline
The solutions studied in these paper are generalizations of the simpler Freund-Rubin vacua found in \cite{Mourad:2016xbk}, where indeed brane nucleation does occur \cite{Antonelli:2019nar}. The peculiar mechanism by which this happens is that, although the branes involved (namely D-branes for orientifold models and NS5-branes for the heterotic model) are extremal when taken as isolated objects, their tension-to-charge ratio is modified by the supersymmetry-breaking background. The effective tension becomes smaller than the charge, leading to a non-zero (albeit exponentially suppressed) decay rate per unit volume and an expanding bubble\footnote{Although not relevant for our purposes in this paper, it is worth noting that expanding bubbles in AdS or quasi-AdS backgrounds lead to stringy realizations \cite{Basile:2020mpt, Danielsson:2022lsl, Basile:2025lek} of dark bubble cosmology \cite{Banerjee:2018qey, Banerjee:2019fzz, Banerjee:2020wix, Basile:2023tvh, Basile:2025lwx}.}. This mechanism is in fact a realization of the weak-gravity conjecture for membranes \cite{Lanza:2020qmt, Lanza:2021udy}, which persists in other non-supersymmetric settings \cite{Basile:2021mkd}.\newline\newline
In the semi-classical limit, the decay rate per unit volume $\gamma \equiv \Gamma/\text{Vol}$ is dominated by the extremized action of a Euclidean brane. In this case, the fluxes present in the geometry can nucleate a NS5-brane wrapped on a three-cycle, whose homology class must therefore take the form $\Sigma_3 = a \, [S^3] + b \, [\widetilde{S}^3]$ for some non-negative integers $a,b$ (so that neither of the two fluxes increases). In the semiclassical limit, the exponent is proportional to
\begin{equation}\label{eq:log_gamma}
    - \log \gamma \sim S_\text{NS5}^E|_\text{min} = \frac{1}{(2\pi)^5\alpha'^3 g_s^2} \int d^6\sigma \sqrt{-g} - \frac{1}{(2\pi)^5\alpha'^3} \int B_6 \, ,
\end{equation}
the (minimized) Euclidean effective action of the NS5-brane worldvolume (parametrized by local coordinates $\sigma$). Here $B_6$ denotes the local two-form gauge potential dual to $B_2$, defined such that its curvature $H_7 = dB_6$ satisfies $H_7 = g_s^{-2} \star H_3$ (the factor of $g_s$ comes from the kinetic term). The background which minimizes the action within the homology class $\Sigma_3$ must minimize the volume of the embedding three-cycle. To see this, we first observe that the Chern-Simons term is invariant under continuous deformations of the embedding cycle. Indeed, on the one hand, the dual curvature $H_7 = dB_6$ has support in the full (Euclidean) $AdS_4$, since $H_3$ is only supported on the internal space. On the other hand, the difference between Chern-Simons terms evaluated on homologous three-cycles (times the external surface of a bubble in $AdS_4$) is given by the integral of $H_7$ over a bounding four-cycle times the bubble surface, which therefore vanishes. As for the DBI term, once again the contribution of external bubble surface only depends on the nucleation radius and factorizes from the internal volume, which is minimized independently. This problem is amenable to the theory of \emph{calibrations} (see e.g. \cite{Gauntlett:2003di, Koerber:2005qi, Martucci:2011dn} for its appearance in string compactifications), which was recently employed in the study of non-supersymmetric solutions and their instabilities \cite{Giri:2021eob, Menet:2023rnt, Menet:2025nbf}. As we show in Appendix \ref{app:calibration}, it turns out that the embedding three-cycle which minimizes the volume in the homology class $\Sigma_3$ is simply $\Gamma_3 = a \, S^3 + b \, \widetilde{S}^3$.

In order to evaluate it in the background, we first proceed in Poincaré coordinates $(z,x^\mu)$ for $AdS_4$, in which the (string-frame) metric reads
\begin{equation}\label{eq:10d_metric_poincare}
    ds^2_{10} = L_A^2 \, \frac{dz^2+dx^2_{2,1}}{z^2} + L^2 d\Omega_3^2 + \widetilde{L}^2 d\widetilde{\Omega}_3^2 \, .
\end{equation}
The embedding coordinates $\sigma$ are chosen to align with $x^\mu$ and cover the three-cycle $\Gamma_3$ at fixed $z$. Pulling the above metric back to the worldvolume, one obtains the DBI term
\begin{equation}\label{eq:DBI_NS5}
    \frac{1}{g_s^2} \Omega_3 (a L^3 + b \widetilde{L}^3) \frac{L_{A}^3}{z^3} \, ,
\end{equation}
where $\Omega_3 = 2\pi^2$ is the volume of the unit three-sphere. The Chern-Simons term, in the normalization such that the naked tension is equal to the charge at $g_s=1$, can be evaluated by Stokes' theorem, obtaining the integral of the curvature $H_7$ over the volume enclosed by the brane. The result is
\begin{equation}\label{eq:CS_NS5}
    -\frac{2\alpha'}{g_s^2} \Omega_3\left(a \, n \left(\frac{\widetilde{L}}{L}\right)^3 + b \, \widetilde{n} \left(\frac{L}{\widetilde{L}}\right)^3\right) \frac{L_{A}^4}{3 z^3} \, ,
\end{equation}
which (in the Lorentzian picture) shows that the brane is subject to a potential controlled by a competition of tension and charge effects. When the brane is extremal, in the sense that its naked tension equals its charge, the resulting force vanishes in supersymmetric backgrounds. However, non-supersymmetric backgrounds effectively renormalize the charge-to-tension extremality ratio of the brane, leading to a non-vanishing force \cite{Antonelli:2019nar, Basile:2021mkd} between the brane and the background. Since the latter is sourced by fluxes, the analysis of \cite{Antonelli:2019nar, Basile:2021mkd} shows that this force can be interpreted as the interaction between the probe brane and the stack whose near-horizon throat produces the geometry in \eqref{eq:10d_metric_poincare}. From \eqref{eq:DBI_NS5} and \eqref{eq:CS_NS5}, it follows that the effective charge-to-tension ratio is
\begin{equation}\label{beta exp}
    \beta=\frac{a \, n \left(\frac{\widetilde{L}}{L}\right)^3 + b \, \widetilde{n} \left(\frac{L}{\widetilde{L}}\right)^3}{a L^3 + b \widetilde{L}^3} \frac{2\alpha'L_{A}}{3} \overset{\widetilde{n} \gg n \gg 1}{\sim} \sqrt{\frac{2}{3}} \frac{4a \widetilde{n}+b n}{2b \widetilde{n}+a n} \, ,
\end{equation}
where in the last step we made use of the closed-form asymptotics for $\widetilde{n} \gg n \gg 1$ \cite{Basile:2020xwi}. The consistency of the above semiclassical computation requires that $\beta$ be of order one in the large-flux limit. We will indeed verify below that this is the case. Moreover, the decay occurs for $\beta > 1$ \cite{Antonelli:2019nar}. From \eqref{eq:DBI_NS5} and \eqref{eq:CS_NS5}, one can glean some physical intuition for this: when $\beta>1$, the repulsive force from the Chern-Simons term overcomes the attractive force from the DBI term. From this perspective, the requirement that $\beta \geq 1$ can be understood as a form of weak gravity conjecture for membranes \cite{Lanza:2020qmt, Basile:2021mkd, Lanza:2021udy}, which has been shown to hold in other non-supersymmetric settings as well \cite{Aparici:2025kjj}. \newline\newline
While sufficient to derive the value of the effective extremality ratio $\beta$, from the perspective of the instanton calculation of the decay rate, Poincaré coordinates are not enough; we need to take into account that the Euclidean brane forms a bubble of finite proper radius $\rho$ in hyperbolic space. The upshot is that the $z^{-3}$ terms in \eqref{eq:DBI_NS5} and \eqref{eq:CS_NS5} are replaced by the area and enclosed volume of a sphere in hyperbolic space \cite{Antonelli:2019nar}, and the Euclidean action in \eqref{eq:log_gamma} is minimized at radius $\rho = L_A / \sqrt{\beta^2-1}$ if $\beta > 1$; otherwise, the action is unbounded and the decay rate is driven to zero, at least within the approximations we made. When $\beta > 1$, the minimized action carries the flux prefactors from \eqref{eq:DBI_NS5} and \eqref{eq:CS_NS5}, times a dimensionless hypergeometric function $B$ involving $\beta$ \cite{Antonelli:2019nar}. The former arises because the tension of the brane must multiply a combination of $AdS_4$ and sphere radii for dimensional reasons, whereas the latter is given by 
\begin{equation}\label{eq:B_factor}
    \begin{aligned}
    B & = \frac{1}{(\beta^2-1)^\frac{p+1}{2}} - \frac{p+1}{2}\beta \int^{\frac{1}{\beta^2-1}}_0 \frac{u^\frac{p}{2}}{\sqrt{1+u}} du \\
    & = \frac{1}{(\beta^2-1)^\frac{p+1}{2}} - \frac{(p+1)\beta}{2(p+2)(\beta^2-1)^{\frac{p}{2}+1}} \, {}_2F_1\lp\frac{1}{2},\frac{p}{2}+1;\frac{p}{2}+2;\frac{1}{1-\beta^2}\rp
    \end{aligned}
\end{equation}
for a general $p$-brane \cite{Antonelli:2019nar}. \newline\newline
Let us first analyze \eqref{beta exp} in the simple case in which $\widetilde{n} \gg n \gg 1$. For $ab \neq 0$ the ratio asymptotes to approximately $1.63 \frac{a}{b}$, which may or may not be larger than 1. Moreover, for $a=0$ the result is proportional to $n/\widetilde{n} \ll 1$ which suppresses the decay, while for $b=0$ the result is instead proportional to the reciprocal ratio. In this case the semiclassical approximation would seem to break down. Does this mean that there exist NS5-brane nucleation channels that are not suppressed? To see what happens, we evaluate \eqref{eq:B_factor} for large $\beta$. In our case $p=2$, since we have an effective two-brane in $AdS_4$, and the integral is $\sim \beta^{-3} \propto \frac{n^3}{\widetilde{n}^3}$. The flux prefactor in $\log \gamma$ is then $L_A^3 L^3/g_s^2 \sim n^4$ leads to an expression which may not be large, depending on whether $\widetilde{n}^3 \ll n^7$; if not, the semi-classical approximation does seem to break down. This means that the decay process, absent some unexpected cancellations, occurs on timescales of the order of $\sqrt{\alpha'}$.\newline\newline
The opposite regime is when fluxes are equal i.e. $n = \widetilde{n} $. Using \eqref{equal flux Lo Loh}, the expression for $\beta$ simplifies to
\begin{equation}
    \beta = \frac{2\alpha'n L_A}{3L^3} = \sqrt{\frac{5}{3}} \, .
\end{equation}
This is of order one and greater than one (in fact, it is the same value found for $AdS_7 \times S^3$ in \cite{Antonelli:2019nar}), meaning that the decay occurs but is semiclassically slow. Concretely, the decay rate per unit volume is exponentially suppressed in a power of the flux number, but due the unbounded volume density toward the conformal boundary of $AdS_4$ nucleation happens immediately at the boundary, reaching a bulk observer in a timescale up to $\mathcal{O}(L_A)$, which is nevertheless much larger than the cutoff of the effective theory. \newline

\begin{figure}[!ht]
    \centering
    \includegraphics[width=0.8\linewidth]{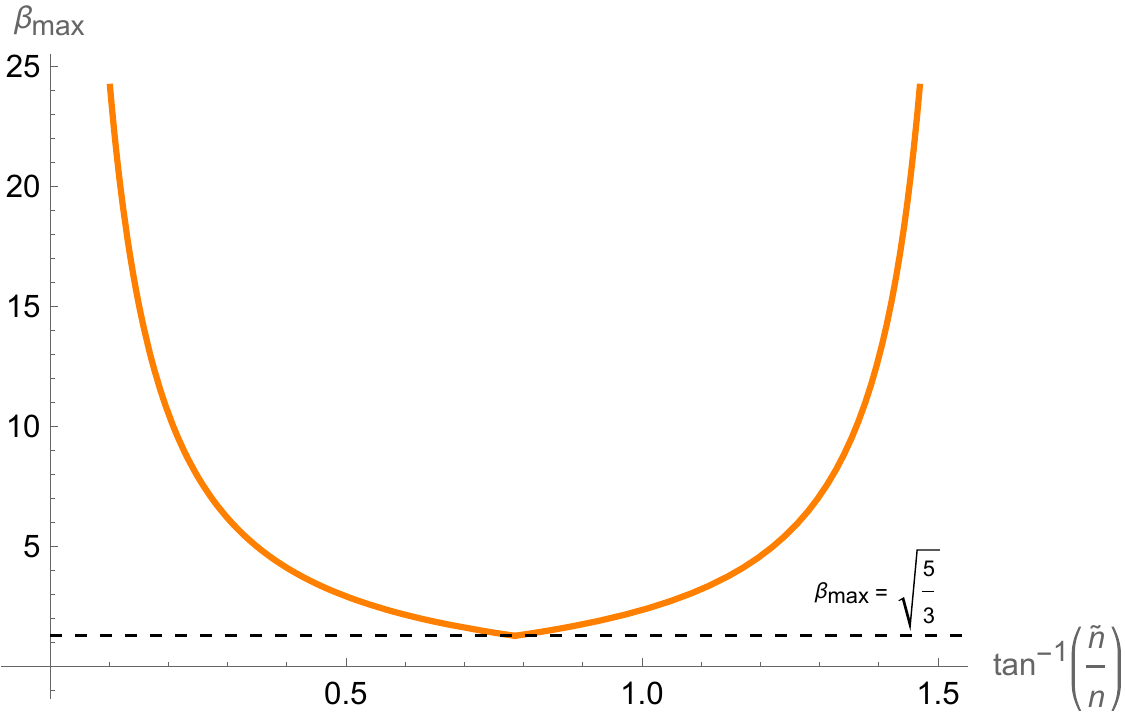}
    \caption{Plot of the extremality parameter $\beta_\text{max}$ extremized over wrapping cycles as a function of the direction in flux space. One can observe that the decay process becomes fast when a hierarchy between fluxes is present. The minimal value, attained only for $n = \widetilde{n}$, is $\sqrt{5/3}$.}
    \label{fig: betamax plot}
\end{figure}

For proportional but unequal fluxes, one may worry that numerics be required, since the general solution for the geometry is not explicit. However, since the radii of the solution scale equally when both fluxes are co-scaled, the extremality ratio $\beta$ only depends on the ratio $\widetilde{n}/n$. In order to probe the dominant decay channel, one ought to maximize $\beta$ with respect to the choice $(a,b)$ of wrapping cycle, yielding the maximal extremality parameter $\beta_\text{max}(\widetilde{n}/n)$. Thanks to these simplifications, these quantities turn out to afford closed-form analytic expressions. In particular, letting
\be
\xi=\frac{\widetilde{L}}{L}\efill \zeta=\frac{a}{b}\efill \m:=\frac{\widetilde{n}}{n}=\xi^3\sqrt{\frac{4+\xi^2}{1+4\xi^2}} \, ,
\ee
one finds
\be\label{beta exp ratios}
\beta=\sqrt{\frac{2}{3}}\frac{\zeta \xi^3\sqrt{1+4\xi^2}+\sqrt{4+\xi^2}}{(\zeta+\xi^3)\sqrt{1+\xi^2}} \, .
\ee
One can, in principle, eliminate $\xi$ in the expression for $\beta$ in favor of $\m$ using the relation between $\m$ and $\xi$ given above. One can also see that $\beta$ is invariant under $\xi\rr \xi^{-1}$, which manifests the exchange symmetry between the two three-spheres. Treating $\zeta$ formally as a continuous parameter, the derivative of $\beta$ with respect to it yields
\be
\frac{\p \beta}{\p \zeta}=\frac{1}{(\zeta+\xi^{3})^{2}}\sqrt{\frac{2}{3(1+\xi^2)}}\ls \xi^6\sqrt{1+4\xi^2}-\sqrt{4+\xi^2}\rs .
\ee
Since the factor in the square brackets is positive for $\xi>1$ and negative for $\xi<1$, $\beta$ is a monotonically increasing or decreasing function of $\zeta$ depending on whether $\xi>1$ or $\xi<1$. If $\xi>1$, $\beta$ increases until it reaches a maximum value as $\zeta\rr\infty$. If $\xi<1$, $\beta$ has a maximum value at $\zeta=0$. Therefore, we have the following result:
$$
\beta_{\text{max}}=\begin{cases}\xi^{3}\sqrt{\frac{2(1+4\xi^2)}{3(1+\xi^2)}}\efill (\xi\geq 1) \, , \\
\frac{1}{\xi^{3}}\sqrt{\frac{2(4+\xi^2)}{3(1+\xi^2)}}\efill (\xi\leq 1) \, .
\end{cases}
$$
The plot in Figure \ref{fig: betamax plot} displays the interpolation between the two limiting cases we studied analytically. The decay process stochastically explores all allowed channels, but when $n \neq \widetilde{n}$ the optimal brane instantons exclusively wrap the cycle threaded by the largest flux. We conclude that the most likely chain of decay processes, at least within our semiclassical analysis, evolves the system toward equal fluxes. As a final comment, in order for the analysis in this section to be relevant, tachyonic modes ought to be projected out first. As in \cite{Basile:2018irz}, this is possible via smooth orbifolds of the internal spheres; for instance, an antipodal $\mathbb{Z}_2$ projection on one of the spheres removes the offending mode, which has odd angular momenta. This operation halves the volume of one of the two factors; however, as we have seen above, this effectively amounts to a rescaling of $\zeta$, which drops out of our expression after maximizing $\beta$.

\section{Discussion and outlook}
\label{sec:Discussion}

In this paper we carried on the search for (meta)stable non-supersymmetric solutions of string theory. Specifically, we studied settings where supersymmetry is absent or broken at the string scale, the simplest of which give rise to effective theories in ten dimensions \cite{Alvarez-Gaume:1986ghj, Dixon:1986iz, Sagnotti:1995ga, Sagnotti:1996qj, Sugimoto:1999tx}. The presence of an exponential potential for the dilaton raises the concern of whether solutions can exist in the relevant regime of validity, i.e. large negative dilaton and small curvatures in string units. Solutions of this type do exist \cite{Mourad:2016xbk, Baykara:2022cwj, Basile:2020xwi, Raucci:2025bev}, thanks to fluxes preventing the strongly warped regimes exhibited by simpler solutions \cite{Dudas:2000ff, Mourad:2021roa, Basile:2022ypo, Raucci:2022jgw, Raucci:2022bjw, Raucci:2023xgx, Mourad:2023loc, Mourad:2024dur, Mourad:2024mpg}. Although generically these solutions are perturbatively unstable \cite{Gubser:2001zr, Basile:2018irz, Fraiman:2023cpa, Robbins:2025wlm}, it is possible to select the internal manifold to remove the offending tachyons \cite{Basile:2018irz} and yield metastable vacua which decay via brane nucleation \cite{Antonelli:2019nar}. Beyond their clear relevance for the non-supersymmetric string landscape, these settings offer a possibility to explore $AdS$ holography of a flavor different from standard $AdS/$CFT \cite{Ooguri:2016pdq, Freivogel:2016qwc, Antonelli:2018qwz, Ghosh:2021lua, Basile:2022zee}.

The case of $AdS_4 \times S^3 \times \widetilde{S}^3$ examined in this paper proved to be richer than its simpler counterparts, due to the presence of two three-form fluxes threading the spheres. Still, the tachyonic spectrum (which only arises for comparable fluxes) is simple enough that it is possible to identify a suitable projection, as in \cite{Basile:2018irz}. Indeed, since both angular momentum quantum numbers are odd for the tachyonic perturbations, an antipodal $\mathbb{Z}_2$ projection of one of the three-spheres removes them. This modifies the global geometry to, say, $S^3 \times \mathbb{R}P^3$. Alternatives are internal manifolds whose Laplacian gaps are large enough to go above the tachyonic modes: insofar as they carry Einstein metrics, the local analysis remains unaffected. The upshot of the combined perturbative and non-perturbative analysis leads us to a richer story for the evolution of the $AdS_4 \times S^3 \times \widetilde{S}^3$ compared to simpler settings: even if the two fluxes are hierarchically different, brane nucleation will tend to bring them closer, and then tachyonic instabilities take over (unless they are projected away at the level of the background). To our knowledge, this is the first explicit construction where this peculiar combination of effects occurs.

Another interesting generalization would be to replace one of the spheres with a compact hyperbolic manifold\footnote{We would like to thank an anonymous SciPost referee for this suggestion.}. This case has been discussed in \cite{Raucci:2025bev}.  Replacing both spheres does not lead to a solution (we need at least some positive curvature contribution), but there are solutions with one sphere and one hyperbolic manifold.  The perturbative stability analysis would depend on the details of the space (and Mostow rigidity might have a role to play), but we expect the non-perturbative analysis to be broadly similar.  It would be interesting to explore this class of solutions further in future work.

Various directions to expand our analysis are possible. Firstly, the geometry can \mbox{feature} toroidal factors. In particular, we expect similar richness from solutions of the form $AdS_3 \times S^3$ $\times \widetilde{S}^3 \times S^1$, where there is an additional radion modulus to take into account. More generally, toroidal moduli significantly complicate the non-perturbative stability analysis, since one has to study the full potential to identify possible local minima and compare tunneling rates within the field space to brane nucleation rates. Another direction of interest is warped compactifications, which introduce significant technical complications but potentially open up a wide new class of non-supersymmetric solutions. In this respect, the no-go theorem of \cite{Basile:2020mpt} prohibits $dS$ solutions, but it is conceivable that non-trivial $AdS$ ones exist. More broadly, an outstanding question is how to concretely compute high-energy properties, say string scattering amplitudes, around non-supersymmetric vacua of this type. In principle, the Fischler-Susskind mechanism ought to reorganize the string-loop expansion around the new quantum vacuum and give rise to well-defined perturbative observables, especially in the heterotic setting where only Kalb-Ramond flux is present. In practice, it seems likely that this will require techniques borne out of string field theory.

Ultimately, the core theoretical question behind these investigation is the existence and consistency of (metastable) non-supersymmetric vacua. Their lifetime seems to be bounded by the $AdS$ timescale, but the detailed physics of their high-energy excitations is largely unknown. If low-energy supersymmetry turns out not to be realized in our universe, the understanding of these signatures, and how they relate to more realistic models with positive dark energy and four-dimensional physics, may be an essential ingredient in connecting string theory to observations.

\section*{Acknowledgments}
The authors would like to thank Oleg Lunin and Jacob McNamara for useful discussions. The work of IB is supported by the Origins Excellence Cluster and the German-Israel-Project (DIP) on Holography and the Swampland.


\appendix

\section{Minimal-volume representative for wrapped NS5-branes}\label{app:calibration}

In this appendix we complement our computation of the decay rate of our $AdS_4 \times S^3 \times \widetilde{S}^3$ solution via NS5-brane nucleation by proving that the embedding cycle we used actually minimizes the volume within its homology class. The wrapping data for an NS5-brane wrapping a three-cycle in $S^3 \times \widetilde{S}^3$ is encoded in two (non-negative\footnote{We recall that, in order for the decay to occur, the three-form fluxes threading each sphere must not increase.}) integers, or equivalently in the homology class
\begin{equation}
    \Sigma_3 = a[S^3] + b[\widetilde{S}^3] \in H_3(S^3 \times \widetilde{S}^3,\mathbb{Z}) \simeq \mathbb{Z} \oplus \mathbb{Z} \, .
\end{equation}
In the main text, we used the fact that the embedding cycle of the brane which represents this homology class must minimize the Euclidean action, which in turn means minimizing volume (since the Chern-Simons term is unaffected by deformations of representatives in $\Sigma_3$). In this appendix we show that the embedding cycle we used, i.e. the straightforward concatenation $a \, S^3 \times \text{pt} + b \, \text{pt} \times \widetilde{S}^3$ of the two sphere factors, does indeed minimize volume within the homology class $\Sigma_3$. This actually remains true replacing the dimension 3 of each factor with any $n > 1$, whereas for $n=1$ we obtain the known result that the volume of least-volume one-cycles in a two-torus is given by a sum-in-quadratures of the form $\sqrt{a^2 L^2+b^2 \widetilde{L}^2}$ with $L, \widetilde{L}$ the respective radii. To this end, we make use of the notion of \emph{calibration} \cite{Gauntlett:2003di, Koerber:2005qi, Martucci:2011dn,Giri:2021eob, Menet:2023rnt, Menet:2025nbf}: a closed $k$-form $\alpha \in Z^k(M)$ on a Riemannian manifold $(M,g)$ is a calibration if at any point $p\in M$ we have $\alpha_p(v_1,\dots,v_k) \leq 1$ for any orthonormal set $\{v_1,\dots,v_k\} \subset T_pM$. An important property of calibrations is that \emph{calibrated manifolds}, i.e. submanifolds of $M$ whose volume form is the pullback of $\alpha$, minimize volume within their homology class. Indeed, letting $j: N \hookrightarrow M$ be $\alpha$-calibrated and $N' \in [N] \in H_k(M)$ we have $j^*\alpha = \text{vol}_N$, and thus
\begin{equation}
    \text{Vol}(N) = \int_N \alpha = \int_{N'} \alpha \leq \int_{N'} \text{vol}_{N'} = \text{Vol}(N') \, .
\end{equation}
Thus, in our case where $M = S^n \times \widetilde{S}^n$ and $k=n$, we only need prove that $\alpha \equiv \text{vol}_{S^n} + \text{vol}_{\widetilde{S}^n}$ is a calibration. \newline

Let $\{e_i\}_{i=1,\dots, n}$ (resp. $\{\widetilde{e}_i\}_{i=1,\dots, n}$) denote orthonormal bases for the tangent spaces of $S^n$ (resp. $\widetilde{S}^n$) at some point $p$. Then, a general orthonormal set $\{ v_i \}_{i=1,\dots, n}$ takes the form
\begin{equation}
    v_i = \sum_{j=1}^n \left(A_{ij} e_j + B_{ij} \widetilde{e}_j\right) ,
\end{equation}
where the $n \times n$ matrices $A=(A_{ij})$ and $B=(B_{ij})$ satisfy the orthonormality condition
\begin{equation}
    A^T A+B^T B = I \, .
\end{equation}
From this condition, it follows that the eigenvalues of $A$ and $B$ are bounded by 1 in absolute value. Indeed, letting $\lambda$ be a (possibly complex) eigenvalue of $A$ with some unit-normalized (possibly complex) eigenvector $v$, we have
\begin{equation}
    |\lambda|^2 = v^\dagger A^T A v = 1 - v^\dagger B^T B v \leq 1 \, ,
\end{equation}
where we used the fact that $A^T = A^\dagger$ since $A$ is real (and analogously for $B$). Therefore, using the fact that $|\det B| = \sqrt{\det B^T B}$, we find
\begin{equation}
    |\alpha_p(v_1,\dots,v_n)| = |\det A+ \det B| \leq |\det A|+\sqrt{\det(I-A^TA)} = \prod_{i=1}^n |\lambda_i| + \prod_{i=1}^n \sqrt{1 - |\lambda_i|^2} \, ,
\end{equation}
where the product is taken over the $n$ (possibly identical) eigenvalues of $A$. Letting $|\lambda_i| \equiv |\cos \theta_i|$, one can easily see that for $n>1$
\begin{equation}\label{eq:cos_sin}
    \prod_{i=1}^n |\cos \theta_i| + \prod_{i=1}^n |\sin \theta_i| \leq 1 \, ,
\end{equation}
whereas for $n=1$ the left-hand side can achieve an upper bound of $\sqrt{2}$. To see this, notice that for $n=2$ one can write \eqref{eq:cos_sin} as the inner product of the two-component vectors $u_i=(|\cos \theta_i|,|\sin\theta_i|)$, $i=1,2$. This inner product is bounded by the Cauchy-Schwarz inequality in $\mathbb{R}^2$ according to $v_1 \cdot v_2 \leq \|v_1\| \, \|v_2\| = 1$. By induction on $n$, assuming that \eqref{eq:cos_sin} is bounded by 1 for $n \leq N$, the expression with $n=N+1$ can be bounded by that with $n=N$ by bounding $|\cos \theta_{N+1}|, |\sin \theta_{N+1}| \leq 1$. This proves the claim: $\alpha$ is a calibration and the embedding cycle, which is $\alpha$-calibrated, thus minimizes volume within its homology class $\Sigma_3$.

\printbibliography
\end{document}